\documentclass[
    aps,
    pra,
    reprint,
    superscriptaddress,
    longbibliography,
    nofootinbib,
    amsmath,
    amssymb,
    floatfix
]{revtex4-2}

\usepackage{graphicx}
\usepackage{xcolor}
\usepackage{bm}
\usepackage{booktabs}
\usepackage{siunitx}
\usepackage{hyperref}
\usepackage[capitalise]{cleveref}
\usepackage{orcidlink}
\graphicspath{{./}}

\crefname{equation}{Eq.}{Eqs.}
\Crefname{equation}{Eq.}{Eqs.}
\crefname{section}{Sec.}{Secs.}
\Crefname{section}{Sec.}{Secs.}
\crefname{subsection}{Sec.}{Secs.}
\Crefname{subsection}{Sec.}{Secs.}
\crefname{subsubsection}{Sec.}{Secs.}
\Crefname{subsubsection}{Sec.}{Secs.}
\crefname{figure}{Fig.}{Figs.}
\Crefname{figure}{Fig.}{Figs.}
\crefname{table}{Table}{Tables}
\Crefname{table}{Table}{Tables}
\crefname{appendix}{Appendix}{Appendices}
\Crefname{appendix}{Appendix}{Appendices}

\definecolor{orangered}{HTML}{FF4500}
\definecolor{crimson}{HTML}{DC143C}
\definecolor{rossoferrari}{HTML}{D9073D}
\definecolor{steelblue}{HTML}{4682B4}
\definecolor{mediumblue}{HTML}{0000CD}
\definecolor{forestgreen}{HTML}{228B22}
\hypersetup{
setpagesize=false,
bookmarksnumbered=true,
urlcolor=steelblue,
citecolor=steelblue,
}

\allowdisplaybreaks[1]

\newcommand{\dd}{\mathrm{d}}
\newcommand{\blambda}{\bm{\lambda}}
\newcommand{\btheta}{\bm{\theta}}
\newcommand{\vaphi}{\varphi}
\newcommand{\E}{\mathbb{E}}
\newcommand{\Prob}{\mathbb{P}}
\newcommand{\Bern}{\mathrm{Bern}}

\begin{document}

\title{Adaptive detection of Rabi signals under composite hypotheses}

\author{So Chigusa\,\orcidlink{0000-0001-6005-4447}}
\affiliation{Center for Theoretical Physics -- a Leinweber Institute, Massachusetts Institute of Technology,\\
77 Massachusetts Avenue, Cambridge, MA 02139, USA}

\begin{abstract}
Motivated by searches for weak coherent drives, we formulate repeated quantum sensing with a fixed shot budget as an asymmetric composite hypothesis test.
Taking Rabi sensing as a concrete example, we benchmark detection power, sensitivity, and Type-II error exponents in the resonant case with unknown signal amplitude and phase.
We compare non-adaptive population and transverse readouts with a myopic Bayesian policy that selects each projective axis by maximizing the expected information gain in one step.
The common decision statistic is a log Bayes factor, with a policy specific threshold calibrated under the null to enforce a common Type-I error.
A weak signal expansion shows that population readout is phase independent but quadratic in amplitude, giving $n^{-1/4}$ sensitivity, whereas transverse readout is linear in amplitude and permits $n^{-1/2}$ sensitivity without adaptation, but is phase-dependent.
In Monte Carlo pseudoexperiments, the adaptive policy exploits posterior information about the unknown direction to guide subsequent readouts; its sensitivity is consistent with $n^{-1/2}$ over the simulated large $n$ range and, at the largest simulated shot counts, outperforms the fixed transverse schedules.
Its phase-averaged effective Type-II exponent also exceeds the non-adaptive references over the simulated range.
These results demonstrate the finite budget value of exploiting nuisance parameter information under calibrated false positive control.
\end{abstract}

\maketitle

\begingroup
\small
\makeatletter
\let\l@subsubsection\@gobbletwo
\makeatother
\tableofcontents
\endgroup

\section{Introduction}
\label{sec:intro}

Quantum sensing uses controllable quantum systems to probe quantities ranging from electromagnetic fields and frequencies to forces and material properties~\cite{DegenEtAl:2017,PirandolaEtAl:2018}.
Much of quantum metrology is formulated as parameter estimation around a known operating point, but some sensing tasks instead ask whether a weak signal is present at all.
Prominent examples include target detection by quantum illumination~\cite{Lloyd:2008nwf,Tan:2008oho} and searches for weakly coupled new physics, including dark matter, with quantum sensors~\cite{ChouEtAl:2023QuantumSensors}.
We focus on the detection of weak coherent drives acting on a two-level sensor, motivated in part by wave-like dark matter searches.

Signal detection is naturally formulated as hypothesis testing.
One approach is asymmetric testing, which fixes the Type-I error (false positive probability) and minimizes the Type-II error (false negative probability), or equivalently maximizes detection power.
When the signal model contains unknown parameters such as an amplitude, frequency, or phase, the alternative hypothesis is composite rather than simple.
In a frequentist search, allowing a range of distinguishable signal models gives background fluctuations more opportunities to resemble a signal, producing a look elsewhere effect~\cite{GrossVitells:2010}.
In Bayesian model comparison, averaging over the alternative prior introduces an Occam penalty when a good fit is confined to a small fraction of its prior mass~\cite{KassRaftery:1995}.
The unknown parameters can also determine which measurements are most informative, linking signal detection to parameter learning.

In a protocol with repeated measurements, this suggests using information gained from earlier outcomes to guide subsequent measurement choices.
Posterior-based control selection is used in Bayesian Hamiltonian learning~\cite{GranadeEtAl:2012}, and real-time adaptation of the readout basis has been demonstrated in single-spin magnetometry~\cite{BonatoEtAl:2016}.
We investigate this adaptive approach with a prescribed budget of $n$ shots.
Each shot prepares a fresh qubit, lets it interact with the signal, measures it, and resets it.
The quantum state is not retained between shots, but the accumulated classical history can be used to update the signal posterior and select later readouts.
Information can thus accumulate over a total duration that greatly exceeds the coherence time of a single probe, with the final decision made after all $n$ shots.

Classical controlled sensing with a fixed sample size has been studied for finite hypothesis sets~\cite{NitinawaratAtiaVeeravalli:2013}, with finite sample bounds and asymptotically optimal strategies established for asymmetric formulations~\cite{KartikEtAl:2022}.
For quantum hypothesis testing with a fixed copy budget, adaptive single copy projective measurements attain the collective minimum error for two known pure states~\cite{AcinEtAl:2005}.
For two known mixed qubit states, optimizing the full adaptive policy can outperform greedy measurement choices at finite copy number~\cite{PhysRevLett.103.220503}, though adaptive gains over optimized fixed local measurements need not extend to the asymptotic error exponent~\cite{HigginsEtAl:2011}.
Adaptive field-presence detection has also been studied under specified Gaussian and non-Gaussian noise models within simple minimum error testing~\cite{WangLiZou:2017}.
For composite alternatives, an initial estimate of the unknown state can determine a subsequent collective asymmetric test~\cite{Fujiki:2025okc}.

Related works consider adaptive testing with different resource assumptions.
Sequential tests allow the sample size to depend on the observed evidence, in both classical controlled sensing~\cite{Chernoff:1959,NitinawaratAtiaVeeravalli:2013} and quantum state discrimination~\cite{MartinezVargasEtAl:2021,LiTanTomamichel:2022}.
For composite hypotheses, this includes tests valid at any stopping time~\cite{ZecchinEtAl:2025} and exponent analyses under expected sample size constraints~\cite{SimpsonEtAl:2026}.
Quantum channel discrimination permits optimization over inputs and intermediate operations, with adaptive advantages depending on the channel family and performance criterion~\cite{HarrowEtAl:2010,SalekHayashiWinter:2022,BerghDattaSalzmann:2023}.

Posterior-based readout adaptation can be applied to signal detection in a range of sensing models.
We focus here on Rabi sensing of a weak transverse drive near resonance, with fixed probe preparation, interrogation time, and shot budget, and adapt only the readout axis.
The unknown parameters are the resonant Rabi angle $\Phi_R$, the detuning $\Delta$, and the transverse phase $\tilde{\varphi}$ that determines the signal direction in the rotating frame.
For a wave-like dark matter signal, these quantities encode the coupling-weighted field amplitude, the mismatch between the dark matter oscillation frequency and the sensor reference frequency, and the local field phase, respectively~\cite{BudkerEtAl2014,FierlingerEtAl2024,ChenEtAl2022HiddenPhoton,Chigusa:2025rqs}.
A recent review of qubit dark matter searches discusses hypothesis testing and highlights the role of phase coherence in comparing population and transverse readout~\cite{Moroi:2026ges}.
The fixed-$\btheta$ model represents one coherent segment over which the amplitude, detuning, and initial phase are treated as constant; integrating beyond the physical field coherence time would require a blockwise or stochastic latent signal model~\cite{CentersEtAl:2021}.
Population readout is insensitive to the unknown phase but responds only quadratically to a weak drive, whereas an aligned transverse measurement has a linear response but requires the informative quadrature to be learned.
We ask whether learning this nuisance direction can improve detection power at a fixed Type-I error.

To address this question, we develop posterior-based readout policies and compare detection power using Bayes factor thresholds calibrated to the same Type-I error.
We establish quantitative power and sensitivity benchmarks on resonance ($\Delta=0$), complemented by a detuned trajectory illustrating phase tracking and a discussion of broadband search penalties.
A calibrated power study for a broadband search remains for future work.
The weak signal analysis gives $n^{-1/4}$ sensitivity for population readout and shows that $n^{-1/2}$ sensitivity is attainable with non-adaptive transverse measurements.
At the largest simulated shot budgets, adaptive readout improves sensitivity over the fixed transverse baselines studied, with scaling consistent with $n^{-1/2}$.

The remainder of the paper is organized as follows.
\cref{sec:setup} defines the Rabi response and composite test with a fixed shot budget, \cref{sec:policies} introduces the measurement policies, \cref{sec:analytic} develops analytic benchmarks, \cref{sec:numerical_results} presents the numerical results, \cref{sec:discussion} discusses limitations and extensions, and \cref{sec:conclusion} concludes the paper.
\cref{app:hypothesis_testing} reviews the information theoretic background, and \cref{app:rabi_crossing} derives the weak signal Rabi benchmarks.
\cref{app:more_rabi_plots} provides additional numerical diagnostics.

\section{Rabi sensing as a composite hypothesis testing}
\label{sec:setup}

Rabi sensing provides a simple setting in which a weak coherent drive must be detected even though its amplitude, detuning, and phase are not known in advance.
Wave-like dark matter searches provide one motivation, with related resonant excitation schemes proposed for neutron spins, transmon qubits, and optically trapped Rydberg atoms, together with quantum-enhanced multisensor extensions~\cite{FierlingerEtAl2024,ChenEtAl2022HiddenPhoton,ChenEtAl2024,Chigusa:2025rqs,ChenEtAl2023QuantumEnhancement,SichanugristEtAl2024}.
The model below isolates the Rabi limit of a single two-level sensor: the signal is a weak transverse drive near resonance, its parameters are unknown, and the detection statistic is defined below.

\subsection{Rabi dynamics and readout probabilities}
\label{subsec:rabi_model}

We model a two-level sensor exposed to a weak classical field near resonance.
In each shot, the qubit is initialized in its ground state $\rho_0$, allowed to interact with the field for a fixed time $T$, and then measured along a controllable readout direction.
The $z$ axis is the qubit energy axis, and the $xy$ plane is the transverse plane in which the drive acts.
The bare qubit Hamiltonian is
\begin{align}
    H_{\rm q}
    &=
    -\frac{\omega_{\rm q}}{2}\sigma_z,
    \label{eq:bare_qubit_hamiltonian}
\end{align}
where $\omega_{\rm q}>0$ is the qubit transition angular frequency and $\sigma_x,\sigma_y,\sigma_z$ are the Pauli matrices.
Taking the signal polarization to define the $x$ axis, we write the laboratory frame signal Hamiltonian at absolute time $\tau$ as
\begin{align}
    H_{\rm sig}(\tau)
    &=
    \Omega\cos(\omega_{\rm s}\tau-\tilde{\varphi})\,\sigma_x,
    \label{eq:lab_signal_hamiltonian}
\end{align}
where $\omega_{\rm s}$ is the signal angular frequency, $\tilde{\varphi}$ specifies its phase at $\tau=0$, and the on-resonance Rabi angular frequency $\Omega$ is proportional to the field amplitude and the relevant transition matrix element.

We define the detuning by $\Delta\equiv\omega_{\rm q}-\omega_{\rm s}$ and collect the physical Hamiltonian parameters as
\begin{align}
    \blambda&=(\Omega,\Delta,\tilde{\varphi}).
    \label{eq:rabi_physical_parameters}
\end{align}
After transforming to the interaction picture generated by $H_{\rm q}$ and applying the rotating wave approximation, detuning appears as a slow phase drift in the transverse drive plane.
The resulting interaction picture drive Hamiltonian is
\begin{align}
    H_{\rm I}(\tau;\blambda)
    &=
    \frac{\Omega}{2}
    \left[
        \cos(\tilde{\varphi}+\Delta\tau)\,\sigma_x
        +\sin(\tilde{\varphi}+\Delta\tau)\,\sigma_y
    \right].
    \label{eq:rabi_hamiltonian}
\end{align}
For a fixed interrogation time $T$, we use the equivalent statistical parameterization
\begin{align}
    \btheta
    &=
    (\Phi_R,\Delta,\tilde{\varphi}),
    \label{eq:rabi_statistical_parameters}\\
    \Phi_R
    &=
    \Omega T.
    \label{eq:rabi_angle}
\end{align}
where $\btheta$ denotes the statistical parameter vector used to label the Rabi response and $\Phi_R$ is the resonant Rabi angle accumulated during one interrogation.
When using this statistical parameterization, $\blambda$ is understood with $\Omega=\Phi_R/T$.

Let $\rho_{\btheta}^{\rm id}(t,T)$ denote the ideal post-interrogation state before relaxation, contrast loss, and readout error.
It is given by
\begin{align}
    \rho_{\btheta}^{\rm id}(t,T)
    &=
    U_{\blambda}(t,T)
    \rho_0
    U_{\blambda}^\dagger(t,T),
    \label{eq:ideal_rabi_state}\\
    U_{\blambda}(t,T)
    &=
    \mathcal{T}
    \exp\left[
        -i\int_0^T\dd s\,H_{\rm I}(t+s;\blambda)
    \right].
    \label{eq:rabi_unitary}
\end{align}
Here $t$ is the laboratory start time of the shot, $s\in[0,T]$ is the time measured from the beginning of that shot, and $\mathcal{T}$ denotes time ordering.
The exact evolution is time-ordered because the transverse drive direction changes during the interrogation.
In the weak drive model used below, we keep the leading Magnus term of this time-dependent drive and replace it by an effective constant drive,
\begin{align}
    U_{\blambda}(t,T)
    &\simeq
    \exp[-i\bar{H}_{\blambda}(t,T)T],
    \label{eq:rabi_effective_unitary}\\
    \bar{H}_{\blambda}(t,T)
    &=
    \frac{\Phi_{\rm eff}}{2T}
    \left(
        \cos\tilde{\varphi}_{\rm eff}\,\sigma_x
        +\sin\tilde{\varphi}_{\rm eff}\,\sigma_y
    \right),
    \label{eq:rabi_effective_hamiltonian}\\
    \Phi_{\rm eff}
    &=
    \Phi_R\operatorname{sinc}\!\left(\frac{\Delta T}{2}\right),
    \label{eq:effective_rabi_phase}\\
    \tilde{\varphi}_{\rm eff}
    &=
    \tilde{\varphi}+\Delta\left(t+\frac{T}{2}\right).
    \label{eq:effective_rabi_phase_angle}
\end{align}
Here $\bar{H}_{\blambda}(t,T)$ is the first Magnus effective Hamiltonian, $\Phi_{\rm eff}$ is the detuning-suppressed effective Rabi angle, $\tilde{\varphi}_{\rm eff}$ is the corresponding effective drive phase.
This approximation neglects higher order terms in $\Phi_R$, including commutator corrections from Hamiltonians at different times.
The readout probabilities used below evaluate \cref{eq:ideal_rabi_state} with the effective unitary in \cref{eq:rabi_effective_unitary}.

We model relaxation, finite contrast, and classical readout error as three separate operations.
Their composition gives the observed readout probabilities.

First, relaxation during the sensing interval is represented by an effective quantum channel $\mathcal{R}_T$.
With the ground state chosen at the north pole, write
\begin{align}
    \rho
    &=
    \frac{1}{2}
    \left(\mathbb{I}+r_x\sigma_x+r_y\sigma_y+r_z\sigma_z\right),
    \label{eq:bloch_vector_convention}\\
    \rho_0
    &=
    \frac{1}{2}(\mathbb{I}+\sigma_z),
    \label{eq:ground_state_convention}
\end{align}
where $\mathbb{I}$ denotes the $2\times2$ identity matrix.
The channel is defined by
\begin{align}
    \mathcal{R}_T(\rho)
    &=
    \frac{1}{2}\left[
        \mathbb{I}
        +\eta_2(T)(r_x\sigma_x+r_y\sigma_y)
    \right.
    \notag\\
    &\hspace{3.6em}\left.
        +\{1-\eta_1(T)(1-r_z)\}\sigma_z
    \right].
    \label{eq:relaxation_channel}
\end{align}
Here $\eta_2(T)$ is the time-averaged transverse attenuation factor and $\eta_1(T)$ is the corresponding population transfer attenuation factor.
Thus the transverse components decay toward zero, the longitudinal component relaxes toward the ground state value $r_z=+1$, and the ground state itself is a fixed point.
For the numerical benchmarks we assume Markovian exponential envelopes~\footnote{For pure dephasing by zero mean Gaussian frequency noise $\delta\omega(t)$, the coherence factor is $W(T)=\left\langle e^{-i\int_0^T\dd t\,\delta\omega(t)}\right\rangle=\exp[-\frac12\int_0^T\dd t\int_0^T\dd t'\,C_{\delta\omega}(t-t')]$ when the noise is Gaussian, with $C_{\delta\omega}(\tau)=\langle\delta\omega(\tau)\delta\omega(0)\rangle$.
In the phenomenological form $W(T)=e^{-(T/\tau_\phi)^\nu}$, $\tau_\phi>0$ is a dephasing timescale and $\nu>0$ is a dimensionless exponent that characterizes the time dependence of the decay.
White noise, $C_{\delta\omega}(\tau)=2\Gamma_\phi\delta(\tau)$, gives $W(T)=e^{-\Gamma_\phi T}$, corresponding to $\nu=1$.
In contrast, quasistatic noise with $C_{\delta\omega}(\tau)\simeq\sigma^2$ over the experimental time gives $W(T)=e^{-\sigma^2T^2/2}$, corresponding to $\nu=2$.
Thus compressed exponential envelopes with $\nu>1$ can arise from finite memory, non-Markovian dephasing, although the precise envelope is noise spectrum dependent.} and use weak drive, time-averaged relaxation factors.
Let $\Gamma_1$ and $\Gamma_2$ be the longitudinal and transverse relaxation rates.
For weak drive, the attenuation factors are expressed as
\begin{align}
    \eta_2(T)
    &=
    \frac{1}{T}\int_0^T\dd \tau\,e^{-\Gamma_2\tau}
    =
    \frac{1-e^{-\Gamma_2T}}{\Gamma_2T},
    \label{eq:eta2_definition}\\
    \eta_1(T)
    &=
    \frac{2}{T^2}
    \int_0^T\dd \tau\,e^{-\Gamma_1(T-\tau)}
    \int_0^\tau\dd \tau'\,e^{-\Gamma_2(\tau-\tau')} .
    \label{eq:eta1_definition}
\end{align}
For $\Gamma_1\ne\Gamma_2$, the second attenuation factor can be written explicitly as
\begin{align}
    \eta_1(T)
    &=
    \frac{2}{T^2\Gamma_2}
    \left[
        \frac{1-e^{-\Gamma_1T}}{\Gamma_1}
        -
        \frac{e^{-\Gamma_2T}-e^{-\Gamma_1T}}{\Gamma_1-\Gamma_2}
    \right],
    \label{eq:eta1_explicit}
\end{align}
For coincident rates, $\Gamma_1=\Gamma_2=\Gamma$,
\begin{align}
    \eta_1(T)
    &=
    \frac{2}{(\Gamma T)^2}
    \left[
        1-(1+\Gamma T)e^{-\Gamma T}
    \right].
    \label{eq:eta1_equal_rates}
\end{align}
\cref{eq:eta1_definition} accounts for coherence generated during the drive, its $T_2$ decay before population transfer, and the subsequent $T_1$ relaxation of the generated population.
In this paper, we use the benchmark choice $\Gamma_1 T=\Gamma_2 T=1$, where
\begin{align}
    \eta_2(T)
    &=
    1-\frac{1}{e},
    \label{eq:eta2_equal_times}\\
    \eta_1(T)
    &=
    2\left(1-\frac{2}{e}\right).
    \label{eq:eta1_equal_times}
\end{align}

Second, finite contrast is represented by a depolarizing channel
\begin{align}
    \mathcal{D}_C(\rho)
    &=
    C\rho+(1-C)\frac{\mathbb{I}}{2},
    \qquad 0\le C\le1.
    \label{eq:contrast_channel}
\end{align}
Here $C$ is a dimensionless contrast parameter; the channel multiplies every Bloch vector component by $C$.
The quantum state entering the ideal projective measurement is therefore
\begin{align}
    \rho_{\btheta}^{\rm q}(t,T)
    &=
    (\mathcal{D}_C\circ\mathcal{R}_T)
    \left(\rho_{\btheta}^{\rm id}(t,T)\right).
    \label{eq:quantum_channel_composition}
\end{align}
The superscript ``q'' indicates that the state includes quantum imperfections but not the subsequent classical readout error.

Finally, readout is modeled by an ideal binary projective measurement followed by a classical symmetric bit flip channel.
For a unit readout direction $\bm{n}$, define the ideal binary projection-valued measure (PVM)
\begin{align}
    M_{r'}(\bm{n})
    &=
    \frac{1}{2}
    \left[\mathbb{I}+r'\,\bm{n}\cdot\bm{\sigma}\right],
    \qquad r'=\pm1,
\end{align}
and the corresponding quantum measurement probability
\begin{align}
    p_{\btheta}^{\rm q}(r'|t,T,\bm{n})
    &=
    \operatorname{Tr}\!\left[
        M_{r'}(\bm{n})\rho_{\btheta}^{\rm q}(t,T)
    \right].
\end{align}
The classical bit flip channel acts on the ideal outcome $r'=\pm1$ and produces the observed outcome $r=\pm1$,
\begin{align}
    \mathcal{B}_{\epsilon}(r|r')
    &=
    (1-\epsilon)\delta_{r,r'}
    +\epsilon\delta_{r,-r'},
    \qquad 0\le\epsilon\le\frac{1}{2}.
    \label{eq:classical_bit_flip}
\end{align}
Here $\epsilon$ is the probability that the classical readout flips the binary outcome, and $\delta$ denotes the Kronecker delta.
The observed outcome probability is obtained by the convolution
\begin{align}
    p_{\btheta}(r|t,T,\bm{n})
    &=
    \sum_{r'=\pm1}
    \mathcal{B}_{\epsilon}(r|r')
    p_{\btheta}^{\rm q}(r'|t,T,\bm{n}).
    \label{eq:full_measurement_convolution}
\end{align}
Therefore,
\begin{align}
    p_{\btheta}(+|t,T,\bm{n})
    &=
    \epsilon+(1-2\epsilon)
    p_{\btheta}^{\rm q}(+|t,T,\bm{n}).
    \label{eq:full_measurement_probability}
\end{align}

In the weak drive Rabi model used below, define $\bm{r}_{\btheta}^{\rm q}(t,T)$ by $\rho_{\btheta}^{\rm q}(t,T)=\{\mathbb{I}+\bm{r}_{\btheta}^{\rm q}(t,T)\cdot\bm{\sigma}\}/2$.
It is
\begin{align}
    \bm{r}_{\btheta}^{\rm q}(t,T)
    &=
    C
    \begin{pmatrix}
        \eta_2(T)\sin\Phi_{\rm eff}\sin\tilde{\varphi}_{\rm eff}\\
        -\eta_2(T)\sin\Phi_{\rm eff}\cos\tilde{\varphi}_{\rm eff}\\
        1-\eta_1(T)(1-\cos\Phi_{\rm eff})
    \end{pmatrix}.
    \label{eq:rabi_quantum_bloch_vector}
\end{align}
The dependence of $\Phi_{\rm eff}$ and $\tilde{\varphi}_{\rm eff}$ on $\btheta$, $t$, and $T$ is left implicit in this expression.
The probability for a general readout direction is therefore
\begin{align}
    p_{\btheta}(+|t,T,\bm{n})
    &=
    \epsilon+\frac{1-2\epsilon}{2}
    \left(1+\bm{n}\cdot\bm{r}_{\btheta}^{\rm q}(t,T)\right).
    \label{eq:general_readout_probability}
\end{align}
The population and transverse readout axes, together with a general unit readout direction, are
\begin{align}
    \bm{n}_z
    &=
    (0,0,1),
    \label{eq:z_readout_axis}\\
    \bm{n}_\perp(\vaphi)
    &=
    (\cos\vaphi,\sin\vaphi,0),
    \label{eq:transverse_readout_axis}\\
    \bm{n}(\theta,\vaphi)
    &=
    (\sin\theta\cos\vaphi,
      \sin\theta\sin\vaphi,
      \cos\theta).
    \label{eq:spherical_readout_axis}
\end{align}
Here $\theta\in[0,\pi]$ is the polar angle measured from the positive $z$ axis and $\vaphi$ is the azimuthal angle in the rotating frame $xy$ plane.
Thus $\bm{n}_\perp(\vaphi)=\bm{n}(\pi/2,\vaphi)$, while $\bm{n}(\theta,\vaphi)$ describes the general projective readout used by the adaptive policies.
Substituting these directions into \cref{eq:general_readout_probability} gives the probabilities used in the simulations:
\begin{align}
    p_{\btheta}(+|t,T,\bm{n}_\perp(\vaphi))
    &=
    \epsilon+\frac{1-2\epsilon}{2}
    \notag\\
    &\quad\times
    \left[
        1-C\eta_2(T)\sin\Phi_{\rm eff}
        \sin(\vaphi-\tilde{\varphi}_{\rm eff})
    \right],
    \label{eq:rabi_transverse_probability}\\
    p_{\btheta}(+|t,T,\bm{n}_z)
    &=
    \epsilon+\frac{1-2\epsilon}{2}
    \notag\\
    &\quad\times
    \left[
        1+C\{1-\eta_1(T)(1-\cos\Phi_{\rm eff})\}
    \right],
    \label{eq:rabi_z_probability}
\end{align}
In later sections, the explicit arguments $t$, $T$, and $\bm{n}$ are absorbed into the control and history notation.

\subsection{Fixed budget composite test and Bayesian evidence}
\label{subsec:composite_hypothesis}
\label{subsec:bayesian_update}

We now turn the single shot readout probabilities into a detection procedure with a fixed shot budget.
The experiment is repeated over many shots to decide whether a signal is present at a controlled false positive rate.
If a Rabi signal is present, we assume that there is a single signal with one fixed but unknown parameter vector $\btheta$ throughout the run.
Our approach is Bayesian: the measurement record updates both the posterior probability for signal presence and the posterior distribution over $\btheta$.
The same posterior can also be used adaptively to choose later readout directions.

The two hypotheses specify the possible data generating models.
The null hypothesis $H_0$ is the no signal model, obtained from the readout model of \cref{subsec:rabi_model} by setting $\Phi_R=0$.
The signal hypothesis $H_1$ is composite: the signal parameter belongs to a search region $\Theta$ and has prior density $\pi_0(\btheta)$.
\begin{align}
    H_0 &: \quad \Phi_R=0,
    \label{eq:null_hypothesis}\\
    H_1 &: \quad \btheta\in\Theta,
        \qquad \btheta\sim\pi_0(\btheta).
    \label{eq:signal_hypothesis}
\end{align}
At $\Phi_R=0$, the readout probabilities do not depend on the phase or detuning, so these parameters are not identifiable under $H_0$.
This is the structure of nuisance parameters present only under the alternative studied in~\cite{Davies:1977}; we therefore calibrate the decision threshold directly under the null rather than relying on the standard $\chi^2$ asymptotics for likelihood ratio tests.
These hypotheses define the models to be compared, while the experimental controls determine which conditional probabilities are sampled from those models.

For a run with $n$ total shots, let $k=0,1,\ldots,n-1$ denote the shot index.
At each shot, the sensor is reinitialized, a control setting $u_k$ is chosen, and a binary outcome is recorded.
We write $R_k\in\{\pm1\}$ for the random outcome of the $k$th shot before observation and $r_k$ for its realized value.
The previous controls and outcomes form the classical history
\begin{align}
    h_k&=(u_0,r_0,\ldots,u_{k-1},r_{k-1}) .
    \label{eq:history_definition}
\end{align}
A policy is a sequence of rules $\mu=\{\mu_k\}$ that chooses the next control from the available information,
\begin{align}
    u_k
    &=
    \mu_k(h_k).
    \label{eq:policy_definition}
\end{align}
For a finite horizon dynamic programming formulation \cite{Bellman1957DynamicProgramming}, the same rule can be written as $u_k=\mu_k(k,S_k)$ in terms of the Bayesian state $S_k$ defined below.
In the main analysis below, all controls except the binary projective readout direction are fixed, and we henceforth write the control as $u_k\equiv\bm{n}_k$.
In a more general protocol, $u_k$ could include state preparation, sensing time, or other controls, as discussed in \cref{subsec:probe_optimization}.
In the simulations instead the interrogation time is fixed, overhead time is neglected, and the $k$th shot starts at $t_k=kT$.
For compact notation, define the one shot probabilities at shot $k$ by
\begin{align}
    p_{\btheta,k}(r|\bm{n})
    &\equiv
    p_{\btheta}(r|t_k,T,\bm{n}),
    \label{eq:controlled_signal_probability}\\
    p_{0,k}(r|\bm{n})
    &\equiv
    \left.
    p_{\btheta,k}(r|\bm{n})
    \right|_{\Phi_R=0}.
    \label{eq:controlled_null_probability}
\end{align}
For the realized $k$th shot, these probabilities are evaluated at $\bm{n}=\bm{n}_k$.

Before selecting the $k$th readout direction, the Bayesian state is
\begin{align}
    S_k
    &=
    \bigl(q_k,\pi_k(\btheta)\bigr),
    \label{eq:bayesian_state}\\
    q_k
    &=
    \Prob(H_1|h_k),
    \label{eq:signal_posterior_probability}
\end{align}
where $q_k$ is the posterior probability that a signal is present and $\pi_k(\btheta)=\Prob{\btheta|H_1,h_k}$ is the parameter posterior conditional on signal presence.
The initial values are $q_0=\Prob(H_1)$ and the prior density $\pi_0(\btheta)$.
In our analysis we take $q_0=1/2$; the numerical choice of $\pi_0(\btheta)$ is specified in \cref{subsec:numerical_procedure}.
The direction $\bm{n}_k$ may be fixed in advance or selected adaptively from the current Bayesian state; concrete choices are specified in \cref{sec:policies}.
After choosing $\bm{n}_k$ and observing $r_k$, the conditional parameter posterior and the signal probability update as
\begin{align}
    \pi_{k+1}(\btheta)
    &=
    \frac{
        \pi_k(\btheta)p_{\btheta,k}(r_k|\bm{n}_k)
    }{
        m_k(r_k|\bm{n}_k)
    },
    \label{eq:posterior_density_update}\\
    q_{k+1}
    &=
    \frac{
        q_k m_k(r_k|\bm{n}_k)
    }{
        (1-q_k)p_{0,k}(r_k|\bm{n}_k)
        +q_k m_k(r_k|\bm{n}_k)
    },
    \label{eq:signal_probability_update}
\end{align}
where
\begin{align}
    m_k(r|\bm{n})
    &=
    \int_\Theta\dd\btheta\,\pi_k(\btheta)
    p_{\btheta,k}(r|\bm{n}).
    \label{eq:predictive_likelihood}
\end{align}
is the posterior predictive probability for the next outcome under the current signal mixture.

The evidence for the signal is summarized by the Bayes factor $B_{10,k}$.
For a fixed policy and signal prior $\pi_0(\btheta)$, $B_{10,k}$ is the likelihood ratio between $H_0$ and $H_1$ for the accumulated history $h_k$.
We use it below as the final test statistic and evaluate performance by the power at fixed Type-I error defined in \cref{eq:calibrated_threshold,eq:calibrated_power}.
The boundary condition is $B_{10,0}=1$, and the posterior odds obey
\begin{align}
    \frac{q_k}{1-q_k}
    &=
    B_{10,k}\,
    \frac{q_0}{1-q_0}.
    \label{eq:posterior_odds_bayes_factor}
\end{align}
Equivalently, the log Bayes factor updates as
\begin{align}
    \log B_{10,k+1}
    &=
    \log B_{10,k}
    +\log m_k(r_k|\bm{n}_k)
    \notag\\
    &\quad
    -\log p_{0,k}(r_k|\bm{n}_k).
    \label{eq:log_bayes_update}
\end{align}

After $n$ shots, we reject $H_0$ when $\log B_{10,n}\ge c$.
This condition is evaluated for the complete adaptive experiment, including the same policy used during data taking.
For a fixed policy $\mu$, let $\Prob_0^\mu$ denote the joint distribution of complete records of $n$ shots generated under $H_0$, and let $\Prob_{\btheta}^\mu$ denote the corresponding law under $H_1$ with fixed signal parameter $\btheta$.
At a generic threshold $c$, the Type-I error is
\begin{align}
    \alpha^\mu(c)
    &=
    \Prob_0^\mu(\log B_{10,n}\ge c).
    \label{eq:type_i_error_main}
\end{align}
The Type-I error is the probability that the full search reports a signal when the data are generated by the null model.\footnote{In high energy terminology, this null probability includes the look elsewhere effect: scanning over many possible signal parameters, can increase the chance that a background only data set passes the final selection.}
For a target Type-I error $\alpha$, we calibrate a separate threshold $c_\alpha^\mu$ for each fixed policy $\mu$ and shot budget $n$ from complete pseudoexperiments under $H_0$, using the same readout policy and final statistic as in the signal search: \footnote{For finite pseudoexperiment samples, the equality is implemented through an empirical upper tail threshold. In the numerical analysis we choose the order statistic corresponding to the target upper tail fraction of the null ensemble and report the resulting empirical Type-I error; exact equality could instead be enforced by randomizing at the boundary.}
\begin{align}
    \Prob_0^\mu(\log B_{10,n}\ge c_\alpha^\mu)
    &=
    \alpha.
    \label{eq:calibrated_threshold}
\end{align}
Both the null record law and the calibrated threshold depend on $\mu$; the shot budget, final statistic, and calibration rule are otherwise held fixed in this notation.
When the policy is clear from context, or when a statement applies to any fixed policy, we suppress the superscript $\mu$.
Because $H_1$ is composite, the calibrated Type-II error and power remain pointwise functions of the true signal parameter.
We define them by
\begin{align}
    \beta^\mu(\btheta;\alpha)
    &\equiv
    \Prob_{\btheta}^\mu(\log B_{10,n}<c_\alpha^\mu),
    \label{eq:type_ii_error_main}\\
    1-\beta^\mu(\btheta;\alpha)
    &=
    \Prob_{\btheta}^\mu(\log B_{10,n}\ge c_\alpha^\mu).
    \label{eq:calibrated_power}
\end{align}
An averaged or worst case power requires an additional aggregation criterion over $\btheta$; below, any such averaging is stated explicitly.

Two common criteria are symmetric discrimination, which minimizes an average error probability, and asymmetric testing, which fixes the Type-I error $\alpha$ and minimizes the Type-II error $\beta$.
The calibrated construction above implements the latter, and below we use the target value $\alpha=0.05$.

The same testing construction can also be applied to other sensing models.
We keep the main text centered on Rabi sensing, mainly because it clearly demonstrates benefits of adaptive approaches on its sensitivity.

\section{Measurement policies}
\label{sec:policies}

We now specify how $\bm{n}_k$ are chosen.
The final test statistic is the calibrated log Bayes factor defined above, but the policies below should be understood as practical readout selection rules rather than solutions of a global optimization problem for $\log B_{10,n}$.
Their performance is judged by the resulting power and sensitivity at fixed $\alpha$.
\cref{subsec:global_optimization} discusses what the genuinely global finite horizon optimization would require.

\subsection{Non-adaptive baselines}
\label{subsec:fixed_policies}

We use three fixed schedules as baselines.
The fixed-$z$ policy performs population readout along the qubit energy axis on every shot; this is robust to an unknown transverse phase but has only a quadratic weak signal response (see \cref{eq:weak_z_response}).
The fixed-$x$ policy measures one transverse quadrature and therefore has blind signal phases near $\tilde{\varphi}_{\rm eff}=0$ and $\pi$ (see \cref{eq:weak_transverse_response}).
The fixed-$x/y$ policy alternates two orthogonal transverse quadratures; it removes exact blind directions at the cost of spending only half of the shots on each quadrature.
Their weak signal behavior is summarized in \cref{sec:analytic}.
In the notation of \cref{eq:policy_definition}, these schedules are
\begin{align}
    \mu_k^{z}(h_k)
    &=
    \bm{n}_z,
    \label{eq:fixed_z_policy}\\
    \mu_k^{x}(h_k)
    &=
    \bm{n}_\perp(0),
    \label{eq:fixed_x_policy}\\
    \mu_k^{x/y}(h_k)
    &=
    \begin{cases}
        \bm{n}_\perp(0), & k\ {\rm even},\\
        \bm{n}_\perp(\pi/2), & k\ {\rm odd}.
    \end{cases}
    \label{eq:fixed_xy_policy}
\end{align}

\subsection{Known signal transverse oracle benchmark}
\label{subsec:oracle}

As a weak signal reference, we use a known signal transverse oracle that is allowed to depend on the true $\btheta=(\Phi_R,\Delta,\tilde{\varphi})$ before the first shot.
It is an oracle within the set of transverse projective readout axes
\begin{align}
    \mathcal{N}_\perp
    &=
    \left\{
        \bm{n}_\perp(\vaphi):0\le\vaphi<\pi
    \right\},
    \label{eq:transverse_readout_set}
\end{align}
where the azimuthal range uses the equivalence of antipodal projective axes up to an exchange of outcome labels.
After $\btheta$ is supplied, the oracle chooses a direction in $\mathcal{N}_\perp$.
Define the transverse unit vector aligned with the known weak signal response at shot $k$ by
\begin{align}
    \bm{e}_{\perp,k}(\btheta)
    &=
    \bm{n}_\perp\!\left(
        \tilde{\varphi}_{\rm eff}(t_k,T;\btheta)+\frac{\pi}{2}
    \right).
    \label{eq:known_signal_transverse_axis}
\end{align}
The aligned transverse oracle is the policy
\begin{align}
    \mu_{\btheta,k}^{{\rm O}_\perp}(h_k)
    &=
    \bm{e}_{\perp,k}(\btheta).
    \label{eq:transverse_oracle_policy}
\end{align}

For two known qubit states with equal prior probabilities, the Helstrom measurement projects along the difference between their Bloch vectors and maximizes the one shot total variation distance, as reviewed in \cref{appsubsec:helstrom}.
Since the leading difference between the signal and null Bloch vectors is parallel to $\bm{e}_{\perp,k}(\btheta)$, the oracle approximates the weak signal Helstrom direction within transverse readout.
Therefore, the oracle provides a useful benchmark for the best possible performance of any adaptive policy, though it is not a practical policy.
The separation between the transverse oracle curve and an adaptive curve measures the cost of learning the unknown signal direction from data.
On the other hand, because the oracle is restricted to transverse readout, it is not an upper bound over all projective measurements and can be outperformed by the fixed-$z$ policy when the signal is sufficiently strong.

\subsection{Adaptive Bayesian policy}
\label{subsec:adaptive_policy}

The adaptive policy selects a readout direction from the current Bayesian state.
The admissible set is the full family of projective readout axes
\begin{align}
    \mathcal{N}
    &=
    \left\{
        \bm{n}(\theta,\vaphi):
        0\le\theta\le\pi,
        \ 0\le\vaphi<\pi
    \right\}.
    \label{eq:full_readout_set}
\end{align}
The policies used here are myopic: at shot $k$ they maximize a one step utility over $\bm{n}\in\mathcal{N}$ using the current state $S_k$, without optimizing the full future control tree.
A seemingly natural myopic objective is the expected Bayes drift, namely the expected one shot increase in the log Bayes factor under the current signal predictive distribution.
For the phase symmetric resonant posterior relevant at initialization, however, this objective selects population readout, which leaves the phase posterior unchanged and thereby locks the strategy into the fixed-$z$ policy (see \cref{subsec:global_optimization} for details).
This failure motivates an explicit learning objective.

Our first utility is the expected information gain about the unknown hypothesis and signal parameters.
Let $\Lambda_k$ denote the latent model before the $k$th shot.
It takes the null value $\Lambda_k=0$ with probability $1-q_k$, and signal values $\Lambda_k=(1,\btheta)$ with density $q_k\pi_k(\btheta)$.
The total predictive distribution before the next shot is
\begin{align}
    \bar p_k(r|\bm{n})
    &=
    (1-q_k)p_{0,k}(r|\bm{n})
    +q_k m_k(r|\bm{n}).
    \label{eq:total_predictive_probability}
\end{align}
We define the information gain utility by
\begin{align}
    U_{{\rm I},k}(\bm{n})
    &=
    I_{S_k}(\Lambda_k;R_k|\bm{n})
    \notag\\
    &=
    h_{\rm Ber}\!\left(\bar p_k(+|\bm{n})\right)
    -(1-q_k)
    h_{\rm Ber}\!\left(p_{0,k}(+|\bm{n})\right)
    \notag\\
    &\quad
    -q_k
    \int_\Theta\dd\btheta\,\pi_k(\btheta)
    h_{\rm Ber}\!\left(p_{\btheta,k}(+|\bm{n})\right),
    \label{eq:information_gain_utility}
\end{align}
Here $I_{S_k}$ denotes mutual information under the joint distribution induced by the current Bayesian state $S_k$ and the candidate direction $\bm{n}$; the second equality is its explicit binary outcome entropy form.
The Bernoulli entropy is $h_{\rm Ber}(p)=-p\log p-(1-p)\log(1-p)$.
Equivalently, $U_{{\rm I},k}$ is the expected information gained from the next outcome about both signal presence and the parameters $\btheta$ under $H_1$.
It can therefore reward measurements that resolve the signal family even when nuisance parameter averaging suppresses the direct signal--null separation.
Because relaxation, contrast loss, and readout error enter the likelihood model, $U_{{\rm I},k}$ optimizes the information that survives these modeled imperfections, although it neither reverses information loss nor guarantees robustness to model misspecification.
Expected information gain is a standard utility in Bayesian experimental design because it selects the next control by the expected reduction of posterior uncertainty without requiring a model specific estimation heuristic~\cite{ChalonerVerdinelli:1995,HuanMarzouk:2013}.
Closely related utilities have been used for robust online Hamiltonian learning and real time adaptive quantum sensing~\cite{GranadeEtAl:2012,WangEtAl:2022Realtime}.

Our second utility directly promotes one shot signal detection through the Helstrom-like separation
\begin{align}
    U_{{\rm H},k}(\bm{n})
    &=
    |m_k(+|\bm{n})-p_{0,k}(+|\bm{n})|.
    \label{eq:helstrom_utility}
\end{align}
For binary readout, this is the total variation distance between the signal mixture and null predictive distributions, as discussed in \cref{app:hypothesis_testing}.
Although $U_{{\rm H},k}$ is evaluated using the same noisy likelihoods and is therefore also noise aware, it first marginalizes over $\btheta$ and does not reward information that only resolves the nuisance parameters.
It can consequently become nearly degenerate when the posterior is broad over nuisance phases; the situation is the same as myopic maximization of the expected Bayes drift discussed in \cref{subsec:global_optimization}.

We use information gain shots as an exploration component and Helstrom-like shots as a direct evidence accumulation component.
The corresponding one step policies select a maximizer of the shot dependent utilities,
\begin{align}
    \mu_k^{\rm H}(h_k)
    &\in
    \arg\max_{\bm{n}\in\mathcal{N}}U_{{\rm H},k}(\bm{n}),
    \label{eq:helstrom_policy}\\
    \mu_k^{\rm I}(h_k)
    &\in
    \arg\max_{\bm{n}\in\mathcal{N}}U_{{\rm I},k}(\bm{n}).
    \label{eq:information_policy}
\end{align}
A fixed tie breaking convention makes each policy single valued.
Another practical policy tried in our analysis is the periodic hybrid
\begin{align}
    \mu_k^{{\rm H}/{\rm I}}(h_k)
    &=
    \begin{cases}
        \mu_k^{\rm I}(h_k), & k\equiv0\pmod L,\\
        \mu_k^{\rm H}(h_k), & \text{otherwise}.
    \end{cases}
    \label{eq:helstrom_information_hybrid_policy}
\end{align}
The information gain shots keep the posterior from remaining broad, while the remaining shots accumulate evidence for $H_1$.
This hybrid is still a myopic policy and is not claimed to maximize the final calibrated power globally.
A global finite horizon policy could combine exploration and detection in a single objective; the periodic hybrid should instead be viewed as an empirical heuristic.
The empirically best myopic strategy among those tested depends strongly on detector performance.
Pure information gain ($L=1$) is favored for the baseline detector profile used in the main numerical benchmarks, whereas a periodic hybrid ($L=8$) is favored for the higher fidelity profile discussed in \cref{appsubsec:higher_fidelity}.

\section{Analytic benchmarks}
\label{sec:analytic}

This section summarizes the analytic benchmarks used to interpret the numerical scans.
Its role is to identify the weak signal mechanisms, the resulting sensitivity exponents, and the finite coefficient costs that distinguish the fixed policies, oracle benchmark, and adaptive policy.
The derivations and closed form coefficients are collected in \cref{app:rabi_crossing}; here we keep only the formulas needed to read the main figures and tables.

\subsection{Weak signal response}
\label{subsec:weak_signal}

Expanding the probabilities in \cref{eq:rabi_transverse_probability,eq:rabi_z_probability} at small $\Phi_{\rm eff}$ separates the two relevant weak signal responses.
Let $p_{z,0}$ and $p_{\perp,0}$ denote the positive outcome probabilities under $H_0$ for population and transverse readout, respectively; their explicit values are given in \cref{eq:app_fixed_z_null_probability,eq:app_transverse_null_probability}.
The leading nonzero probability shifts are
\begin{align}
    p_{\btheta}(+|t,T,\bm{n}_z)-p_{z,0}
    &\simeq
    -a_z\Phi_{\rm eff}^2,
    \label{eq:weak_z_response}\\
    p_{\btheta}(+|t,T,\bm{n}_\perp(\vaphi))-p_{\perp,0}
    &\simeq
    -b_\perp\Phi_{\rm eff}
    \sin(\vaphi-\tilde{\varphi}_{\rm eff}).
    \label{eq:weak_transverse_response}
\end{align}
The coefficients $a_z$ and $b_\perp$ are defined in \cref{eq:app_fixed_z_response_coefficient,eq:app_transverse_response_coefficient}.
The corresponding derivations are given in \cref{eq:app_fixed_z_quadratic_response,eq:app_transverse_linear_response}.
Thus the population readout response is phase independent but quadratic, whereas transverse readout is linear and maximized by alignment with the signal~\cite{Moroi:2026ges}.
Realizing the transverse gain therefore requires knowing, scanning, or learning the unknown phase.
This distinction controls both the error exponent at weak signal and the sensitivity scaling derived below.

\subsection{Weak signal sensitivity scaling}
\label{subsec:sensitivity_scaling}

Because the response depends on the signal amplitude through $\Phi_{\rm eff}$, we express the weak signal sensitivity in terms of this effective angle.
For a target power $P$ and Type-I error $\alpha$, let $\Phi^{\rm req}(n;\alpha,P)$ denote the smallest value of $\Phi_{\rm eff}$ that reaches power $P$ with $n$ shots at test size $\alpha$.
If the probability shift behaves as $\delta p\propto\Phi_{\rm eff}^{\ell}$, where $\ell$ is the leading response order, the signal to noise ratio in the weak signal limit scales as $\sqrt{n}\,\Phi_{\rm eff}^{\ell}$.
Holding the power fixed therefore gives
\begin{align}
    \Phi^{\rm req}(n;\alpha,P)
    &\propto
    n^{-1/(2\ell)}.
    \label{eq:generic_sensitivity_scaling}
\end{align}
Consequently,
\begin{align}
    \Phi_z^{\rm req}(n;\alpha,P)
    &=
    A_z(\alpha,P)n^{-1/4}
    \label{eq:main_fixed_z_sensitivity}
    \\
    \Phi_\perp^{\rm req}(n;\alpha,P)
    &=
    A_\perp(\alpha,P)n^{-1/2},
    \label{eq:main_transverse_sensitivity}
\end{align}
for the fixed-$z$ policy and a generic transverse strategy, respectively.
Among the transverse schedules, the transverse oracle should realize the second scaling with the smallest coefficient within transverse readout.
An adaptive policy realizes the same exponent provided that a non-vanishing fraction of its selected readouts retain a linear response with a finite, non-zero root mean square coefficient.
The constants $A_z$ and $A_\perp$ depend on the chosen Type-I error, target power, detector parameters, and on how the transverse nuisance phase is handled.
Their definitions are given by \cref{eq:app_fixed_z_coefficient,eq:app_oracle_coefficient,eq:app_fixed_x_coefficient,eq:app_xy_coefficient} for fixed-$z$, the transverse oracle, fixed-$x$, and fixed-$x/y$, respectively.
For the baseline benchmark parameters defined in \cref{eq:baseline_detector_profile_1,eq:baseline_detector_profile_2,eq:baseline_detector_profile_3}, the coefficients relevant to the numerical target powers are summarized in \cref{tab:main_sensitivity_coefficients}.

\begin{table}[tb]
\centering
\caption{Weak signal coefficients in $\Phi_j^{\rm req}(n;\alpha,P)=A_j(\alpha,P)n^{-\gamma_j}$ for the benchmark setup at $\alpha=0.05$.
Here $j\in\{z,{\rm O}_\perp,x,xy\}$ labels the listed readout strategies.
The exponents are $\gamma_z=1/4$ and $\gamma_{{\rm O}_\perp}=\gamma_x=\gamma_{xy}=1/2$.}
\label{tab:main_sensitivity_coefficients}
\setlength{\tabcolsep}{3.5pt}
\begin{tabular}{@{}crrrr@{}}
\toprule
$P$ & $A_z$ & $A_{{\rm O}_\perp}$ & $A_x$ & $A_{xy}$\\
\midrule
$0.50$ & $1.123$ & $2.655$ & $4.810$  & $5.082$\\
$0.70$ & $1.289$ & $3.501$ & $7.248$  & $6.335$\\
$0.90$ & $1.498$ & $4.724$ & $20.29$  & $8.120$\\
\bottomrule
\end{tabular}
\end{table}

\subsection{Sensitivity crossing with the fixed-\texorpdfstring{$z$}{z} policy}
\label{subsec:crossing_estimate_main}

The different exponents imply that a transverse strategy eventually overtakes the fixed-$z$ policy in the weak signal asymptotic regime.
Equating \cref{eq:main_fixed_z_sensitivity,eq:main_transverse_sensitivity} gives
\begin{align}
    n_{z:\perp}(\alpha,P)
    &=
    \left[
        \frac{A_\perp(\alpha,P)}{A_z(\alpha,P)}
    \right]^4,
    \label{eq:main_crossing_estimate}
\end{align}
for the crossing between the fixed-$z$ policy and a generic transverse strategy.
For $n<n_{z:\perp}$, the fixed-$z$ policy is more sensitive, whereas for $n>n_{z:\perp}$ the transverse strategy is more sensitive.
This expression is useful because all detector dependence can be made explicit, while the composite phase cost enters only through a dimensionless transverse displacement coefficient.
The general detector dependent expression and its strategy specific forms are derived in \cref{eq:app_crossing_general_explicit} and \crefrange{eq:app_crossing_oracle_explicit}{eq:app_crossing_xy_explicit}, respectively.
For the same benchmark parameters as \cref{tab:main_sensitivity_coefficients}, the estimated crossing points are summarized in \cref{tab:main_crossings}.

\begin{table}[tb]
\centering
\caption{Estimated shot counts at which each listed transverse strategy overtakes the fixed-$z$ policy for the benchmark setup at $\alpha=0.05$.
The upper and lower blocks use the asymptotic Gaussian and finite background Poisson predictions for this policy, respectively.}
\label{tab:main_crossings}
\setlength{\tabcolsep}{3pt}
\begin{tabular}{@{}crrrr@{}}
\toprule
$P$ & $n_{z:{\rm O}_\perp}$ & $n_{z:x}$ & $n_{z:xy}$ & $n_{z:\mathrm{adapt}}$\\
\midrule
\multicolumn{5}{c}{Asymptotic Gaussian fixed-$z$ prediction}\\
$0.50$ & $3.13{\times}10^1$ & $3.37{\times}10^2$ & $4.20{\times}10^2$ & $2.29{\times}10^2$\\
$0.70$ & $5.44{\times}10^1$ & $9.98{\times}10^2$ & $5.83{\times}10^2$ & $3.63{\times}10^2$\\
$0.90$ & $9.89{\times}10^1$ & $3.37{\times}10^4$ & $8.64{\times}10^2$ & $5.79{\times}10^2$\\
\midrule
\multicolumn{5}{c}{Finite background Poisson fixed-$z$ prediction}\\
$0.50$ & $7.88$ & $2.32{\times}10^2$ & $3.03{\times}10^2$ & $1.47{\times}10^2$\\
$0.70$ & $7.90$ & $7.20{\times}10^2$ & $3.77{\times}10^2$ & $2.12{\times}10^2$\\
$0.90$ & $7.50$ & $3.08{\times}10^4$ & $4.89{\times}10^2$ & $2.86{\times}10^2$\\
\bottomrule
\end{tabular}
\end{table}

The oracle, fixed-$x$, and fixed-$x/y$ entries are obtained by intersecting their analytic transverse sensitivities with the indicated fixed-$z$ prediction.
The adaptive entries instead use the numerical results in \cref{subsec:numerical_sensitivity}.
Because the inferred adaptive crossings lie below the fit interval, they should be interpreted as extrapolations rather than directly resolved numerical crossings.
The representative $P=0.7$ sensitivity curves and fit are shown in \cref{fig:rabi_discovery_sensitivity_p07}, together with further discussion of their behavior at finite $n$.
The fitted adaptive crossing lies below the analytic fixed-$x/y$ crossing, consistent with posterior learning improving over the non-adaptive two quadrature benchmark.

\cref{eq:main_crossing_estimate} should nevertheless be interpreted as an asymptotic guide.
The Gaussian background approximation underlying $A_z$ requires $n(1-p_{z,0})\gg1$.
At small shot counts or low effective readout error rates, such that $n(1-p_{z,0})\lesssim O(1)$, the fixed-$z$ test is closer to a low count binomial or Poisson problem, and the apparent finite range scaling can be steeper than the asymptotic $n^{-1/4}$ law.
The lower block of \cref{tab:main_crossings} shows that this correction at finite $n$ shifts every crossing to a smaller shot count.

\subsection{Fixed signal Type-II error exponents}
\label{subsec:error_exponents}

For fixed signal parameters and fixed Type-I error, an asymptotic test is expected to have
\begin{align}
    \beta_n(\btheta)
    &=\exp[-n\,\xi(\btheta)+o(n)] .
    \label{eq:type_two_exponent}
\end{align}
For Bernoulli distributions with positive outcome probabilities $p$ and $q$, define the Kullback-Leibler (KL) divergence by
\begin{align}
    D_{\rm Ber}(p\Vert q)
    &=
    p\log\frac{p}{q}
    +(1-p)\log\frac{1-p}{1-q}.
    \label{eq:bernoulli_divergence_definition}
\end{align}
For an independent and identically distributed (i.i.d.) Bernoulli record, the Stein exponent is
\begin{align}
    \xi
    &=
    D_{\rm Ber}(p_0\Vert p_1).
    \label{eq:bernoulli_stein_exponent}
\end{align}
For an adaptive experiment the record is not i.i.d.; the corresponding relative entropy is a sum of history-conditioned one shot divergences.
A composite alternative further requires care about whether performance is defined pointwise, in the worst case, or after prior averaging.
In our numerical analysis, we therefore compare the measured slopes at finite $n$ with analytic fixed policy references, without assuming that a fitted adaptive slope is already the rigorous composite Stein exponent.
The relation between adaptive record information and the composite Bayes factor is discussed in \cref{appsubsec:policy_connection}.

\section{Numerical results}
\label{sec:numerical_results}

In this section, we show our numerical procedure and results.\footnote{OpenAI Codex using GPT-5.6 Sol (\texttt{gpt-5.6-sol}) and ChatGPT were used throughout this work for scientific discussion, development of the physical and statistical formulation and analysis strategy, analytic derivations and checks, code development and debugging, execution of Monte Carlo simulations, and interpretation of the results.
The author directed the AI-assisted work, reviewed and revised its outputs, verified the reported analytic and numerical results, and takes full responsibility for the work.}

\subsection{Numerical procedure}
\label{subsec:numerical_procedure}

The composite signal prior is discretized on a particle grid, and each pseudoexperiment updates its posterior and log Bayes factor shot by shot.
At every saved shot count, a separate null ensemble for each policy $\mu$ determines $c_\alpha^\mu$, and for each fixed signal parameter $\btheta$ a corresponding signal ensemble estimates the pointwise power $1-\beta^\mu(\btheta;\alpha)$ in \cref{eq:calibrated_power}.
The main numerical results use the baseline detector profile
\begin{align}
    C
    &=
    0.99,
    \label{eq:baseline_detector_profile_1}
    \\
    \epsilon
    &=
    0.005,
    \label{eq:baseline_detector_profile_2}
    \\
    T_1
    &=
    T_2
    =
    T,
    \label{eq:baseline_detector_profile_3}
\end{align}
and the pure information gain policy $L=1$.
Defining the effective symmetric binary readout fidelity $F_{\rm ro}$ as the probability of reporting the correct outcome for either ideal measurement eigenstate, the combined contrast and bit flip model gives $F_{\rm ro}=[1+(1-2\epsilon)C]/2\simeq0.990$.
For computational efficiency, all numerical results throughout this section use the resonant model with $\Delta=0$ and identify $(\Phi_R, \tilde{\varphi})$ with $(\Phi_{\rm eff}, \tilde{\varphi}_{\rm eff})$; effects of detuning are discussed in \cref{subsec:detuning}.
Under this restriction the signal parameter is two dimensional, $\btheta=(\Phi_R,\tilde{\varphi})$.
The numerical prior $\pi_0(\btheta)$ in \cref{eq:signal_hypothesis} is log uniform in amplitude over $0.02\le\Phi_R\le1.2$\footnote{
    The log uniform amplitude prior is a numerical choice rather than a physical signal model.
    It assigns equal prior weight to equal intervals of $\log\Phi_R$, which is convenient for sensitivity scans displayed on a logarithmic amplitude axis.
} and uniform in phase over $0\le\tilde{\varphi}<2\pi$:
\begin{align}
    \pi_0(\Phi_R,\tilde{\varphi})
    &=
    \frac{1}{2\pi\log(1.2/0.02)}\frac{1}{\Phi_R},
    \label{eq:numerical_signal_prior}
\end{align}
with zero density outside this region.

\subsection{Posterior learning}
\label{subsec:posterior_learning}

\begin{figure}[t]
\centering
\includegraphics[width=\columnwidth]{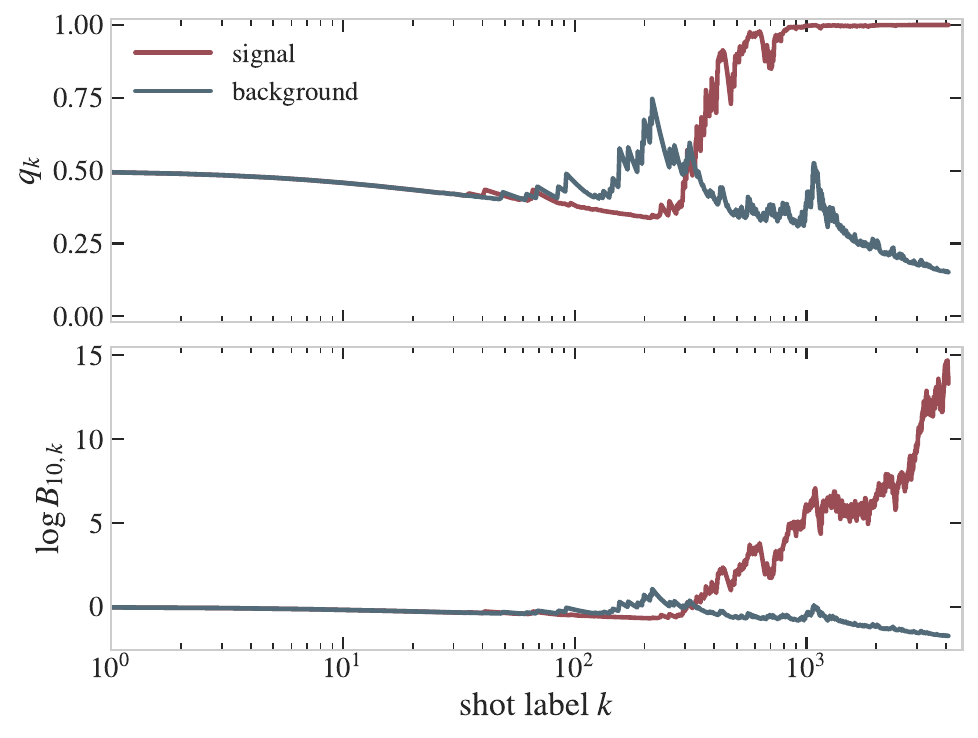}
\caption{Posterior signal probability (upper panel) and log Bayes factor (lower panel) along representative resonant signal and background trajectories, with $(\Phi_R,\tilde{\varphi})=(0.19,0.7)$ injected in the signal trajectory.}
\label{fig:rabi_posterior_learning_curves}
\end{figure}

To visualize what the adaptive policy learns, we compare one record generated at the representative resonant signal point $\btheta_{\rm true}=(\Phi_R,\tilde{\varphi})=(0.19,0.7)$ with one record generated under $H_0$.
The evolution of $q_k$ and $\log B_{10,k}$ is shown in \cref{fig:rabi_posterior_learning_curves}.
The red and blue curves show the two diagnostics along the signal and background records, respectively.
Along the signal trajectory, the evidence begins to favor $H_1$ after a few hundred shots and subsequently drives $q_k$ toward unity, whereas the background trajectory leaves $\log B_{10,k}$ near or below zero at the end of the run.
The blue posterior probability nevertheless exhibits transient peaks, which may be viewed as fake signals produced by finite sample fluctuations in the imperfect readout.
They do not produce sustained growth of the log Bayes factor because successive fluctuations favor mutually inconsistent signal parameters and therefore cannot accumulate coherently as evidence for one signal hypothesis.
The non-monotonic intermediate evolution is expected for a single realization and emphasizes that the posterior probability at an intermediate shot should not be identified with the fixed budget test statistic.

The onset of this evidence growth can be checked approximately using the local Fisher geometry of the signal state.
For the resonant benchmark, where $\Phi_{\rm eff}=\Phi_R$, \cref{eq:rabi_null_qfi,eq:rabi_null_transverse_cfi} give the single shot quantum Fisher information (QFI) of $\rho_{\btheta}^{\rm q}$ with respect to $\Phi_R$ at the null as $F_{\mathrm{Q}}^{(\Phi_R)}(0)=C^2\eta_2^2(T)\simeq0.392$, while an optimally aligned transverse measurement followed by the classical readout channel has local classical Fisher information (CFI) $F_{\mathrm{C}}^{(\Phi_R)}(0)=(1-2\epsilon)^2F_{\mathrm{Q}}^{(\Phi_R)}(0)\simeq0.384$.
Let $P_{\Phi_R}$ and $P_0$ denote the one shot outcome distributions for this aligned readout at the injected amplitude and under the null, respectively.
Let $D_{\rm KL}$ denote the KL divergence defined in \cref{eq:classical_kl}.
Writing $B_{10,k}^{\rm pt}$ for the Bayes factor against this known direction point alternative, independence of the repeated outcomes gives
\begin{align}
    \mathbb{E}_{\Phi_R}
    \left[
        \log B_{10,k}^{\rm pt}
    \right]
    &=
    kD_{\rm KL}(P_{\Phi_R}\Vert P_0)
    \notag\\
    &=
    \frac{k\Phi_R^2}{2}
    F_{\rm C}^{(\Phi_R)}(0)
    +k\,o(\Phi_R^2).
    \label{eq:local_expected_log_point_bayes}
\end{align}
The first equality is the expected log likelihood ratio identity, and the second follows from the local KL--CFI expansion in \cref{eq:local_forward_kl_fisher_expansion}.
Using this idealized growth rate as a proxy for the full composite Bayes factor, $q_0=1/2$ and $q_k=0.99$, or equivalently $\log B_{10,k}=\log 99$, give the rough estimate $k\simeq6.6\times10^2$ for $\Phi_R=0.19$.
For the displayed realization, $q_{512}=0.902$ and $q_{1024}=0.997$.
The onset of sustained evidence growth is therefore compatible with the local estimate at the order of magnitude level.

The above estimate neglects corrections at finite $\Phi_R$, marginalization over the unknown amplitude and phase, and the history dependence of the adaptive readout axes; record to record fluctuations further obscure the comparison for a single trajectory.
\cref{appsubsec:local_fisher_information} derives the local KL--CFI relation and its finite signal limitation, while \cref{appsubsec:policy_connection} connects the adaptive readout utilities and complete record information to the composite Bayes factor.

\begin{figure}[t]
\centering
\includegraphics[width=\columnwidth]{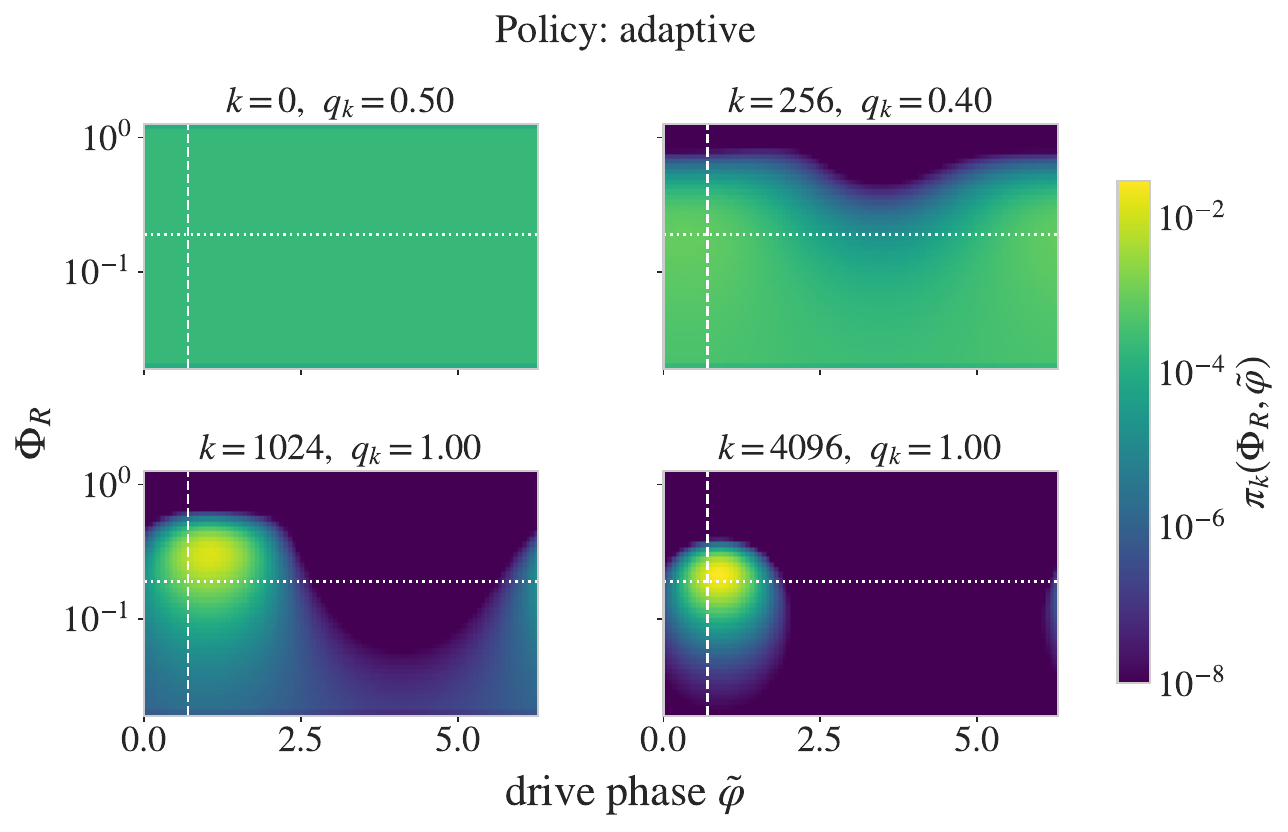}
\caption{Conditional $H_1$ posterior density in the $(\Phi_R,\tilde{\varphi})$ plane   along the resonant adaptive signal trajectory, with the corresponding $q_k$ reported above each panel.
The dashed and dotted lines mark the injected phase and amplitude, respectively.}
\label{fig:rabi_posterior_learning_heatmaps}
\end{figure}

The four posterior snapshots in \cref{fig:rabi_posterior_learning_heatmaps} show the evolution along the same signal trajectory from $k=0$ (before the first shot) to $k=4096$.
Yellow regions carry the largest conditional $H_1$ posterior density and purple regions the smallest; the white dashed and dotted lines mark the injected $\tilde{\varphi}$ and $\Phi_R$, respectively.
The initially broad distribution first contracts in amplitude and subsequently resolves the drive phase, ultimately localizing around the injected values.

\subsection{Fixed budget detection performance}
\label{subsec:fixed_budget}

At a fixed injected amplitude, we average the pointwise power in \cref{eq:calibrated_power} over the nuisance phase using its conditional prior.
For the numerical prior in \cref{eq:numerical_signal_prior}, the conditional phase distribution is uniform, and we denote the resulting phase-averaged power by
\begin{align}
    \overline{P}^\mu(\Phi_R;\alpha)
    &\equiv
    \int_0^{2\pi}\frac{\dd\tilde{\varphi}}{2\pi}
    \left[
        1-\beta^\mu((\Phi_R,\tilde{\varphi});\alpha)
    \right].
    \label{eq:phase_composite_detection_power}
\end{align}
Here $\mu$ labels the readout policy and its corresponding calibrated test, whereas $(\Phi_R,\tilde{\varphi})$ labels the fixed signal model used to generate a record.
For a different phase prior, the measure $\dd\tilde{\varphi}/(2\pi)$ in \cref{eq:phase_composite_detection_power} would be replaced by the corresponding conditional measure at fixed $\Phi_R$.

\begin{figure}[t]
\centering
\includegraphics[width=\columnwidth]{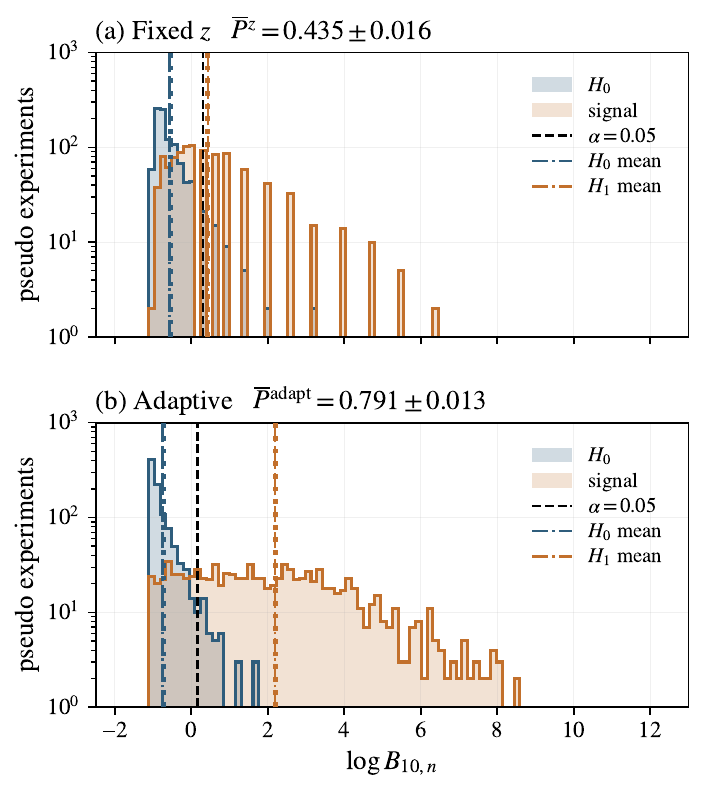}
\caption{Final log Bayes factor distributions under $H_0$ and signal records generated at fixed $\Phi_R=0.19$ with uniformly sampled phase, for the fixed-$z$ policy (upper panel) and the pure information gain adaptive policy (lower panel) at $n=1024$.
The black dashed line in each panel is the policy specific threshold calibrated to $\alpha\simeq0.05$.
Dotted and dash-dotted lines show the mean estimates from the KL identities in \cref{appsubsec:policy_connection} and the sample means, respectively, with colors matching the histograms.}
\label{fig:rabi_log_bayes_histograms}
\end{figure}

At the representative baseline benchmark $\Phi_R=0.19$ and $n=1024$, \cref{fig:rabi_log_bayes_histograms} compares the distributions of $\log B_{10,n}$ for the fixed-$z$ and adaptive policies.
For each policy, the distributions are constructed from $10^3$ pseudoexperiments under $H_0$ and $10^3$ signal pseudoexperiments at fixed $\Phi_R=0.19$, with $\tilde{\varphi}$ sampled uniformly.
The Bayes factor continues to compare $H_0$ with the full composite hypothesis $H_1$; fixing $\Phi_R$ only selects the signal points at which the resulting test's power is evaluated.
Blue and orange histograms in \cref{fig:rabi_log_bayes_histograms} denote the null and fixed amplitude, phase-averaged signal ensembles, respectively, and the black dashed line in each panel denotes the policy specific detection threshold.
The adaptive policy shifts more of the signal distribution above its threshold and reaches a power of $0.79$, compared with $0.44$ for the fixed-$z$ policy.
The comb structure of the fixed-$z$ histogram arises because the integer count of positive outcomes is sufficient for the resonant fixed-$z$ likelihood: all records with the same count therefore yield the same value of $\log B_{10,n}$.\footnote{
    We calibrate the fixed-$z$ threshold directly with the discrete $H_0$ ensemble and evaluate the power without smoothing the histogram or applying a continuum approximation. The resulting empirical Type-I error is $0.051$.
}

\begin{table}[t]
\centering
\caption{Phase-averaged power at $\Phi_R=0.19$, $n=1024$, and calibrated $\alpha\simeq0.05$ for the baseline detector profile.
Each entry is estimated from $10^3$ phase-averaged signal pseudoexperiments.}
\label{tab:rabi_fixed_budget_power}
\begin{tabular}{lc}
\toprule
policy & power $\overline{P}^\mu$\\
\midrule
fixed-\(z\) & \(0.44\)\\
fixed-\(x\) & \(0.60\)\\
fixed-\(x/y\) & \(0.67\)\\
adaptive (information gain, $L=1$) & \(0.79\)\\
\bottomrule
\end{tabular}
\end{table}

The powers of all four policies are summarized in \cref{tab:rabi_fixed_budget_power}.
For every policy, the threshold is calibrated independently with the corresponding $H_0$ ensemble.
Thus the baseline profile preserves a large adaptive gain over the fixed-$z$ policy and the non-adaptive transverse policies at this benchmark.
As a cross check that the use of $\log B_{10}$ does not artificially reduce the apparent performance of the fixed policies, \cref{app:count-based} evaluates conventional count-based tests analytically.
Their powers agree closely with the corresponding Bayes factor results in \cref{tab:rabi_fixed_budget_power}.

\subsection{Phase-averaged power and sensitivity scaling}
\label{subsec:numerical_sensitivity}

We suppress the superscript $\mu$ when referring to one policy curve in the figures below.
For each shot count we scan $\Phi_R$, construct the phase-averaged power curve, and interpolate the amplitude at which a specified target power is reached.
To compare different shot counts on the scale natural to a transverse measurement, we define
\begin{align}
    \rho
    &=
    \Phi_R\sqrt{n}.
    \label{eq:transverse_scaling_variable}
\end{align}

\begin{figure}[t]
\centering
\includegraphics[width=\columnwidth]{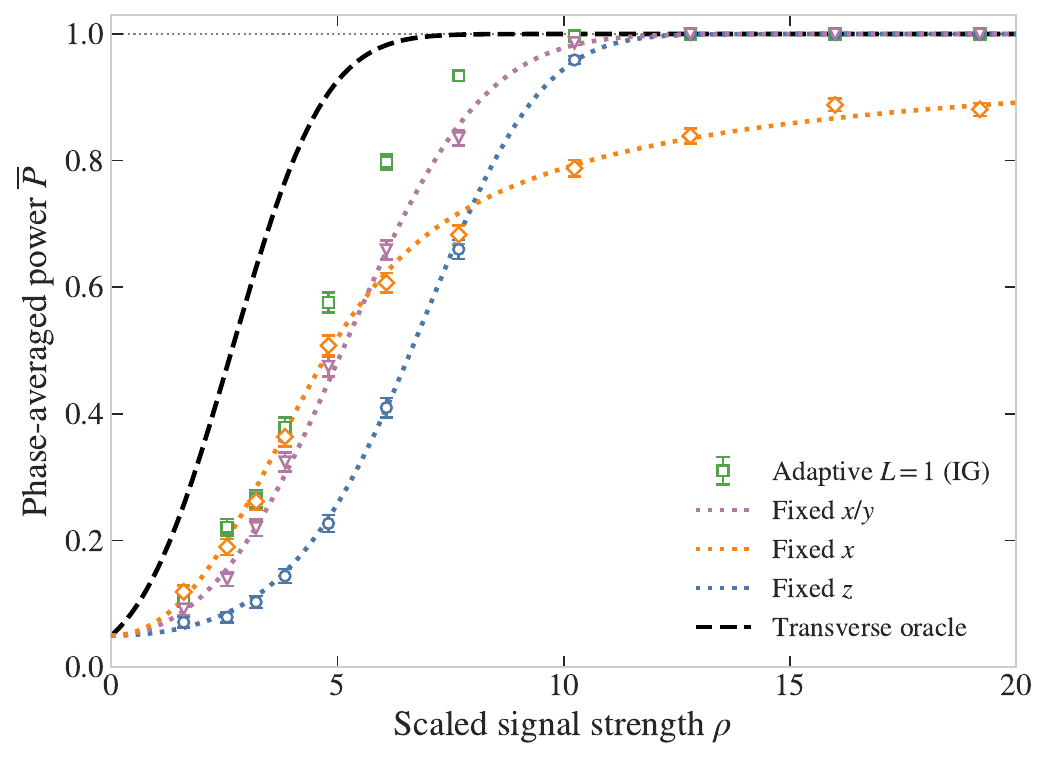}
\caption{Phase-averaged power versus the scaled signal strength $\rho=\Phi_R\sqrt n$ for the baseline detector profile at $n=1024$ and $\alpha\simeq0.05$.
Markers show Monte Carlo estimates, and the fixed policy curves show the corresponding analytic benchmarks.}
\label{fig:rabi_discovery_probability_n1024}
\end{figure}

\Cref{fig:rabi_discovery_probability_n1024} shows the resulting phase-averaged power curves for the baseline detector profile at $n=1024$.
For the transverse oracle, fixed-$x$, fixed-$x/y$, and a successful adaptive policy, \cref{eq:main_transverse_sensitivity} predicts that a constant value of $\rho$ gives a constant $\overline{P}$ in the weak signal limit.
Once the adaptive policy has enough shots to use its posterior to maintain a non-vanishing root mean square linear response, its curves and data points therefore become approximately independent of $n$ when plotted against $\rho$.
Among the transverse baselines, fixed-$x$ approaches unit phase-averaged power especially slowly because its linear response vanishes for signal phases near the orthogonal quadrature.
These single axis blind spots remain difficult to detect until much larger $\rho$, whereas alternating fixed-$x/y$ readout retains sensitivity to both quadratures.
In contrast, \cref{eq:main_fixed_z_sensitivity} gives $\Phi_z^{\rm req}\sqrt{n}\propto n^{1/4}$, so the fixed-$z$ curve moves progressively to the right and performs increasingly poorly relative to the transverse strategies as $n$ grows.

\begin{figure}[t]
    \centering
    \includegraphics[width=\columnwidth]{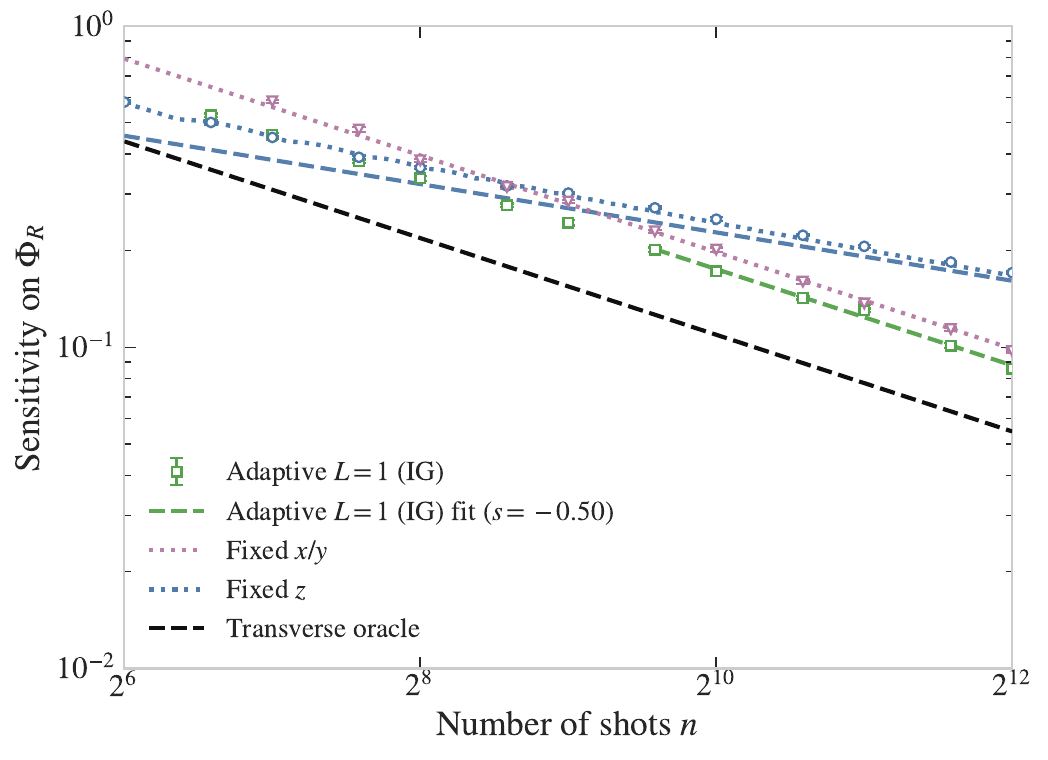}
\caption{Sensitivity to the Rabi amplitude at phase-averaged power $\overline{P}=0.7$ and $\alpha\simeq0.05$ as a function of the shot count.
Monte Carlo sensitivities are compared with the adaptive tail fit and analytic fixed policy benchmarks.}
\label{fig:rabi_discovery_sensitivity_p07}
\end{figure}

The sensitivity is shown directly at the representative intermediate phase-averaged power $\overline{P}=0.7$ in \cref{fig:rabi_discovery_sensitivity_p07}.
The green markers are Monte Carlo sensitivities for the pure information gain policy $L=1$.
The green dashed line is a tail fit to the adaptive sequence.
The fitted slope is $-0.50$, consistent with the $n^{-1/2}$ scaling in \cref{eq:main_transverse_sensitivity}.
The blue dashed curve is the asymptotic fixed-$z$ result $A_z(\alpha,P)n^{-1/4}$ in \cref{eq:main_fixed_z_sensitivity}, while the blue dotted curve is its finite background Poisson counterpart.
The purple dotted curve is $A_{xy}(\alpha,P)n^{-1/2}$, and the black dashed curve is the oracle result $A_{{\rm O}_\perp}(\alpha,P)n^{-1/2}$.
Fixed-$x$ is omitted for clarity because its phase blind spots make its sensitivity coefficient particularly dependent on the chosen target $\overline{P}$, as seen in \cref{tab:main_sensitivity_coefficients}.
At small $n$, the adaptive sensitivity approximately follows fixed-$z$.
It then bends onto an $n^{-1/2}$-like trajectory and becomes the most sensitive non-oracle strategy shown, outperforming fixed-$x/y$ while remaining above the transverse oracle, reflecting the finite budget cost of learning the nuisance direction.

\subsection{Type-II error exponents}
\label{subsec:numerical_exponents}

For comparison with the fixed signal rates in \cref{subsec:error_exponents}, we fix $\alpha=0.05$, suppress it from the notation below, and define the phase-averaged Type-II error at fixed Rabi amplitude by
\begin{align}
    \beta_n^\mu(\Phi_R)
    &=
    1-\overline{P}^\mu(\Phi_R;\alpha).
    \label{eq:numerical_phase_averaged_beta}
\end{align}
The phase average is taken before the logarithm: $\beta_n^\mu$ is the miss probability for an ensemble with a uniform phase drawn once per record.
Because the simulated $\beta_n^\mu$ is resolved only over a finite range of shot counts, we quantify its decay by the finite window effective exponent
\begin{align}
    \xi_{\rm eff}^{\mu,[n_a,n_b]}(\Phi_R)
    &=
    \frac{\log\beta_{n_a}^\mu(\Phi_R)-\log\beta_{n_b}^\mu(\Phi_R)}{n_b-n_a},
    \label{eq:finite_window_effective_exponent}
\end{align}
where $n_a<n_b$ are the endpoints of the chosen shot count window.
This secant slope measures the decay of the phase-averaged miss probability over that window.
If pointwise exponential decay holds with sufficiently uniform convergence over phase, the asymptotic rate of the phase mixture is the essential infimum of the pointwise rates.

\begin{figure}[t]
\centering
\includegraphics[width=\columnwidth]{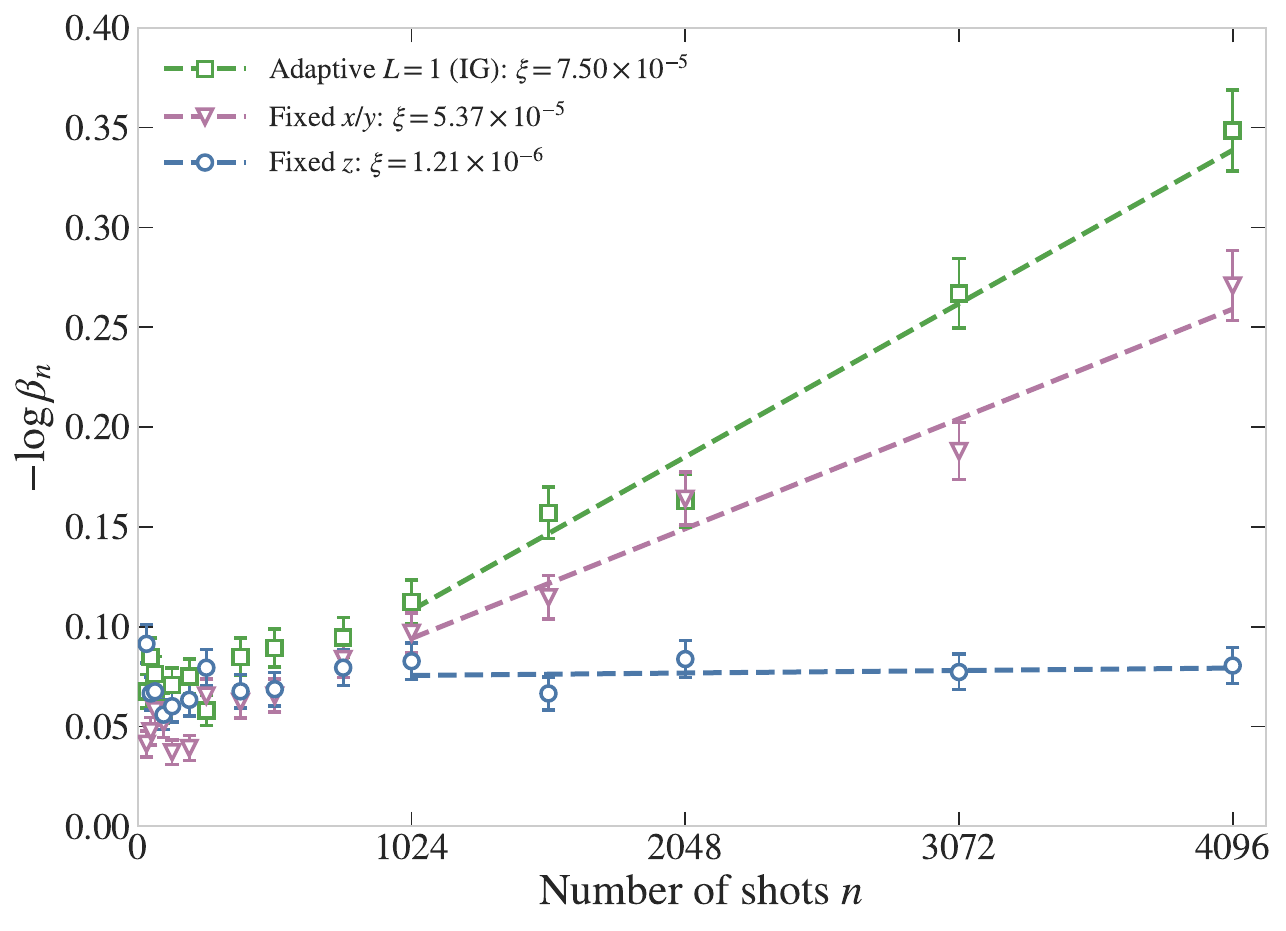}
\caption{Negative logarithm of the phase-averaged Type-II error, $-\log\beta_n$, versus shot count at $\Phi_R=0.05$ and $\alpha=0.05$.
Markers show the Monte Carlo estimates.
Dashed lines are linear fits over $1024\le n\le4096$ using resolved points, with the fitted slopes reported in the legend.}
\label{fig:rabi_beta_minus_log}
\end{figure}

\Cref{fig:rabi_beta_minus_log} first shows the underlying decay at the representative weak signal $\Phi_R=0.05$.
The thresholds are recalibrated at every shot count to maintain $\alpha=0.05$, and $-\log\beta_n$ becomes approximately linear over the resolved range of large $n$.
The fitted adaptive slope is larger than those for fixed-$x/y$ and fixed-$z$, demonstrating a faster finite range suppression of the phase-averaged Type-II error.
The approximate linearity describes the resolved finite window and does not establish an asymptotic adaptive exponent.

\begin{figure}[t]
\centering
\includegraphics[width=\columnwidth]{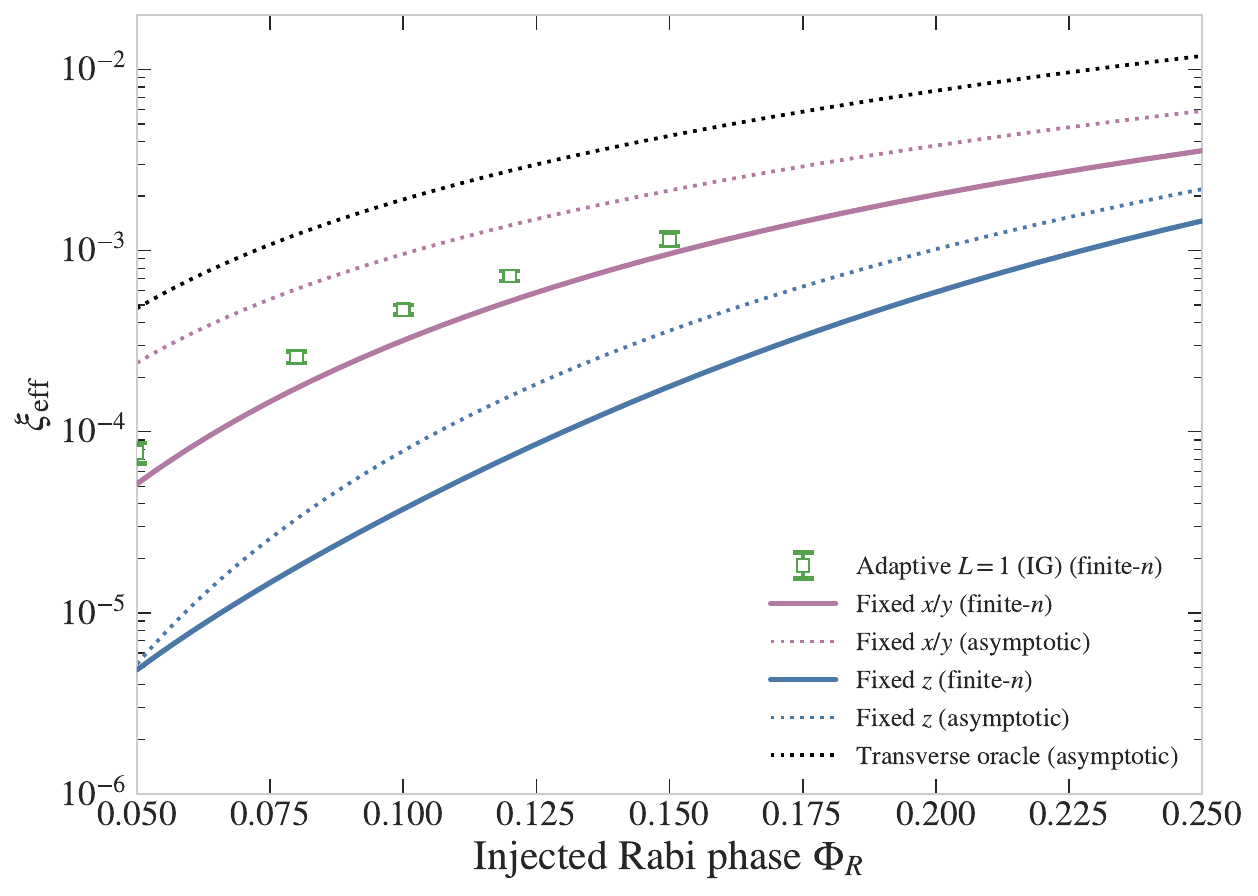}
\caption{Finite window Type-II effective exponents at $\alpha=0.05$, evaluated between $n_a=1024$ and $n_b=4096$.
Green error bars show adaptive estimates with the endpoint uncertainty ranges described in the text, and solid curves are finite window count-based benchmarks.
Dotted curves show asymptotic KL references, using the minimum over phase for fixed-$x/y$ and the known phase for the transverse oracle.
Only amplitudes for which the adaptive error remains numerically resolved at both endpoints are shown.}
\label{fig:rabi_beta_relative_entropy}
\end{figure}

\Cref{fig:rabi_beta_relative_entropy} compares \cref{eq:finite_window_effective_exponent} using the endpoint pair $(n_a,n_b)=(1024,4096)$ for the baseline detector profile.
The solid blue curve is the finite window prediction from an exact size randomized binomial test for fixed-$z$, while the solid purple curve is the corresponding Gaussian non-central $\chi_2^2$ prediction for fixed-$x/y$.
The adaptive estimates use the calibrated Bayes factor, whereas the fixed policy curves use count-based tests as in \cref{app:count-based}; the shot window and target Type-I error are common.
The error bars span the minimum and maximum secant slopes obtained from the marginal $68\%$ Jeffreys intervals at the two endpoints, with the calibrated thresholds held fixed.

The dotted fixed policy curves give the per shot Bernoulli KL, minimized over phase for fixed-$x/y$; the transverse oracle reference assumes that the signal direction is supplied.
These asymptotic references are not finite budget bounds on the adaptive data.
The exact fixed-$z$ test approaches its KL rate, whereas the Gaussian fixed-$x/y$ calculation reproduces the Bernoulli KL rate only to leading order in the weak signal.

Every resolved adaptive point lies above both fixed policy benchmarks for the same finite window, with the fixed-$x/y$ comparison subject to its Gaussian approximation.
The gaps between the solid and dotted curves include finite sample effects and, for fixed-$x/y$, the difference in approximation.

\section{Discussion}
\label{sec:discussion}

Having established the resonant fixed budget benchmark, we now discuss the assumptions that delimit it and the extensions required for a more general sensing protocol.
We consider unknown detuning in \cref{subsec:detuning}, global control in \cref{subsec:global_optimization}, joint optimization of probe and measurement in \cref{subsec:probe_optimization}, and computational constraints in \cref{subsec:real_time}.

\subsection{Detuning}
\label{subsec:detuning}

As a check that the adaptive policy can learn a time dependent signal direction, we repeat the trajectory diagnostic with detuning included as an unknown parameter using the higher fidelity profile defined in \cref{appsubsec:higher_fidelity}.
The readout azimuth in \cref{fig:rabi_posterior_learning_detuned_angles} tracks the phase accumulated through the inferred detuning.
The late time alignment with the drifting reference directions demonstrates successful posterior learning of the signal phase at non-zero detuning.
The numerical prior, grid, injected trajectory, and readout angle conventions are detailed in \cref{appsubsec:detuned_trajectory}.

A broadband detuning search enlarges the latent signal space.
Maintaining a fixed global Type-I error therefore raises the rejection threshold, while resolving the enlarged posterior also increases the numerical cost even though the per shot likelihood model is unchanged.
For a detuning band of fixed physical width $W_\Delta$, let $N_{\Delta,{\rm eff}}$ denote the number of statistically distinct detuning cells resolved over the total record duration $nT$.
Up to constants set by the window and sampling pattern, it scales as
\begin{align}
    N_{\Delta,{\rm eff}}
    &=
    O(W_\Delta nT).
    \label{eq:effective_detuning_trials}
\end{align}
A numerical detuning grid must contain at least a comparable number of points, so a range of width $O(T^{-1})$ requires $O(n)$ points.
In a Gaussian matched filter approximation, calibrating the global Type-I error over these cells gives the transverse sensitivity the logarithmic search penalty
\begin{align}
    \Phi_\perp^{\rm req}(n;\alpha,P)
    &\sim
    \sqrt{\frac{\log N_{\Delta,{\rm eff}}}{n}},
    \label{eq:detuning_search_sensitivity}
\end{align}
up to coefficients that depend on the policy and target power.
Thus a fixed detuning bandwidth preserves the transverse $n^{-1/2}$ power law exponent but adds a slowly varying $\sqrt{\log n}$ factor.

In frequentist terms, the factor $\log N_{\Delta,{\rm eff}}$ is the look elsewhere effect: under $H_0$, each additional resolvable detuning cell provides another opportunity for a background fluctuation to resemble a signal, so the global rejection threshold must increase.
In Bayesian terms, the same scaling is an Occam penalty: spreading the alternative prior over $N_{\Delta,{\rm eff}}$ distinguishable cells reduces the prior mass assigned to the cell supported by the data by a factor of order $N_{\Delta,{\rm eff}}^{-1}$.

For the full Nyquist range $|\Delta|<\pi T^{-1}$ at $n=4096$, the ideal Fourier bin estimate $N_{\Delta,{\rm eff}}=n$ gives a sensitivity penalty of approximately 1.87 at $\alpha=0.05$ and $\overline{P}=0.7$.
The $\sqrt{\log n}$ scaling applies parametrically to any transverse strategy that preserves coherent $n^{-1/2}$ sensitivity by searching or marginalizing over detuning, including Fourier-analyzed fixed-$x/y$ and adaptive readout.
At leading order, population readout does not search the temporal phase drift, so the broadband penalty delays its crossing with a transverse strategy.
If this penalty is treated as a local multiplicative increase of the transverse sensitivity coefficient, the fourth power dependence in \cref{eq:main_crossing_estimate} increases the crossing shot count by roughly one order of magnitude, although the asymptotic $\sqrt{\log n/n}$ sensitivity eventually overtakes the fixed-$z$ scaling $n^{-1/4}$.

A detuned sensitivity curve would require null ensembles for threshold calibration and signal ensembles over $(\Phi_R,\Delta,\tilde{\varphi})$ at several values of $n$, with a fixed physical detuning prior as $n$ varies.
We leave this substantially larger ensemble calculation for future work.

\subsection{Beyond myopic control}
\label{subsec:global_optimization}

The failure of accumulated one step optima to produce a globally optimal policy can be seen directly in the present Rabi problem.
A one step objective that appears naturally aligned with evidence accumulation is the expected Bayes drift
\begin{align}
    U_{\rm B}(\bm n;S_k)
    &\equiv
    \mathbb{E}_{R_k\sim m_k}
    \left[
        \log B_{10,k+1}-\log B_{10,k}
    \right]
    \notag\\
    &=
    D_{\rm KL}\!\left(
        m_k(\cdot|\bm n)
        \middle\Vert
        p_{0,k}(\cdot|\bm n)
    \right).
    \label{eq:myopic_bayes_drift}
\end{align}
The second equality follows immediately from the update in \cref{eq:log_bayes_update}, with the KL divergence defined in \cref{eq:classical_kl}.
Suppose that the current resonant posterior is uniform in the unknown drive phase,
$\pi_k(\Phi_R,\tilde{\varphi})=\pi_k(\Phi_R)/(2\pi)$.
Phase averaging \cref{eq:rabi_quantum_bloch_vector} gives the predictive signal and null Bloch vectors
\begin{align}
    \overline{\bm r}_{1,k}
    &=
    C\!\left(
        0,0,
        1-\eta_1
        \left\langle1-\cos\Phi_R\right\rangle_{\pi_k}
    \right),
    \notag\\
    \bm r_0
    &=
    C(0,0,1).
    \label{eq:phase_averaged_predictive_bloch_vectors}
\end{align}
Here $\langle\cdot\rangle_{\pi_k}$ denotes averaging over the marginal amplitude posterior $\pi_k(\Phi_R)$.
These two states commute and have the $z$ basis as their common eigenbasis.
By data processing, any other binary projective readout is a stochastic coarse graining of this common eigenbasis measurement and cannot increase the measured relative entropy in \cref{eq:myopic_bayes_drift}; the symmetric readout error channel used here leaves the same optimum.
Consequently, the myopic Bayes drift policy selects $\bm n=\pm\bm n_z$.
The population readout likelihood is independent of $\tilde{\varphi}$, so its outcome leaves the phase posterior uniform.
The same argument therefore applies at the next shot and inductively enforces the fixed-$z$ policy, which has been shown not to maximize the expectation value of $\log B_{10,n}$.

For the global objective of maximizing the expected final log Bayes factor under $H_1$, a finite horizon policy can instead be written as a Bellman recursion on the posterior state,
\begin{align}
    V_n(S)
    &=
    \log B_{10}(S),
    \label{eq:log_bayes_terminal_reward}\\
    V_k(S)
    &=
    \max_u\sum_r m_k(r|u,S)
    V_{k+1}\!\left(\mathcal{F}(S,u,r)\right).
    \label{eq:log_bayes_bellman_recursion}
\end{align}
Here $V_k(S)$ is the optimal conditional expectation of the terminal log Bayes factor after $k$ shots when the current Bayesian state is $S$, and $\mathcal{F}(S,u,r)$ is the Bayesian state update map after outcome $r$.
At the final decision step, \cref{eq:log_bayes_terminal_reward,eq:log_bayes_bellman_recursion} reduce to maximizing \cref{eq:myopic_bayes_drift}; at earlier steps, the value function additionally accounts for how the measurement changes all future controls.
If the objective is instead the prior average of the calibrated detection power $\Prob_{\btheta}^\mu(\log B_{10,n}\ge c_\alpha^\mu)$ in \cref{eq:calibrated_power}, the terminal reward is one when $H_0$ is rejected and zero otherwise.
The policy dependent null threshold $c_\alpha^\mu$ must be calibrated as part of the outer optimization.

Solving the Bellman equation exactly is difficult because the posterior is continuous and the control tree grows exponentially.
Truncated lookahead and rollout approximate the Bellman tree directly.
Complementary approaches optimize a restricted policy offline using machine learning or reinforcement learning methods, as explored in adaptive quantum metrology, multiparameter estimation, and quantum multiple hypothesis testing~\cite{PhysRevLett.107.233601,Cimini:2022krl,Belliardo:2023vzb,Brandsen:2020hvo}.
On the experimental side, adaptive learning has been combined with multiparameter photonic metrology under limited resources~\cite{PhysRevResearch.5.013138}.
For a broader review of machine learning in quantum metrology, see Ref.~\cite{Huang:2025mtw}.

\subsection{Joint optimization of probe and measurement}
\label{subsec:probe_optimization}

The present study fixes the initial state and optimizes only the final measurement.
A natural extension is to optimize the probe preparation, sensing time, intermediate controls, and final measurement jointly.
Because the hypotheses then specify alternative quantum evolutions acting on an optimized input, this extension is naturally viewed as a resource-constrained composite channel discrimination problem, where adaptive strategies can be advantageous~\cite{HarrowEtAl:2010,SalekHayashiWinter:2022,BerghDattaSalzmann:2023}.
Variational quantum algorithms offer practical ansatzes for jointly training probe preparation and measurement or for optimizing channel discrimination circuits with intermediate quantum operations~\cite{KardashinEtAl:2022,Subramanian:2023rbv}.
Related work on NV sensors combines detection and parameter estimation with reinforcement learning for adaptive sensing control~\cite{Taherpour:2025oxv}.
Adapting these methods to the present setting, including nuisance parameter averaging and calibration at fixed global Type-I error, is a possible direction for future work.

\subsection{Computational cost and real time implementation}
\label{subsec:real_time}

For a particle representation, the per shot cost of a myopic scan is roughly
\begin{align}
    \text{cost per shot}
    &\sim
    \mathcal{O}(N_{\rm part}N_{\rm cand}).
    \label{eq:cost_per_shot}
\end{align}
Here $N_{\rm part}$ is the number of particles representing the signal parameter posterior and $N_{\rm cand}$ is the number of candidate controls evaluated per shot.
The two factors in \cref{eq:cost_per_shot} suggest complementary routes to reducing the online computational burden.
Adaptive posterior representations, including posterior compression, adaptive grids, and sequential Monte Carlo, can lower the required $N_{\rm part}$.
Rejection filtering in Bayesian phase estimation provides an example of reducing the memory required for posterior updates~\cite{PhysRevLett.117.010503}.
Local control optimization can reduce $N_{\rm cand}$, and an offline-trained policy can eliminate the online candidate scan altogether.
GPU acceleration and asynchronous operation have been investigated to reduce or hide the overhead of Bayesian design~\cite{WangEtAl:2022Realtime}, while neural network heuristics have been proposed as fast substitutes for online experiment design optimization~\cite{FidererEtAl:2021}.

Reducing the online computational burden is not sufficient; the resulting latency must also fit within the physical shot cycle.
Let $t_{\rm fb}$ denote the dead time required to update the posterior and select the next measurement control.
The overhead free timing $t_k=kT$ assumed in the simulations is a good approximation when $t_{\rm fb}\ll T$, so that the adaptive calculation is completed within a small fraction of one interrogation time.
For a superconducting qubit implementation with coherence and interrogation times on the scale of $\unit{\micro\second}$, this motivates a feedback path shorter than $\unit{\micro\second}$, although the precise requirement is set by the physical shot cycle.
A digital signal processing system implemented on a field programmable gate array has demonstrated a feedback trigger latency of \qty{110}{\nano\second} and a complete feedback loop latency of \qty{352}{\nano\second}~\cite{SalatheEtAl:2018}.
In a real time deep reinforcement learning implementation, a minimum feedback cycle duration of \qty{767}{\nano\second} is reported, including a \qty{48}{\nano\second} contribution from latency-optimized neural network execution~\cite{Reuer:2022efh}.
As a scale comparison outside this domain, the CMS Level-1 trigger, implemented in custom hardware, makes an accept/reject decision within approximately \qty{4}{\micro\second} at the nominal \qty{40}{\mega\hertz} bunch crossing rate~\cite{CMS:2016ngn}.

\section{Conclusion}
\label{sec:conclusion}

We have formulated and analyzed a fixed horizon protocol based on repeated reset and measurement for asymmetric testing of a composite signal hypothesis.
Within a coherent record segment, one unknown signal parameter vector is shared across all shots, while each readout axis may be chosen causally from the accumulated classical history.
We use the complete record Bayes factor as the test statistic and calibrate a separate null threshold for each policy, so that fixed and adaptive strategies are compared at a common Type-I error.
This construction separates the shot local utility used for measurement selection from the final performance criterion of calibrated finite budget power.
The accompanying record level analysis clarifies the distinct roles of mutual information, total variation, and conditional KL divergence, and explains why a seemingly detection-oriented greedy Bayes drift rule can become locked into phase insensitive population readout.

For the resonant Rabi benchmark with unknown amplitude and transverse phase, the weak signal expansion identifies a quadratic, phase independent population response and a linear, direction dependent transverse response.
We derived detector dependent sensitivity coefficients and crossover estimates for the fixed policies and the known direction transverse oracle.
In particular, fixed-$z$ readout has $n^{-1/4}$ sensitivity, whereas the phase robust fixed-$x/y$ schedule already has $n^{-1/2}$ sensitivity.
The main role of adaptivity is to use posterior information about the nuisance direction to improve the finite coefficient relative to fixed transverse schedules.

For the baseline detector profile, the pure information gain policy was the best among the tested myopic schedules.
At the representative point $\Phi_R=0.19$, $n=1024$, and $\alpha\simeq0.05$, its phase-averaged power is approximately $0.79$, compared with $0.67$ for fixed-$x/y$ and $0.44$ for fixed-$z$.
Across the simulated large $n$ range, its sensitivity is consistent with $n^{-1/2}$ and, at the largest simulated shot counts, improves upon the fixed transverse schedules considered.
Count-based cross-checks give powers consistent with the fixed policy Bayes factor results, indicating that this advantage is not an artifact of an unnecessarily weak baseline statistic.
Over the window $1024\leq n\leq4096$, the phase-averaged finite window effective Type-II exponent exceeds the finite $n$ fixed policy references at every resolved amplitude.

These quantitative conclusions are restricted to the chosen resonant, phase-averaged benchmark and do not constitute either a worst case composite guarantee or a universal asymptotic adaptive advantage.
The preferred myopic policy also depends on detector profile and shot budget: for the higher fidelity profile, a periodic information gain/Helstrom hybrid was favored among the tested schedules at a representative target power.
With nonzero detuning, a narrow prior trajectory demonstrates tracking of the drifting signal direction, and a Gaussian matched filter argument suggests the parametric sensitivity penalty $\sqrt{\log n}$ ; a fully calibrated broadband power study remains for future work.
Further extensions include globally optimizing the finite horizon policy against the final calibrated objective, enlarging the control space to include probe preparation and interrogation time, and developing low latency and model robust posterior updates.

\begin{acknowledgments}
The author thanks Jesse Thaler and Soonwon Choi for helpful comments.
The author thanks Hajime Fukuda and Zachary Bogorad for useful discussions on sensitivity scaling.
OpenAI systems disclosed in \cref{subsec:numerical_procedure} also assisted with drafting and revising the manuscript.
All AI-assisted text was reviewed and edited by the author, who takes full responsibility for the content of the manuscript.
The author acknowledges support from the Simons Foundation.
This material is based upon work supported by the U.S. Department of Energy, Office of Science, National Quantum Information Science Research Centers, Co-design Center for Quantum Advantage (C2QA) under Contract No.~DE-SC0012704 (Subaward No.~390034).
\end{acknowledgments}

\appendix

\section{Classical and quantum distinguishability}
\label[appendix]{app:hypothesis_testing}

The purpose of this appendix is to place the distinguishability measures used in the main text on a common operational footing and to identify which of them are attainable under the measurement restrictions of the sensing protocol.
We track how distinguishability is transferred from a quantum state to the classical record and connect the weak drive limit to Fisher information.
The standard operational correspondences used throughout the appendix are summarized in \cref{tab:distinguishability_summary}.
The first three subsections review standard results for known simple hypotheses; \cref{appsubsec:local_fisher_information} gives their local Fisher information expansions, and \cref{appsubsec:policy_connection} connects these measures to adaptive readout and complete record detection.

\begin{table*}[t]
\centering
\caption{Optimal distinguishability measures for the three operational tasks discussed in this appendix.}
\label{tab:distinguishability_summary}
\begin{tabular}{lcc}
\toprule
Operational task & Classical optimum & Quantum optimum\\
\midrule
Single shot minimum error discrimination
&
$D_{\rm TV}(P_0,P_1)$
&
$D_{\rm tr}(\rho_0,\rho_1)$
\\
i.i.d.\ exponent at fixed $\alpha$,
individual one copy measurement
&
$D_{\rm KL}(P_0\Vert P_1)$
&
$D_{\rm meas}(\rho_0\Vert\rho_1)$
\\
i.i.d.\ exponent at fixed $\alpha$,
collective measurement on $n$ copies
&
$D_{\rm KL}(P_0\Vert P_1)$
&
$D(\rho_0\Vert\rho_1)$
\\
\bottomrule
\end{tabular}
\end{table*}

\subsection{Hypothesis testing conventions}
\label[appendix]{appsubsec:testing_conventions}

When discussing a pair of simple classical hypotheses, $P_0$ and $P_1$ denote the distributions under $H_0$ and $H_1$, respectively.
For the complete record generated by a fixed policy $\mu$ at a fixed signal point $\btheta$, they specialize to $\Prob_0^\mu$ and $\Prob_{\btheta}^\mu$ entering \cref{eq:type_i_error_main,eq:type_ii_error_main}, respectively.
The quantum counterparts are denoted by $\rho_0$ and $\rho_1$.
The decision rule underlying \cref{eq:type_i_error_main,eq:type_ii_error_main} selects $H_1$ when $\log B_{10,n}\ge c$ and selects $H_0$ otherwise; the test at fixed $\alpha$ uses $c=c_\alpha^\mu$ as specified in \cref{eq:calibrated_threshold}.

For $i\in\{0,1\}$, a positive operator-valued measure (POVM) $\mathsf{M}=\{M_r\}_{r\in\mathcal{J}}$ with outcome space $\mathcal{J}$ maps $\rho_i$ to the classical outcome distribution
\begin{align}
    p_i^{\mathsf{M}}(r)
    &=
    \operatorname{Tr}(M_r\rho_i),
    \label{eq:povm_induced_distribution}\\
    M_r
    &\ge0,
    \label{eq:povm_positivity}\\
    \sum_rM_r
    &=
    \mathbb{I}.
    \label{eq:povm_completeness}
\end{align}
$P_i$ denotes an arbitrary classical distribution under $H_i$, whereas $p_i^{\mathsf{M}}$ is specifically the distribution induced by measuring $\rho_i$ with $\mathsf{M}$.
Different POVMs generally produce different classical distinguishability from the same pair of quantum states.

\subsection{Single shot discrimination}
\label[appendix]{appsubsec:helstrom}

We first consider minimum error discrimination between two simple hypotheses from a single classical observation or a single copy of a quantum state, assuming equal prior probabilities for $H_0$ and $H_1$.
For classical distributions, the optimal decision rule selects the hypothesis with the larger likelihood for the observed outcome, and its minimum average error is given by \cref{eq:classical_bayes_error} in terms of the total variation distance.
For quantum states, the optimization also includes the choice of measurement; the Helstrom theorem gives \cref{eq:helstrom_error} in terms of the trace distance, with the optimum attained by the measurement in \cref{eq:helstrom_pvm} \cite{Helstrom:1976}.
The definitions and short derivations of these standard results are given below.

\subsubsection{Classical total variation distance}

For two distributions on a discrete outcome space $\mathcal{J}$, the total variation distance is
\begin{align}
    D_{\rm TV}(P_0,P_1)
    &=
    \sup_{A\subseteq\mathcal{J}}|P_0(A)-P_1(A)|
    \notag\\
    &=
    \frac12\sum_{r\in\mathcal{J}}
    |P_0(r)-P_1(r)|.
    \label{eq:total_variation_definitions}
\end{align}
To see the second equality, define $d_r=P_0(r)-P_1(r)$ and $A_+=\{r:d_r>0\}$.
Since $\sum_r d_r=0$, the positive and negative parts have equal total magnitude:
\begin{align}
    \sum_{r\in A_+}d_r
    &=
    -\sum_{r\notin A_+}d_r
    =
    \frac12\sum_r|d_r|.
\end{align}
The event $A_+$ maximizes $P_0(A)-P_1(A)$, while its complement gives the opposite sign.
For Bernoulli distributions the expression reduces to
\begin{align}
    D_{\rm TV}\bigl(\Bern(p_0),\Bern(p_1)\bigr)
    &=|p_0-p_1|.
    \label{eq:bernoulli_total_variation}
\end{align}

It follows that the minimum average error for the classical single shot problem is
\begin{align}
    P_{\rm err}^{\rm cl}
    &=
    \frac12\left[1-D_{\rm TV}(P_0,P_1)\right].
    \label{eq:classical_bayes_error}
\end{align}

\subsubsection{Quantum trace distance}

The corresponding quantum distinguishability measure is the trace distance
\begin{align}
    D_{\rm tr}(\rho_0,\rho_1)
    &=
    \frac12\|\rho_0-\rho_1\|_1,
    \label{eq:trace_distance}\\
    \|X\|_1
    &=
    \operatorname{Tr}\sqrt{X^\dagger X}.
    \label{eq:trace_norm}
\end{align}
It equals the largest total variation distance obtainable by measuring the two states:
\begin{align}
    D_{\rm tr}(\rho_0,\rho_1)
    &=
    \max_{\mathsf{M}}
    D_{\rm TV}
    \left(p_0^{\mathsf{M}},p_1^{\mathsf{M}}\right).
    \label{eq:trace_distance_measured_tv}
\end{align}
For completeness, write $\Delta_\rho=\rho_0-\rho_1=\Delta_{\rho,+}-\Delta_{\rho,-}$, where $\Delta_{\rho,\pm}\ge0$ have orthogonal support.
For any POVM,
\begin{align}
    \frac12\sum_r|\operatorname{Tr}(M_r\Delta_\rho)|
    &\le
    \frac12\operatorname{Tr}(\Delta_{\rho,+}+\Delta_{\rho,-})
    =
    \frac12\|\Delta_\rho\|_1.
\end{align}
Equality is obtained by the binary projective measurement
\begin{align}
    \mathsf{M}_{\rm H}
    &=\{\Pi_+,\mathbb{I}-\Pi_+\},
    \label{eq:helstrom_pvm}
\end{align}
where $\Pi_+$ projects onto the positive eigenspace of $\Delta_\rho$.
Thus, for discrimination between two states, optimization over all POVMs is exhausted by the binary projection-valued measure (PVM) in \cref{eq:helstrom_pvm}.
The resulting Helstrom error is
\begin{align}
    P_{\rm err}^{\rm Hel}
    &=
    \frac12\left[1-D_{\rm tr}(\rho_0,\rho_1)\right].
    \label{eq:helstrom_error}
\end{align}
For unequal prior probabilities $w_0$ and $w_1$ assigned to $H_0$ and $H_1$, respectively, the same construction is applied to $w_0\rho_0-w_1\rho_1$ \cite{Helstrom:1976}.

\subsection{Repeated asymmetric testing}
\label[appendix]{appsubsec:stein}

We next consider asymmetric tests in which the Type-I error is fixed and the Type-II error is minimized over repeated observations or copies.
For i.i.d. classical observations, the classical Stein lemma gives \cref{eq:classical_stein_lemma} in terms of the Kullback--Leibler divergence.
For i.i.d. quantum states measured one copy at a time, the best exponent is governed by the measured relative entropy.
With unrestricted collective measurements on all copies, the quantum Stein lemma instead gives the Umegaki quantum relative entropy.
The distinction clarifies which state space distinguishability is retained by the one copy measurement record used in this work.

\subsubsection{Classical Kullback--Leibler divergence}

For classical distributions, the KL divergence is defined by
\begin{align}
    D_{\rm KL}(P_0\Vert P_1)
    &=
    \sum_rP_0(r)\log\frac{P_0(r)}{P_1(r)}.
    \label{eq:classical_kl}
\end{align}
Terms with $P_0(r)=0$ are taken to be zero, and the divergence is $+\infty$ if $P_0(r)>0$ while $P_1(r)=0$ for any $r$.
For any fixed $0<\alpha<1$, the classical Stein lemma \cite{CoverThomas:2006} gives the optimal Type-II error for $n$ i.i.d.\ observations as
\begin{align}
    \beta_n^*(\alpha)
    &=
    \exp\left[
        -nD_{\rm KL}(P_0\Vert P_1)+o(n)
    \right].
    \label{eq:classical_stein_lemma}
\end{align}

\subsubsection{Quantum relative entropies}

The Umegaki quantum relative entropy is
\begin{align}
    D(\rho_0\Vert\rho_1)
    &=
    \operatorname{Tr}\!\left[
        \rho_0(\log\rho_0-\log\rho_1)
    \right].
    \label{eq:quantum_relative_entropy}
\end{align}
It is finite when $\operatorname{supp}\rho_0\subseteq\operatorname{supp}\rho_1$ and is defined as $+\infty$ otherwise.
A POVM $\mathsf{M}$ defines the quantum to classical measurement channel
\begin{align}
    \mathcal{M}_{\mathsf{M}}(\rho_i)
    &=
    \sum_{r\in\mathcal{J}}
    p_i^{\mathsf{M}}(r)|r\rangle\langle r|,
    \label{eq:povm_measurement_channel}
\end{align}
where $\{|r\rangle\}$ is an orthonormal basis of a classical outcome register.
Because the channel outputs are diagonal in this basis, their quantum relative entropy is the classical KL divergence,
\begin{align}
    D\left(
        \mathcal{M}_{\mathsf{M}}(\rho_0)
        \Vert
        \mathcal{M}_{\mathsf{M}}(\rho_1)
    \right)
    &=
    D_{\rm KL}\left(
        p_0^{\mathsf{M}}\Vert p_1^{\mathsf{M}}
    \right).
    \label{eq:measured_quantum_to_classical_relative_entropy}
\end{align}
The data processing inequality for $\mathcal{M}_{\mathsf{M}}$ therefore gives
\begin{align}
    D_{\rm KL}\left(
        p_0^{\mathsf{M}}\Vert p_1^{\mathsf{M}}
    \right)
    &\le
    D(\rho_0\Vert\rho_1)
    \label{eq:measurement_data_processing}
\end{align}
for every measurement $\mathsf{M}$.

The best KL divergence obtainable from a single copy measurement is the measured relative entropy
\begin{align}
    D_{\rm meas}(\rho_0\Vert\rho_1)
    &=
    \sup_{\mathsf{M}}
    D_{\rm KL}\left(
        p_0^{\mathsf{M}}\Vert p_1^{\mathsf{M}}
    \right).
    \label{eq:measured_relative_entropy}
\end{align}
In finite dimension, the supremum in \cref{eq:measured_relative_entropy} is unchanged if the measurement is restricted to a rank one PVM:
\begin{align}
    D_{\rm meas}(\rho_0\Vert\rho_1)
    &=
    \sup_{\mathsf{M}\,\text{rank one PVM}}
    D_{\rm KL}\left(
        p_0^{\mathsf{M}}\Vert p_1^{\mathsf{M}}
    \right).
    \label{eq:measured_relative_entropy_pvm}
\end{align}
Unlike the binary PVM sufficient for the Helstrom problem, this rank one PVM generally has $d$ outcomes, where $d$ is the Hilbert space dimension of the system treated as one copy.
Thus $d=2$ for one qubit and $d=2^m$ for a register of $m$ qubits treated as a single copy.
Moreover, when the relative entropy is finite,
\begin{align}
    D_{\rm meas}(\rho_0\Vert\rho_1)
    &\le D(\rho_0\Vert\rho_1),
    \label{eq:measured_vs_quantum_relative_entropy}
\end{align}
For full rank states, the inequality is strict if the states do not commute, whereas measuring a common eigenbasis gives equality when $[\rho_0,\rho_1]=0$.
These properties are reviewed and proven in Ref.~\cite{Berta:2017myy}.

For repeated copies measured individually with the same $\mathsf{M}$, applying the classical Stein lemma to $P_i=p_i^{\mathsf{M}}$ gives
\begin{align}
    \beta_n^{\mathsf{M}}(\alpha)
    &=
    \exp\left[
        -nD_{\rm KL}\left(
            p_0^{\mathsf{M}}\Vert p_1^{\mathsf{M}}
        \right)
        +o(n)
    \right].
    \label{eq:fixed_measurement_stein_exponent}
\end{align}
Optimizing the single copy measurement therefore gives
\begin{align}
    \beta_n^{\rm local}(\alpha)
    &=
    \exp\left[
        -nD_{\rm meas}(\rho_0\Vert\rho_1)+o(n)
    \right].
    \label{eq:local_stein_exponent}
\end{align}

Suppose instead that a binary measurement may act collectively on the full states $\rho_i^{\otimes n}$.
Let $T_n$ be the test operator associated with deciding $H_0$, so that $0\le T_n\le\mathbb{I}_n$, where $\mathbb{I}_n$ is the identity on the Hilbert space of $n$ copies.
Its Type-I and Type-II errors are
\begin{align}
    \alpha_n(T_n)
    &=
    \operatorname{Tr}\!\left[
        (\mathbb{I}_n-T_n)\rho_0^{\otimes n}
    \right],
    \label{eq:collective_quantum_type_i_error}\\
    \beta_n(T_n)
    &=
    \operatorname{Tr}\!\left[
        T_n\rho_1^{\otimes n}
    \right].
    \label{eq:collective_quantum_type_ii_error}
\end{align}
Let $\beta_{n,\rm coll}^*(\alpha)$ denote the minimum of \cref{eq:collective_quantum_type_ii_error} over all such $T_n$ satisfying $\alpha_n(T_n)\le\alpha$.
For every fixed $0<\alpha<1$ and finite $D(\rho_0\Vert\rho_1)$, the quantum Stein lemma gives \cite{HiaiPetz:1991,OgawaNagaoka:2000}
\begin{align}
    \beta_{n,\rm coll}^*(\alpha)
    &=
    \exp\left[
        -nD(\rho_0\Vert\rho_1)+o(n)
    \right].
    \label{eq:quantum_stein_lemma}
\end{align}
The optimizing measurement may depend on $n$ and need not decompose into measurements of the individual copies.
Thus the Umegaki relative entropy is not merely a data processing bound for one copy: it is the attainable collective measurement exponent for a known i.i.d. pair.
The gap $D-D_{\rm meas}$ is therefore the maximal asymptotic gain in the exponent made possible by replacing individual readout of each copy with collective measurements.

For composite quantum hypotheses defined by convex mixtures of tensor powers from two state families, the collective Stein exponent can require a regularized relative entropy formula~\cite{BertaBrandaoHirche:2021}.
These collective benchmarks concern joint measurements on many copies; \cref{appsubsec:policy_connection} treats the classical records produced by individual shots with feedback.

\subsection{Small signal expansion and Fisher information}
\label[appendix]{appsubsec:local_fisher_information}

The distinguishability measures above compare two finitely separated hypotheses, whereas Fisher information describes their local separation within a smooth parametric family.
To connect the two descriptions without assuming a particular physical response law, fix the nuisance parameters and let $\zeta$ be a scalar local response coordinate, with $\zeta=0$ under the null.
Let $P_\zeta$ denote the corresponding classical distribution and let $P_0$ denote its value at the null.
The coordinate $\zeta$ need not be the physical signal amplitude: if the leading probability shift is proportional to $A^{\ell}$, where $A$ is an amplitude and $\ell\ge1$ is the order of its first non-zero response coefficient, one may choose $\zeta=A^{\ell}$ so that the leading response is linear in $\zeta$.
The generic notation $A$ keeps this discussion independent of the Rabi model; in that application it may be identified with $\Phi_{\rm eff}$, or with $\Phi_R$ on resonance.
This choice is measurement dependent, and Fisher information is parameterization dependent.
If $\zeta$ is restricted to be non-negative, derivatives at the null are understood as one-sided derivatives.

Assume that the outcome space is finite, that $P_\zeta$ is smooth at $\zeta=0$, and that $P_0(r)>0$ for every outcome retained in the model.
Define the local derivative and the classical Fisher information (CFI) with respect to $\zeta$ by
\begin{align}
    \dot{P}_0(r)
    &\equiv
    \left.
        \frac{\partial P_\zeta(r)}{\partial\zeta}
    \right|_{\zeta=0},
    \label{eq:local_classical_derivative}\\
    F_{\rm C}^{(\zeta)}(0)
    &\equiv
    \sum_r\frac{\dot{P}_0(r)^2}{P_0(r)}.
    \label{eq:local_classical_fisher_information}
\end{align}
Normalization implies $\sum_r\dot{P}_0(r)=0$, so the linear term in either ordering of the KL divergence vanishes.
The local expansions are
\begin{align}
    D_{\rm TV}(P_0,P_\zeta)
    &=
    \frac{|\zeta|}{2}
    \sum_r|\dot{P}_0(r)|
    +O(\zeta^2),
    \label{eq:local_tv_expansion}\\
    D_{\rm KL}(P_0\Vert P_\zeta)
    &=
    \frac{\zeta^2}{2}
    F_{\rm C}^{(\zeta)}(0)
    +o(\zeta^2).
    \label{eq:local_kl_fisher_expansion}\\
    D_{\rm KL}(P_\zeta\Vert P_0)
    &=
    \frac{\zeta^2}{2}
    F_{\rm C}^{(\zeta)}(0)
    +o(\zeta^2).
    \label{eq:local_forward_kl_fisher_expansion}
\end{align}
Thus both KL orderings have the same local quadratic curvature given by the CFI, while the leading coefficient of total variation is an $L^1$ norm of the probability derivative rather than CFI itself.
The Cauchy--Schwarz inequality nevertheless gives
\begin{align}
    \left[
        \sum_r|\dot{P}_0(r)|
    \right]^2
    &\le
    F_{\rm C}^{(\zeta)}(0),
    \label{eq:local_tv_fisher_bound}
\end{align}
which locally relates the linear total variation response to the Fisher information scale.
At finite $\zeta$, the neglected higher order terms depend on the full distributions, so the CFI alone does not determine either total variation or KL divergence.

The quantum counterpart applies when the state family $\rho_\zeta$ itself is smooth in the same coordinate, with tangent operator
\begin{align}
    \dot{\rho}_0
    &\equiv
    \left.
        \frac{\partial\rho_\zeta}{\partial\zeta}
    \right|_{\zeta=0}.
    \label{eq:local_quantum_derivative}
\end{align}
The symmetric logarithmic derivative (SLD) $L_0$ and the corresponding quantum Fisher information (QFI) are defined by
\begin{align}
    \dot{\rho}_0
    &=
    \frac12(L_0\rho_0+\rho_0L_0),
    \label{eq:sld_definition}\\
    F_{\rm Q}^{(\zeta)}(0)
    &\equiv
    \operatorname{Tr}(\rho_0L_0^2).
    \label{eq:local_quantum_fisher_information}
\end{align}
For a POVM $\mathsf{M}$, the induced CFI is
\begin{align}
    F_{{\rm C},\mathsf{M}}^{(\zeta)}(0)
    &=
    \sum_r
    \frac{
        [\operatorname{Tr}(M_r\dot{\rho}_0)]^2
    }{
        \operatorname{Tr}(M_r\rho_0)
    }.
    \label{eq:measured_classical_fisher_information}
\end{align}
For a scalar parameter at a fixed point, optimization over measurements gives
\begin{align}
    F_{{\rm C},\mathsf{M}}^{(\zeta)}(0)
    &\le
    F_{\rm Q}^{(\zeta)}(0),
    \label{eq:cfi_qfi_bound}
\end{align}
with equality attained by the PVM defined by the spectral projections of $L_0$ \cite{BraunsteinCaves:1994}.
The trace distance has the local expansion
\begin{align}
    D_{\rm tr}(\rho_0,\rho_\zeta)
    &=
    \frac{|\zeta|}{2}
    \|\dot{\rho}_0\|_1
    +O(\zeta^2),
    \label{eq:local_trace_distance_expansion}
\end{align}
and \cref{eq:trace_distance_measured_tv,eq:local_tv_fisher_bound,eq:cfi_qfi_bound} imply
\begin{align}
    \|\dot{\rho}_0\|_1^2
    &\le
    F_{\rm Q}^{(\zeta)}(0).
    \label{eq:local_trace_qfi_bound}
\end{align}
Hence QFI bounds the local linear growth of trace distance, but does not generally equal its coefficient.

For full rank $\rho_0$, applying \cref{eq:local_kl_fisher_expansion} to each measurement and optimizing the leading term gives
\begin{align}
    D_{\rm meas}(\rho_0\Vert\rho_\zeta)
    &=
    \frac{\zeta^2}{2}
    F_{\rm Q}^{(\zeta)}(0)
    +o(\zeta^2).
    \label{eq:local_measured_relative_entropy_expansion}
\end{align}
Here we used the restriction of the measured relative entropy optimization to rank one PVMs and the Braunstein--Caves optimization of the local CFI over measurements~\cite{Berta:2017myy,BraunsteinCaves:1994}.
The corresponding expansion of the Umegaki quantum relative entropy involves a different quantum Fisher metric.
Define the Bogoliubov--Kubo--Mori (BKM) Fisher information by
\begin{align}
    F_{\rm BKM}^{(\zeta)}(0)
    &\equiv
    \left.
        \frac{\partial^2}{\partial\zeta^2}
        D(\rho_0\Vert\rho_\zeta)
    \right|_{\zeta=0}.
    \label{eq:bkm_fisher_information}
\end{align}
The relative entropy and the two quantum Fisher metrics then satisfy~\cite{Petz:1994canonical}
\begin{align}
    D(\rho_0\Vert\rho_\zeta)
    &=
    \frac{\zeta^2}{2}
    F_{\rm BKM}^{(\zeta)}(0)
    +o(\zeta^2),
    \label{eq:local_quantum_relative_entropy_expansion}\\
    F_{\rm BKM}^{(\zeta)}(0)
    &\ge
    F_{\rm Q}^{(\zeta)}(0).
    \label{eq:bkm_sld_ordering}
\end{align}
The BKM and SLD quantities coincide for a locally commuting family with $[\rho_0, \dot{\rho}_0]=0$, but differ in general~\cite{Petz:1999xrh}.

\subsection{Information measures for adaptive readout}
\label[appendix]{appsubsec:policy_connection}

For a fixed causal policy $\mu$, each shot produces a classical outcome before the qubit is reset and the next readout is selected.
For a fixed shot budget $n$, let $\Prob_0^\mu$ and $\Prob_{\btheta}^\mu$ denote the complete record laws under the null and at a fixed signal point, respectively.
The signal prior defines the mixture
\begin{align}
    \overline{\Prob}_1^\mu
    &=
    \int\dd\btheta\,\pi_0(\btheta)\Prob_{\btheta}^\mu.
    \label{eq:adaptive_mixture_record}
\end{align}
The same $\btheta$ is shared across all shots, so this mixture generally differs from a product of one shot distributions averaged over the initial prior.
Its conditional outcome distribution is the posterior predictive probability $m_k$ in \cref{eq:predictive_likelihood}.
All record laws use the same policy, including its dependence on the observed history.

Since $B_{10,n}=\dd\overline{\Prob}_1^\mu/\dd\Prob_0^\mu$, the conditional KL chain rule gives
\begin{align}
    \mathbb{E}_0[-\log B_{10,n}]
    &=
    D_{\rm KL}\left(
        \Prob_0^\mu\Vert\overline{\Prob}_1^\mu
    \right)
    \notag\\
    &=
    \sum_{k=0}^{n-1}\mathbb{E}_0
    \left[D_{\rm KL}(p_{0,k}\Vert m_k)\right].
    \label{eq:composite_null_record_kl}
\end{align}
Here $\mathbb{E}_0$ averages over records generated under $H_0$, and $p_{0,k}$ and $m_k$ are evaluated at the random history and selected readout before shot $k$.
The Fisher expansion in \cref{eq:local_kl_fisher_expansion} applies locally to each conditional KL term for a specified smooth response family; the record information also depends on the history distribution induced by the policy.

At a given history, the utilities in \cref{eq:helstrom_utility,eq:information_gain_utility} describe the next measurement.
Let $Z\in\{0,1\}$ indicate signal presence, with current probability $\Prob(Z=1|h_k)=q_k$.
For a candidate readout $u$, total variation and the mutual information chain rule give
\begin{align}
    U_{{\rm H},k}(u)
    &=
    D_{\rm TV}\left(p_{0,k}(\cdot|u),m_k(\cdot|u)\right),
    \label{eq:policy_tv_utility}\\
    U_{{\rm I},k}(u)
    &=
    I_{S_k}(Z;R_k|u)
    \notag\\
    &\quad+
    q_k I_{S_k}(\btheta;R_k|Z=1,u).
    \label{eq:policy_information_utility_decomposition}
\end{align}
The Helstrom-like utility describes equal weight discrimination of the current predictive pair and does not include the posterior weight $q_k$.
The two information gain terms quantify learning about signal presence and, conditional on $H_1$, the common signal parameters.
Each mutual information is an average KL divergence from a conditional outcome distribution to its predictive mixture.
Consequently, parameter learning can be informative even when $m_k$ is close to $p_{0,k}$, and can sharpen predictions for later shots.

The expected evidence at a fixed true signal point obeys a complementary identity,
\begin{align}
    \mathbb{E}_{\btheta}[\log B_{10,n}]
    &=
    D_{\rm KL}\left(
        \Prob_{\btheta}^\mu\Vert\Prob_0^\mu
    \right)
    \notag\\
    &\quad-
    D_{\rm KL}\left(
        \Prob_{\btheta}^\mu\Vert\overline{\Prob}_1^\mu
    \right),
    \label{eq:composite_expected_log_bayes_decomposition}
\end{align}
where $\mathbb{E}_{\btheta}$ averages over records under $\Prob_{\btheta}^\mu$.
The first term measures separation from the null, while the second is the predictive mismatch between the true signal point and the mixture.
Posterior learning changes both future readouts and predictions, so maximizing the current utility need not maximize the expected final evidence or detection power.

These mean identities do not determine the distribution of $\log B_{10,n}$ relative to the policy specific null threshold.
The numerical tests use this same full prior mixture in the Bayes factor, but the Type-II errors in \cref{subsec:numerical_exponents} are averaged over phase at a fixed injected amplitude.
Their finite window slopes do not establish an asymptotic exponent for the error averaged over the full signal prior; identifying such a rate requires additional information spectrum conditions~\cite{Han:2000general,NagaokaHayashi:2007}.

For \cref{fig:rabi_log_bayes_histograms}, we estimate the means in \cref{eq:composite_null_record_kl,eq:composite_expected_log_bayes_decomposition} by averaging the corresponding sums of conditional KL divergences, evaluated before each outcome, over the same $10^3$ records used for each histogram.
Their agreement with the sample means within the Monte Carlo errors of the paired differences provides an internal consistency check of the evidence accumulation.

\section{Weak signal Rabi estimates}
\label[appendix]{app:rabi_crossing}

This appendix gives the analytic estimates used to interpret the sensitivity plots in \cref{sec:analytic}.
The purpose is to separate three effects: the quadratic weak signal response of population readout, the linear response of transverse readout, and the cost of not knowing the transverse phase.
The main analysis treats $H_1$ as a composite hypothesis, whereas the sensitivity curves condition on an injected amplitude and average the pointwise power over the nuisance phase prior according to \cref{eq:phase_composite_detection_power}.
Accordingly, the scalar $P$ below denotes the target phase-averaged power for an implementable policy that is not supplied the phase, except for the known direction oracle, where it denotes phase-conditioned power after the true phase has been supplied to the measurement.
These fixed amplitude estimates isolate the weak signal response coefficients without replacing the full prior-marginalized test.

\subsection{Population readout}
\label[appendix]{appsubsec:fixed_z_estimate}

The population readout probability in \cref{eq:rabi_z_probability} depends on the signal through $\Phi_{\rm eff}$ but is independent of the shot start time $t$ and the effective phase $\tilde{\varphi}_{\rm eff}$.
Its pointwise power at fixed $\Phi_{\rm eff}$ therefore equals the phase-averaged power $\overline P^z$ in \cref{eq:phase_composite_detection_power}.
Thus the fixed amplitude calculation below does not assume that the phase is known, even though no explicit phase integral appears.
Define the readout visibility factor
\begin{align}
    v_\epsilon
    &\equiv
    1-2\epsilon.
    \label{eq:app_readout_visibility}
\end{align}
Define the positive outcome probabilities for population readout under the signal and null by
\begin{align}
    p_{z,\btheta}
    &\equiv
    p_{\btheta}(+|t,T,\bm{n}_z),
    \label{eq:app_fixed_z_signal_probability}\\
    p_{z,0}
    &\equiv
    \left.
    p_{z,\btheta}
    \right|_{\Phi_R=0}
    =
    \frac{1+v_\epsilon C}{2}.
    \label{eq:app_fixed_z_null_probability}
\end{align}
The weak signal expansion is
\begin{align}
    p_{z,\btheta}-p_{z,0}
    &\simeq
    -a_z\Phi_{\rm eff}^2,
    \label{eq:app_fixed_z_quadratic_response}
\end{align}
where
\begin{align}
    a_z
    &=
    \frac{v_\epsilon C\eta_1}{4}.
    \label{eq:app_fixed_z_response_coefficient}
\end{align}
Here $p_{z,0}$ is the positive outcome probability under the null hypothesis for population readout, including the effects of readout error and imperfect contrast.
The probability is locally flat in the signed signal amplitude $\Phi_R$ at fixed detuning, and its amplitude CFI therefore vanishes:
\begin{align}
    F_{{\rm C},z}^{(\Phi_R)}(0)
    &=
    0.
    \label{eq:rabi_null_population_cfi}
\end{align}
To parameterize the first non-zero response, define the signal power coordinate
\begin{align}
    \zeta_z
    &\equiv
    \Phi_R^2.
    \label{eq:rabi_population_response_coordinate}
\end{align}
The expansion in \cref{eq:app_fixed_z_quadratic_response} then gives
\begin{align}
    \left.
        \frac{\partial p_{z,\btheta}}{\partial\zeta_z}
    \right|_{\zeta_z=0}
    &=
    -a_z
    \operatorname{sinc}^2\!\left(\frac{\Delta T}{2}\right),
    \label{eq:rabi_population_power_derivative}\\
    F_{{\rm C},z}^{(\zeta_z)}(0)
    &=
    \frac{a_z^2}{p_{z,0}(1-p_{z,0})}
    \operatorname{sinc}^4\!\left(\frac{\Delta T}{2}\right).
    \label{eq:rabi_population_power_cfi}
\end{align}
For the imperfect readout model used here, $0<p_{z,0}<1$, so the population response is regular in $\zeta_z$.
The local expansions in \cref{app:hypothesis_testing} consequently give
\begin{align}
    D_{\rm TV}^{z}
    &=
    O(\Phi_R^2),
    \label{eq:rabi_population_tv_order}\\
    D_{\rm KL}^{z}
    &=
    O(\Phi_R^4).
    \label{eq:rabi_population_kl_order}
\end{align}

Let $K_z$ be the number of positive outcomes in $n$ fixed-$z$ shots.
\cref{eq:rabi_z_probability} then gives
\begin{align}
    K_z
    &\sim
    {\rm Binom}\!\left(n,p_{z,\btheta}\right),
\end{align}
where ${\rm Binom}(n,p)$ denotes the binomial distribution with $n$ trials and success probability $p$.
The Gaussian approximation requires both $np_{z,0}\gg1$ and $n(1-p_{z,0})\gg1$; for the present north pole convention, the latter is the restrictive condition.
The standard deviation of the null count distribution is then approximately
\begin{align}
    \sigma_0
    &\simeq
    \sqrt{n p_{z,0}(1-p_{z,0})}.
\end{align}
For fixed $n$, Type-I error $\alpha$, and target power $P$, let $\Phi_z^{\rm req}(n;\alpha,P)$ be the smallest value of $\Phi_{\rm eff}$ that satisfies the Gaussian fixed-$z$ detection condition.
Let $Z\sim N(0,1)$ be a standard normal random variable and let $z_q$ denote its $q$ quantile, defined by
\begin{align}
    \Prob[Z\le z_q]
    &=
    q.
\end{align}
Define the Gaussian displacement sum
\begin{align}
    g(\alpha,P)
    &\equiv
    z_{1-\alpha}+z_P.
    \label{eq:app_gaussian_displacement_sum}
\end{align}
A one-sided lower tail Gaussian test with Type-I error $\alpha$ gives
\begin{align}
    n a_z \left[\Phi_z^{\rm req}(n;\alpha,P)\right]^2
    &=
    g(\alpha,P)
    \sqrt{n p_{z,0}(1-p_{z,0})}.
\end{align}
Solving for $\Phi_z^{\rm req}(n;\alpha,P)$ gives
\begin{align}
    \Phi_z^{\rm req}(n;\alpha,P)
    &=
    A_z(\alpha,P)n^{-1/4},
    \label{eq:app_fixed_z_sensitivity}
    \\
    A_z(\alpha,P)
    &=
    \left[
        \frac{
        g(\alpha,P)
        \sqrt{p_{z,0}(1-p_{z,0})}
        }{a_z}
    \right]^{1/2}.
    \label{eq:app_fixed_z_coefficient}
\end{align}
If $n(1-p_{z,0})$ is not large, the integer threshold and the low count negative outcome distribution must instead be treated with an exact binomial test or a Poisson approximation.
In that regime the observed finite range slope can differ visibly from the asymptotic $n^{-1/4}$ law.
The full Bayes factor test still marginalizes over the amplitude prior; phase independence only makes the additional nuisance phase average trivial for this fixed amplitude benchmark.

For the finite background Poisson approximation, define the rare negative outcome count $N_z\equiv n-K_z$.
The weak signal model gives
\begin{align}
    N_z\mid H_0
    &\sim
    {\rm Pois}\!\left(n[1-p_{z,0}]\right),
    \notag\\
    N_z\mid\btheta
    &\sim
    {\rm Pois}\!\left(n[1-p_{z,0}+a_z\Phi_{\rm eff}^2]\right).
    \label{eq:app_fixed_z_poisson_model}
\end{align}
Here ${\rm Pois}(m)$ denotes the Poisson distribution with mean $m$.
The finite background sensitivity $\Phi_{z,{\rm Pois}}^{\rm req}(n;\alpha,P)$ shown in \cref{fig:rabi_discovery_sensitivity_p07} and used for the lower block of \cref{tab:main_crossings} is obtained by numerically solving for power $P$ with an upper tail test in $N_z$, randomized at the boundary to attain size $\alpha$.
This construction retains the discrete rare count background but remains a weak signal Poisson approximation; the exact binomial model is more faithful at very small $n$.

\subsection{Known direction transverse oracle}
\label[appendix]{appsubsec:oracle_estimate}

If the effective transverse phase is known, the readout can be chosen so that $\sin(\vaphi-\tilde{\varphi}_{\rm eff})=1$ on every shot, as in \cref{eq:known_signal_transverse_axis}.
This deliberately converts the composite phase problem into a phase-conditioned simple alternative and should be read only as an oracle benchmark.
Because the readout is realigned separately for every true phase, its pointwise power is phase independent, and averaging this phase-indexed oracle family over the phase prior returns the same numerical power $P$.
That average is not the power of a single admissible policy for the composite problem and includes neither the cost of learning the phase nor the search penalty associated with it.

The estimates below follow from the transverse readout probability in \cref{eq:rabi_transverse_probability}.
Define the positive outcome probability for a general transverse direction and its null value by
\begin{align}
    p_{\perp,\btheta}(\vaphi)
    &\equiv
    p_{\btheta}(+|t,T,\bm{n}_\perp(\vaphi)),
    \label{eq:app_transverse_signal_probability}\\
    p_{\perp,0}
    &\equiv
    \left.
    p_{\perp,\btheta}(\vaphi)
    \right|_{\Phi_R=0}
    =
    \frac12.
    \label{eq:app_transverse_null_probability}
\end{align}
This value follows because under the null condition $\Phi_R=0$ the Bloch vector has no component along any transverse readout axis, and the symmetric readout error leaves an unbiased binary probability unchanged.
The transverse linear response coefficient and the resulting weak signal expansion for an arbitrary transverse readout axis are
\begin{align}
    b_\perp
    &=
    \frac{v_\epsilon C\eta_2}{2}.
    \label{eq:app_transverse_response_coefficient}\\
    p_{\perp,\btheta}(\vaphi)-p_{\perp,0}
    &\simeq
    -b_\perp\Phi_{\rm eff}
    \sin(\vaphi-\tilde{\varphi}_{\rm eff}).
    \label{eq:app_transverse_linear_response}
\end{align}
The aligned oracle is the specialization $\vaphi=\tilde{\varphi}_{\rm eff}+\pi/2$.
For the corresponding quantum state, choose the signed local coordinate $\zeta_\perp\equiv\Phi_R$ at fixed detuning and signal phase.
Then the quantum and classical Fisher information are given respectively as
\begin{align}
    F_{\rm Q}^{(\zeta_\perp)}(0)
    &=
    C^2\eta_2^2
    \operatorname{sinc}^2\!\left(\frac{\Delta T}{2}\right),
    \label{eq:rabi_null_qfi}\\
    F_{{\rm C},\perp}^{(\zeta_\perp)}(0;\vaphi)
    &=
    (1-2\epsilon)^2
    F_{\rm Q}^{(\zeta_\perp)}(0)
    \sin^2(\vaphi-\tilde{\varphi}_{\rm eff}).
    \label{eq:rabi_null_transverse_cfi}
\end{align}
An aligned transverse measurement therefore saturates the quantum bound \cref{eq:local_trace_qfi_bound} before classical readout error, while the bit flip channel reduces its CFI by $(1-2\epsilon)^2$.
For any non-blind transverse direction, the local distinguishability scales as
\begin{align}
    D_{\rm TV}^{\perp}
    &=
    O(|\Phi_R|),
    \label{eq:rabi_transverse_tv_order}\\
    D_{\rm KL}^{\perp}
    &=
    O(\Phi_R^2).
    \label{eq:rabi_transverse_kl_order}
\end{align}
Thus, at fixed detuning with a non-zero leading response, the cumulative local KL divergence from $n$ informative shots scales as $n\Phi_R^2$ for transverse readout, compared with $n\Phi_R^4$ for population readout, reproducing the $n^{-1/2}$ and $n^{-1/4}$ sensitivity laws in \cref{eq:main_transverse_sensitivity,eq:main_fixed_z_sensitivity}.
Equivalently, both readouts have $D_{\rm TV}=O(|\zeta|)$ and $D_{\rm KL}=O(\zeta^2)$ in their respective regular response coordinates, $\zeta_\perp=\Phi_R$ and $\zeta_z=\Phi_R^2$.

Let $K_\perp$ be the number of positive outcomes in $n$ oracle transverse shots.
Then
\begin{align}
    K_\perp
    &\sim
    {\rm Binom}\!\left(
        n,
        p_{\perp,\btheta}\!\left(
            \tilde{\varphi}_{\rm eff}+\frac{\pi}{2}
        \right)
    \right).
\end{align}
Under the null hypothesis, $K_\perp$ has mean $n/2$ and standard deviation $\sqrt{n}/2$.
Under the signal hypothesis, its mean is shifted by $-n b_\perp\Phi_{\rm eff}$ at leading order.
The signal-induced mean shift measured in units of the null standard deviation is therefore $2b_\perp\Phi_{\rm eff}\sqrt n$.
For a one-sided lower tail Gaussian test with Type-I error $\alpha$, the rejection threshold lies $z_{1-\alpha}$ null standard deviations below the null mean.
To obtain power $P$, the signal mean must lie an additional $z_P$ standard deviations beyond this threshold in the same direction, giving
\begin{align}
    2b_\perp\Phi_{{\rm O}_\perp}^{\rm req}(n;\alpha,P)\sqrt n
    &=
    g(\alpha,P).
\end{align}
Solving this condition gives
\begin{align}
    \Phi_{{\rm O}_\perp}^{\rm req}(n;\alpha,P)
    &=
    A_{{\rm O}_\perp}(\alpha,P)n^{-1/2},
    \label{eq:app_oracle_sensitivity}
    \\
    A_{{\rm O}_\perp}(\alpha,P)
    &=
    \frac{g(\alpha,P)}{2b_\perp}.
    \label{eq:app_oracle_coefficient}
\end{align}
The signal--null Bloch vector difference is transverse at $O(\Phi_R)$, while its longitudinal component begins at $O(\Phi_R^2)$.
The transverse oracle therefore approaches the globally optimal known signal projective measurement as $\Phi_R\to0$, although the finite signal optimum can acquire a longitudinal component.

\subsection{Fixed transverse readout}
\label[appendix]{appsubsec:unknown_phase_estimate}

We now derive the weak signal sensitivities of the implementable fixed-$x/y$ and fixed-$x$ policies.
Neither policy is supplied the signal phase, so their sensitivity target is the phase-averaged power $\overline P^\mu$ defined in \cref{eq:phase_composite_detection_power}.

A simple phase robust transverse benchmark is the fixed-$x/y$ quadrature test.
This estimate uses the transverse readout probability in \cref{eq:rabi_transverse_probability} with $\vaphi=0$ for the $x$ axis and $\vaphi=\pi/2$ for the $y$ axis.
At shot $k$, the effective phase is $\tilde{\varphi}_{{\rm eff},k}\equiv\tilde{\varphi}_{\rm eff}(t_k,T;\btheta)$, as defined in \cref{eq:effective_rabi_phase_angle}.
Because all shots use the same interrogation time $T$, $\Phi_{\rm eff}$ in \cref{eq:effective_rabi_phase} is independent of $k$, whereas $\tilde{\varphi}_{{\rm eff},k}$ drifts when $\Delta\ne0$.
Define the corresponding shot-dependent positive outcome probabilities by
\begin{align}
    p_{x,\btheta,k}
    &\equiv
    p_{\btheta}(+|t_k,T,\bm{n}_\perp(0)),
    \label{eq:app_fixed_x_signal_probability}\\
    p_{y,\btheta,k}
    &\equiv
    p_{\btheta}\!\left(+|t_k,T,\bm{n}_\perp\!\left(\frac{\pi}{2}\right)\right),
    \label{eq:app_fixed_y_signal_probability}\\
    p_{x,0}
    &\equiv
    \left.p_{x,\btheta,k}\right|_{\Phi_R=0}
    =
    \frac12,\\
    p_{y,0}
    &\equiv
    \left.p_{y,\btheta,k}\right|_{\Phi_R=0}
    =
    \frac12.
    \label{eq:app_fixed_xy_null_probabilities}
\end{align}
Let $\mathcal{K}_x$ and $\mathcal{K}_y$ be the sets of shots assigned to the $x$ and $y$ axes, respectively, with $|\mathcal{K}_x|=|\mathcal{K}_y|=n/2$, and let $K_x$ and $K_y$ be their positive outcome counts.
For non-zero detuning, the probabilities vary with $k$, so these counts are generally Poisson-binomial rather than binomial under a fixed signal point.
Using the weak signal expansion \cref{eq:app_transverse_linear_response}, the shotwise probability shifts are
\begin{align}
    p_{x,\btheta,k}-p_{x,0}
    &\simeq
    b_\perp\Phi_{\rm eff}
    \sin\tilde{\varphi}_{{\rm eff},k},
    \label{eq:app_fixed_x_shotwise_shift}\\
    p_{y,\btheta,k}-p_{y,0}
    &\simeq
    -b_\perp\Phi_{\rm eff}
    \cos\tilde{\varphi}_{{\rm eff},k}.
    \label{eq:app_fixed_y_shotwise_shift}
\end{align}
For this local estimate, we temporarily treat the detuning as known and use the corresponding sine and cosine templates to combine the time-resolved outcomes.
Because each null Bernoulli outcome has variance $1/4$, a score test with time-resolved outcomes from two quadratures has the local non-centrality
\begin{align}
    \lambda_{xy}(\btheta)
    &=
    4b_\perp^2\Phi_{\rm eff}^2
    \left[
        \sum_{k\in\mathcal{K}_x}
        \sin^2\tilde{\varphi}_{{\rm eff},k}
        +
        \sum_{k\in\mathcal{K}_y}
        \cos^2\tilde{\varphi}_{{\rm eff},k}
    \right].
    \label{eq:app_xy_noncentrality_general}
\end{align}
For the alternating fixed-$x/y$ schedule, the bracket in \cref{eq:app_xy_noncentrality_general} is $n/2+O(1)$ away from sampling aliases and is exactly $n/2$ at resonance.
Thus the leading local non-centrality, including the detuning suppression in $\Phi_{\rm eff}$, is
\begin{align}
    \lambda_{xy}
    &=
    2b_\perp^2n\Phi_{\rm eff}^2
    \left[1+O\!\left(n^{-1}\right)\right].
    \label{eq:app_xy_noncentrality}
\end{align}

For $\Delta\ne0$, the shotwise phases $\tilde{\varphi}_{{\rm eff},k}$ drift, so their signed contributions can cancel in the total counts for both fixed-$x$ and fixed-$x/y$ readout.
A detuned search should therefore retain the time-tagged record and use the full likelihood or, in the weak signal limit, an equivalent Fourier or matched filter statistic.
In the full detuned search, the detuning is unknown and must be searched over or marginalized, with the resulting search penalty and threshold calibration discussed in \cref{subsec:detuning}.

At resonance, the two counts $K_x$ and $K_y$ are sufficient for this local test.
Let $\E_0$ and ${\rm Var}_0$ denote expectation value and variance under the null hypothesis, respectively.
Their null moments are
\begin{align}
    \E_0[K_x]
    &=
    \E_0[K_y]
    =
    \frac{n}{4},\\
    {\rm Var}_0(K_x)
    &=
    {\rm Var}_0(K_y)
    =
    \frac{n}{8}.
\end{align}
The corresponding quadratic statistic is
\begin{align}
    Q_{xy}
    &=
    \frac{(K_x-\E_0[K_x])^2}{\mathrm{Var}_0(K_x)}
    +
    \frac{(K_y-\E_0[K_y])^2}{\mathrm{Var}_0(K_y)}.
    \label{eq:app_xy_statistic}
\end{align}
For large $n$, $Q_{xy}$ is approximately distributed as a central chi-square variable with two degrees of freedom under the null and as a non-central chi-square variable with non-centrality $\lambda_{xy}$ under a fixed signal.

Let $X_0$ be a central chi-square random variable with two degrees of freedom:
\begin{align}
    X_0
    &\sim
    \chi^2_2 .
\end{align}
The critical value $\chi^2_{2,1-\alpha}$ is the $(1-\alpha)$ quantile of $X_0$:
\begin{align}
    \Prob\!\left[
        X_0
        \le
        \chi^2_{2,1-\alpha}
    \right]
    &=
    1-\alpha .
\end{align}
Let $X_\lambda$ be a non-central chi-square random variable with two degrees of freedom and non-centrality $\lambda$:
\begin{align}
    X_\lambda
    &\sim
    \chi^2_2(\lambda).
\end{align}
Within the leading approximation in \cref{eq:app_xy_noncentrality}, the non-centrality is independent of the physical phase $\tilde{\varphi}$, so the pointwise power is the same at every phase and its uniform phase average is
\begin{align}
    \overline P^{x/y}(\Phi_R;\alpha)
    &\simeq
    \Prob\!\left[
        X_{\lambda_{xy}}
        >
        \chi^2_{2,1-\alpha}
    \right].
    \label{eq:app_xy_phase_averaged_power}
\end{align}
The value $\lambda_{xy}(\alpha,P)$ is determined by setting this phase-averaged power equal to the target $P$:
\begin{align}
    \Prob\!\left[
        X_{\lambda_{xy}}
        >
        \chi^2_{2,1-\alpha}
    \right]
    &=
    P .
    \label{eq:app_lambda_xy_definition}
\end{align}
Then
\begin{align}
    \Phi_{xy}^{\rm req}(n;\alpha,P)
    &=
    A_{xy}(\alpha,P)n^{-1/2},
    \label{eq:app_xy_sensitivity}
    \\
    A_{xy}(\alpha,P)
    &=
    \sqrt{\frac{\lambda_{xy}(\alpha,P)}{2b_\perp^2}}.
    \label{eq:app_xy_coefficient}
\end{align}

At resonance, a single fixed transverse axis has the same $n^{-1/2}$ exponent after phase averaging, but its coefficient is controlled by phases for which the signal is nearly orthogonal to the chosen readout axis.
For fixed-$x$ readout, the sign of the mean shift depends on the unknown phase, so both positive and negative count excursions can indicate a signal.
We therefore use a two-sided Gaussian test.
Define the standardized signal amplitude by
\begin{align}
    \rho_x
    &=
    2b_\perp\Phi_{\rm eff}\sqrt n.
\end{align}
For a given physical phase $\tilde{\varphi}$, the standardized Gaussian mean shift is then $\rho_x\sin\tilde{\varphi}$.
A two-sided test rejects when the standardized count lies outside $[-z_{1-\alpha/2},z_{1-\alpha/2}]$.
For the uniform physical phase prior on $[0,2\pi)$, the phase-averaged power is
\begin{align}
    \overline P^x(\rho_x;\alpha)
    &=
    \int_0^{2\pi}\frac{\dd\tilde{\varphi}}{2\pi}
    \Bigl[
        1-\Phi_{\rm N}\!\left(z_{1-\alpha/2}-\rho_x\sin\tilde{\varphi}\right)
    \notag\\
    &\hspace{6.5em}
        +\Phi_{\rm N}\!\left(-z_{1-\alpha/2}-\rho_x\sin\tilde{\varphi}\right)
    \Bigr],
    \label{eq:app_fixed_x_phase_averaged_power}
\end{align}
where $\overline P^x$ is the fixed-$x$ specialization of \cref{eq:phase_composite_detection_power}, expressed in terms of the scaled local amplitude $\rho_x$, and $\Phi_{\rm N}$ is the standard normal cumulative distribution function.
If $\rho_x(\alpha,P)$ solves $\overline P^x(\rho_x;\alpha)=P$, then
\begin{align}
    \Phi_x^{\rm req}(n;\alpha,P)
    &=
    A_x(\alpha,P)n^{-1/2},
    \label{eq:app_fixed_x_sensitivity}
    \\
    A_x(\alpha,P)
    &=
    \frac{\rho_x(\alpha,P)}{2b_\perp}.
    \label{eq:app_fixed_x_coefficient}
\end{align}
At high target power, $A_x$ grows rapidly in the phase stationary coherent regime because phases near the fixed axis blind spots must still be detected.

\subsection{Crossing estimates}
\label[appendix]{appsubsec:crossing_estimates}

This subsection converts the weak signal coefficients derived above into estimates of the shot number at which a transverse strategy overtakes the fixed-$z$ policy.
The comparison is asymptotic: the fixed-$z$ policy has $\Phi_z^{\rm req}\propto n^{-1/4}$, whereas the transverse benchmarks have $\Phi_\perp^{\rm req}\propto n^{-1/2}$.
We therefore solve for the crossing point of these two sensitivity curves in terms of the detector parameters and the target power.

For a generic transverse strategy, define the required dimensionless displacement amplitude by
\begin{align}
    \rho_\perp(\alpha,P)
    &\equiv
    2b_\perp A_\perp(\alpha,P),
    \label{eq:app_transverse_required_displacement}
\end{align}
or more explicitly,
\begin{align}
    \rho_{{\rm O}_\perp}(\alpha,P)
    &=
    g(\alpha,P),
    \\
    \rho_x(\alpha,P)
    &:\
    \overline P^x(\rho_x;\alpha)=P,
    \\
    \rho_{xy}(\alpha,P)
    &=
    \sqrt{2\lambda_{xy}(\alpha,P)}.
\end{align}
This quantity is independent of $n$ in the transverse weak signal scaling and contains the strategy-dependent threshold and nuisance phase cost.
The crossing shot count $n_{z:\perp}(\alpha,P)$ is defined by equality of the required fixed-$z$ and transverse signal strengths:
\begin{align}
    \Phi_z^{\rm req}(n_{z:\perp};\alpha,P)
    &=
    \Phi_\perp^{\rm req}(n_{z:\perp};\alpha,P),
    \label{eq:app_crossing_condition}
\end{align}
which gives
\begin{align}
    n_{z:\perp}(\alpha,P)
    &=
    \left[
        \frac{A_\perp(\alpha,P)}{A_z(\alpha,P)}
    \right]^4
    \notag\\
    &=
    \frac{
        \rho_\perp^4(\alpha,P)\eta_1^2
    }{
        4v_\epsilon^2C^2\eta_2^4
        g^2(\alpha,P)
        (1-v_\epsilon^2C^2)
    }.
    \label{eq:app_crossing_general_explicit}
\end{align}
For the strategies considered here,
\begin{align}
    n_{z:{\rm O}_\perp}
    &=
    \frac{
        g^2\,\eta_1^2
    }{
        4v_\epsilon^2C^2\eta_2^4(1-v_\epsilon^2C^2)
    }.
    \label{eq:app_crossing_oracle_explicit}
    \\
    n_{z:x}
    &=
    \frac{
        \rho_x^4\,\eta_1^2
    }{
        4v_\epsilon^2C^2\eta_2^4g^2(1-v_\epsilon^2C^2)
    }.
    \label{eq:app_crossing_x_explicit}
    \\
    n_{z:xy}
    &=
    \frac{
        \lambda_{xy}^2\,\eta_1^2
    }{
        v_\epsilon^2C^2\eta_2^4g^2(1-v_\epsilon^2C^2)
    }.
    \label{eq:app_crossing_xy_explicit}
\end{align}
where the arguments $(\alpha,P)$ of $g$, $\rho_x$, and $\lambda_{xy}$ are suppressed.
Strong transverse coherence moves the crossing rapidly to smaller $n$ through the $\eta_2^{-4}$ dependence, while larger $\eta_1$ makes the fixed-$z$ policy more competitive.
The apparent divergence as $v_\epsilon C\to1$ is an artifact of the Gaussian background derivation, because then $1-p_{z,0}\to0$.
The resulting Gaussian crossing estimates are reported in the upper block of \cref{tab:main_crossings}.

The corresponding finite background crossings are obtained numerically by intersecting $\Phi_{z,{\rm Pois}}^{\rm req}(n;\alpha,P)$ from \cref{appsubsec:fixed_z_estimate} with the three transverse sensitivities in \cref{eq:app_oracle_sensitivity,eq:app_fixed_x_sensitivity,eq:app_xy_sensitivity}.
For the adaptive entry, the transverse sensitivity is replaced by the fitted adaptive curve.
The resulting Poisson predictions are reported in the lower block of \cref{tab:main_crossings}.

\section{Additional Rabi sensing diagnostics}
\label[appendix]{app:more_rabi_plots}

This appendix supplements the posterior learning discussion in \cref{subsec:posterior_learning}.
\cref{appsubsec:more_posteriors} shows posterior evolution under the null hypothesis and the fixed policies.
\cref{app:count-based} then checks the fixed policy Bayes factor powers against conventional count-based tests.
\cref{appsubsec:higher_fidelity} demonstrates successful learning of signal properties with a higher fidelity detector profile.
Detuning is neglected, $\Delta=0$, throughout this appendix, until its effect is discussed in \cref{appsubsec:detuned_trajectory}.

\subsection{Background and fixed policy posteriors}
\label[appendix]{appsubsec:more_posteriors}

This subsection compares posterior evolution for an adaptive $H_0$ record and for signal records measured with the fixed-$z$, fixed-$x$, and fixed-$x/y$ policies.
The comparison shows which aspects of the signal amplitude and phase each policy can identify.

\begin{figure}[t]
\centering
\includegraphics[width=\columnwidth]{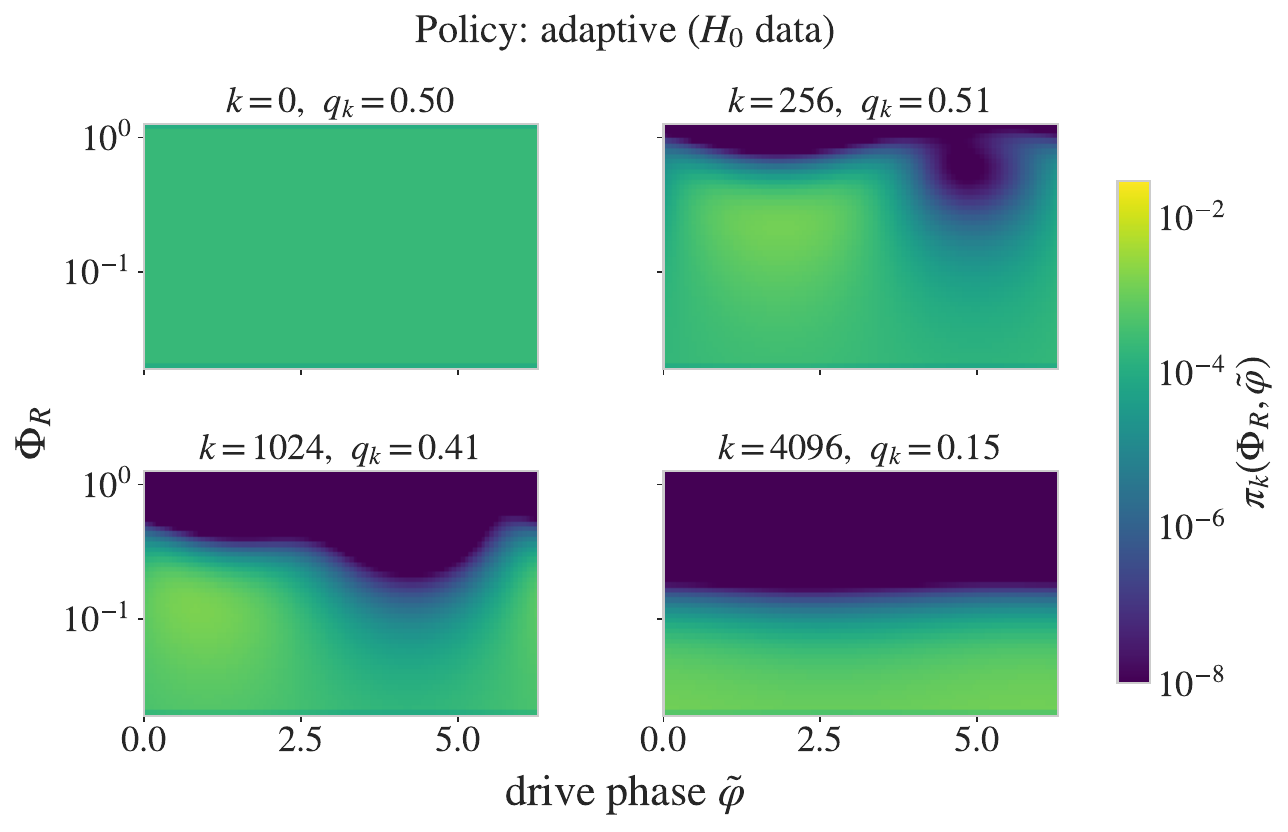}
\caption{Conditional $H_1$ posterior density for an illustrative adaptive measurement record generated under $H_0$.
The trajectory reaches a transient maximum $q_k=0.75$ at $k=216$; at the nearest common snapshot, $k=256$, the background fluctuation produces a localized finite amplitude mode.
The mode weakens and shifts toward lower amplitude by $k=1024$, and by $k=4096$ it is replaced by an approximately phase uniform posterior concentrated near the low amplitude boundary, with $q_k=0.15$.}
\label{fig:rabi_posterior_learning_heatmaps_background}
\end{figure}

\begin{figure}[t]
\centering
\includegraphics[width=\columnwidth]{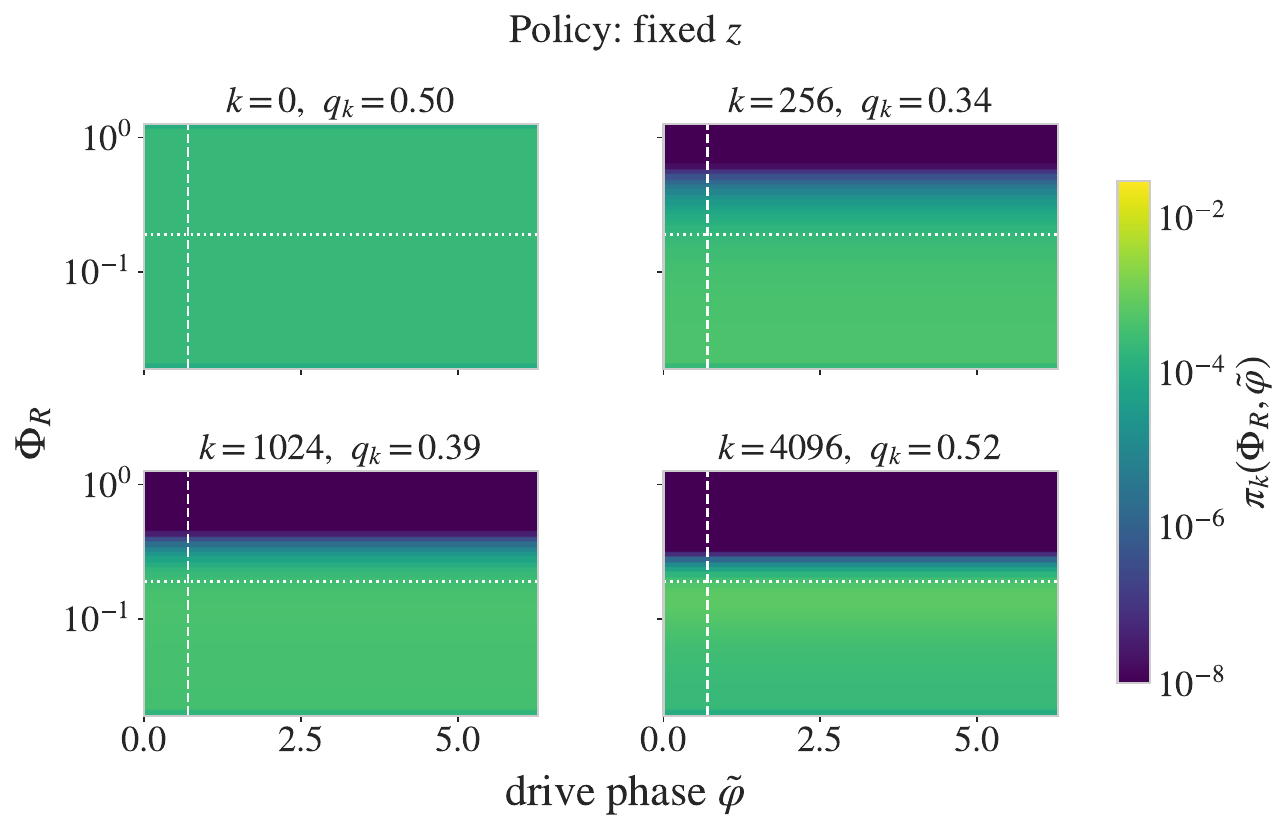}
\caption{Conditional $H_1$ posterior density for a representative resonant signal record measured with the fixed-$z$ policy.
The dashed and dotted lines mark the injected phase $\tilde{\varphi}=0.7$ and amplitude $\Phi_R=0.19$, respectively.
The phase independent measurement constrains the amplitude while leaving the drive phase unconstrained.}
\label{fig:rabi_posterior_learning_fixed_z}
\end{figure}

\begin{figure}[t]
\centering
\includegraphics[width=\columnwidth]{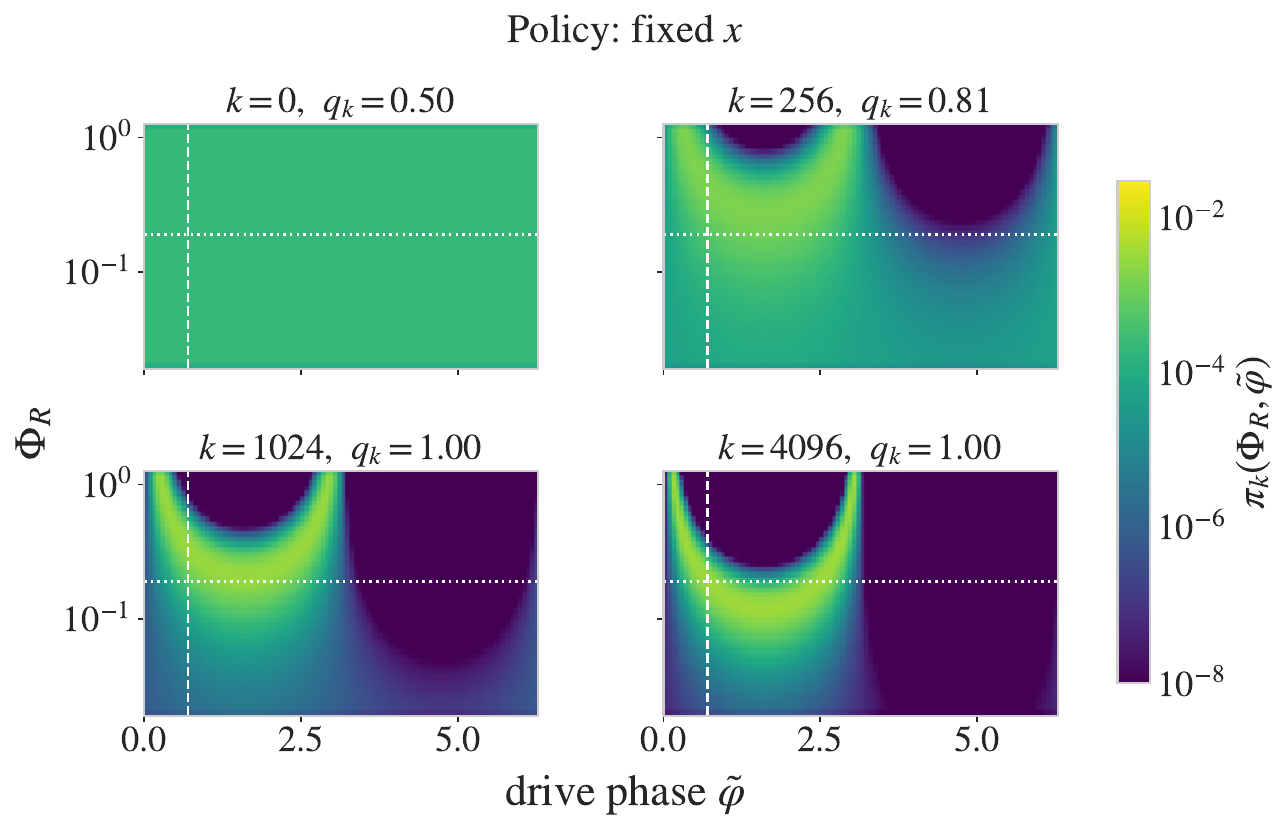}
\caption{Same as \cref{fig:rabi_posterior_learning_fixed_z} but for the fixed-$x$ policy.
Use of only a single quadrature shows the phase degeneracy.}
\label{fig:rabi_posterior_learning_fixed_x}
\end{figure}

\begin{figure}[t]
\centering
\includegraphics[width=\columnwidth]{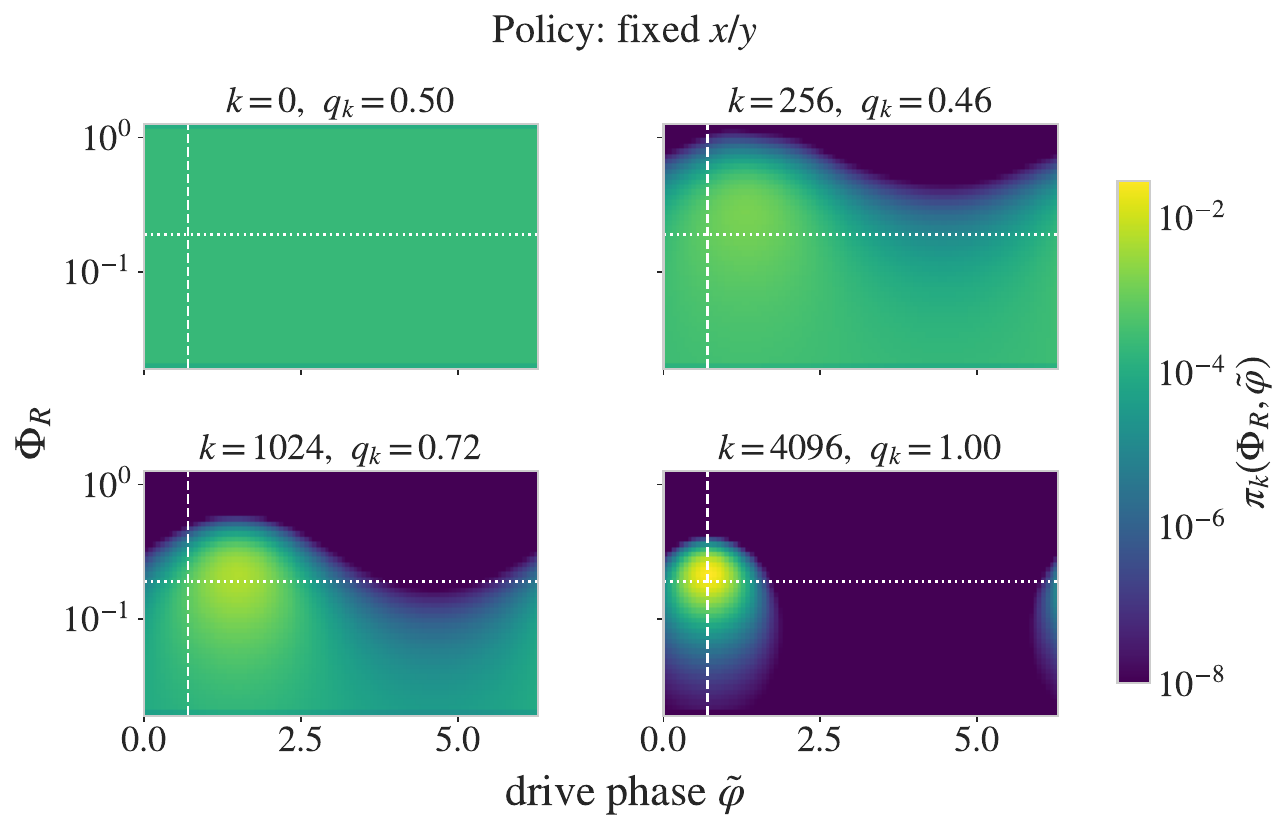}
\caption{Same as \cref{fig:rabi_posterior_learning_fixed_z} but for the fixed-$x/y$ policy.
Alternating transverse quadratures jointly localize the amplitude and phase without adaptive feedback.}
\label{fig:rabi_posterior_learning_fixed_xy}
\end{figure}

\cref{fig:rabi_posterior_learning_heatmaps_background} shows the conditional $H_1$ posterior for the adaptive $H_0$ record; yellow and purple indicate high and low posterior density, respectively, here and in the subsequent heat maps.
The transient finite amplitude mode corresponds to the fake signal peak at $k=216$ along the same background record in \cref{fig:rabi_posterior_learning_curves}.
This trajectory illustrates the look elsewhere mechanism within a single record: among the many amplitude--phase templates, one can temporarily match a background fluctuation, but later outcomes remove the localized mode when they fail to reinforce its prediction.

\crefrange{fig:rabi_posterior_learning_fixed_z}{fig:rabi_posterior_learning_fixed_xy} show posterior snapshots for the three fixed policies.
The white dashed and dotted lines mark the injected phase and amplitude, respectively.
Fixed-$z$ learns the amplitude but retains the full phase degeneracy, as follows from its phase independent response.
Fixed-$x$ develops the amplitude--phase ridge and reflection degeneracy implied by its dependence on $\sin\Phi_R\sin\tilde{\varphi}$, while alternating fixed-$x/y$ resolves these ambiguities by probing both transverse quadratures.

\subsection{Cross check with count-based tests}
\label{app:count-based}

To verify that the Bayes factor analysis does not unduly degrade fixed policy performance, we compare the powers in \cref{tab:rabi_fixed_budget_power} with analytic tests formed directly from the outcome counts.
All comparisons use the same baseline benchmark, $n=1024$, $\Phi_R=0.19$, and $\alpha=0.05$.
The quoted Bayes factor uncertainties below are binomial Monte Carlo standard errors from $10^3$ signal pseudoexperiments.

For the fixed-$z$ policy, the exact binomial model in \cref{appsubsec:fixed_z_estimate} gives $p_{z,0}=0.99005$ and $p_{\btheta}(+|t,T,\bm{n}_z)=0.98539$ at $\Phi_R=0.19$.
An exact test of size $0.05$ in the lower tail rejects for $K_z<1008$ and rejects with probability $0.805$ when $K_z=1008$.
Its power is $0.409$, consistent with the Bayes factor power $0.435\pm0.016$.

For fixed-$x$, an exact binomial version of the phase-averaged two-sided test in \cref{eq:app_fixed_x_phase_averaged_power}, with boundary randomization to attain size $0.05$, gives power $0.624$.
This is consistent with the fixed-$x$ Bayes factor power $0.603\pm0.015$.

For fixed-$x/y$, retaining the two counts separately gives the quadratic statistic $Q_{xy}$ in \cref{eq:app_xy_statistic}.
\cref{eq:app_xy_noncentrality,eq:app_lambda_xy_definition} give $\lambda_{xy}=7.01$ and power $0.656$ for the same benchmark, consistent with the Bayes factor result $0.673\pm0.015$.
The agreement confirms that the adaptive advantage over fixed-$x/y$ is not caused by an unnecessarily weak test statistic; simply combining $K_x$ and $K_y$ would discard the signed quadrature displacement.

\subsection{Learning signals with higher fidelity detector profile}
\label[appendix]{appsubsec:higher_fidelity}

This subsection collects diagnostics obtained with the higher fidelity detector profile
\begin{align}
    C
    &=
    0.999,
    \\
    \epsilon
    &=
    0.0005,
    \\
    T_1
    &=
    T_2
    =
    T.
    \label{eq:higher_fidelity_detector_profile}
\end{align}
The corresponding effective symmetric readout error probability is
\begin{align}
    1-F_{\rm ro}
    &=
    9.995\times10^{-4}.
    \label{eq:higher_fidelity_readout_error}
\end{align}
Relative to the baseline profile in \cref{eq:baseline_detector_profile_1,eq:baseline_detector_profile_2,eq:baseline_detector_profile_3}, this reduces the effective readout error from $9.950\times10^{-3}$.
It also changes the preferred adaptive utility allocation.

The period $L$ fixes the exploration--detection allocation of the hybrid policy: one shot per period maximizes information gain, while the remaining $L-1$ shots use the Helstrom-like detection utility.
Among the periods tested for the higher fidelity profile, $L=8$ was favored by the representative $\overline{P}=0.7$ sensitivity and is used below.

\begin{figure}[t]
\centering
\includegraphics[width=\columnwidth]{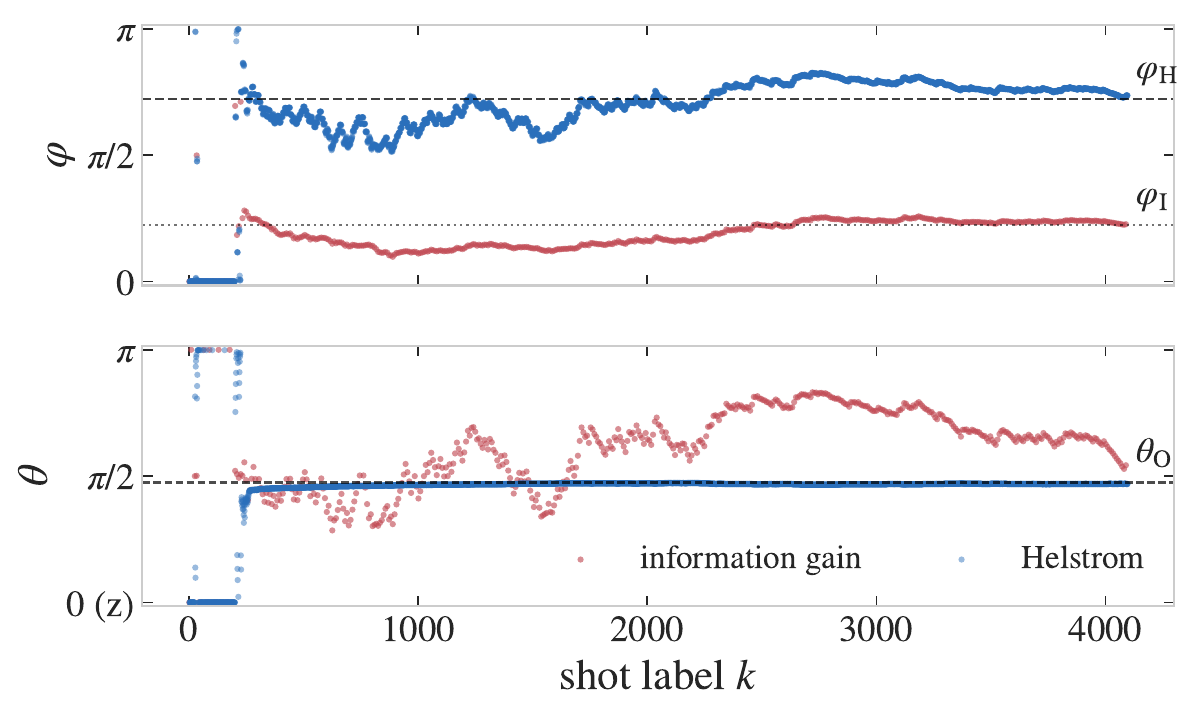}
\caption{Readout directions selected along a representative higher fidelity resonant signal trajectory at $(\Phi_R,\tilde{\varphi})=(0.19,0.7)$, with the azimuthal and polar angles shown in the upper and lower panels, respectively.
Red and blue points denote information gain and Helstrom shots.}
\label{fig:rabi_posterior_learning_angles}
\end{figure}

A representative resonant $L=8$ trajectory illustrates how the two utilities divide their roles.
The upper panel of \cref{fig:rabi_posterior_learning_angles} shows the readout azimuth: red points denote the periodic information gain shots, and blue points denote the intervening Helstrom shots.
Let $\sigma_{\tilde{\varphi},k}^2$ denote the local phase variance after localization.
The exact mutual information utility in \cref{eq:information_gain_utility} then has the narrow posterior expansion $U_{{\rm I},k}\simeq\sigma_{\tilde{\varphi},k}^2\mathcal I_{\tilde{\varphi}}/2$, where $\mathcal I_{\tilde{\varphi}}$ is the binary CFI obtained from the weak signal probability in \cref{eq:weak_transverse_response}.
Together with \cref{eq:helstrom_utility}, this gives
\begin{align}
    U_{{\rm H},k}(\vaphi)
    &\propto
    \left|\sin(\vaphi-\tilde{\varphi})\right|,
    \\
    \mathcal I_{\tilde{\varphi}}(\vaphi)
    &=
    \frac{
        [\partial_{\tilde{\varphi}}
        p_{\btheta}(+|t,T,\bm{n}_\perp(\vaphi))]^2
    }{
        p_{\btheta}(+|t,T,\bm{n}_\perp(\vaphi))
        [1-p_{\btheta}(+|t,T,\bm{n}_\perp(\vaphi))]
    }
    \notag\\
    &\simeq
    4b_\perp^2\Phi_R^2\cos^2(\vaphi-\tilde{\varphi}).
\end{align}
These expressions are maximized by
\begin{align}
    \vaphi_{\rm H}
    &=
    \tilde{\varphi}+\frac{\pi}{2}
    \pmod{\pi},
    \label{eq:localized_helstrom_azimuth}\\
    \vaphi_{\rm I}
    &=
    \tilde{\varphi}
    \pmod{\pi}.
    \label{eq:localized_information_azimuth}
\end{align}
Thus the Helstrom alignment maximizes signal--null separation, whereas the information gain alignment maximizes local sensitivity to the unknown phase.
The gray dotted and black dashed horizontal lines in the upper panel of \cref{fig:rabi_posterior_learning_angles} mark $\vaphi_{\rm I}$ and $\vaphi_{\rm H}$, respectively.

The lower panel of \cref{fig:rabi_posterior_learning_angles} shows the polar angle $\theta$ of the readout direction in the spherical parameterization defined by \cref{eq:spherical_readout_axis}.
For a known resonant signal, the Helstrom PVM in \cref{eq:helstrom_pvm} has an unoriented readout axis parallel to the difference between the signal and null Bloch vectors, whose acute polar angle branch is
\begin{align}
    \theta_{\rm O}(\Phi_R)
    &=
    \arctan\!\left[
        \frac{\eta_2(T)\sin\Phi_R}
        {\eta_1(T)(1-\cos\Phi_R)}
    \right].
    \label{eq:finite_signal_oracle_polar_angle}
\end{align}
For $\Phi_R=0.19$, the attenuation factors in \cref{eq:eta2_equal_times,eq:eta1_equal_times} give $\theta_{\rm O}\simeq1.49$, which agrees with the late Helstrom points.
This is close to $\pi/2$, as expected from the transverse Helstrom direction in the weak signal limit.

\begin{figure}[t]
\centering
\includegraphics[width=\columnwidth]{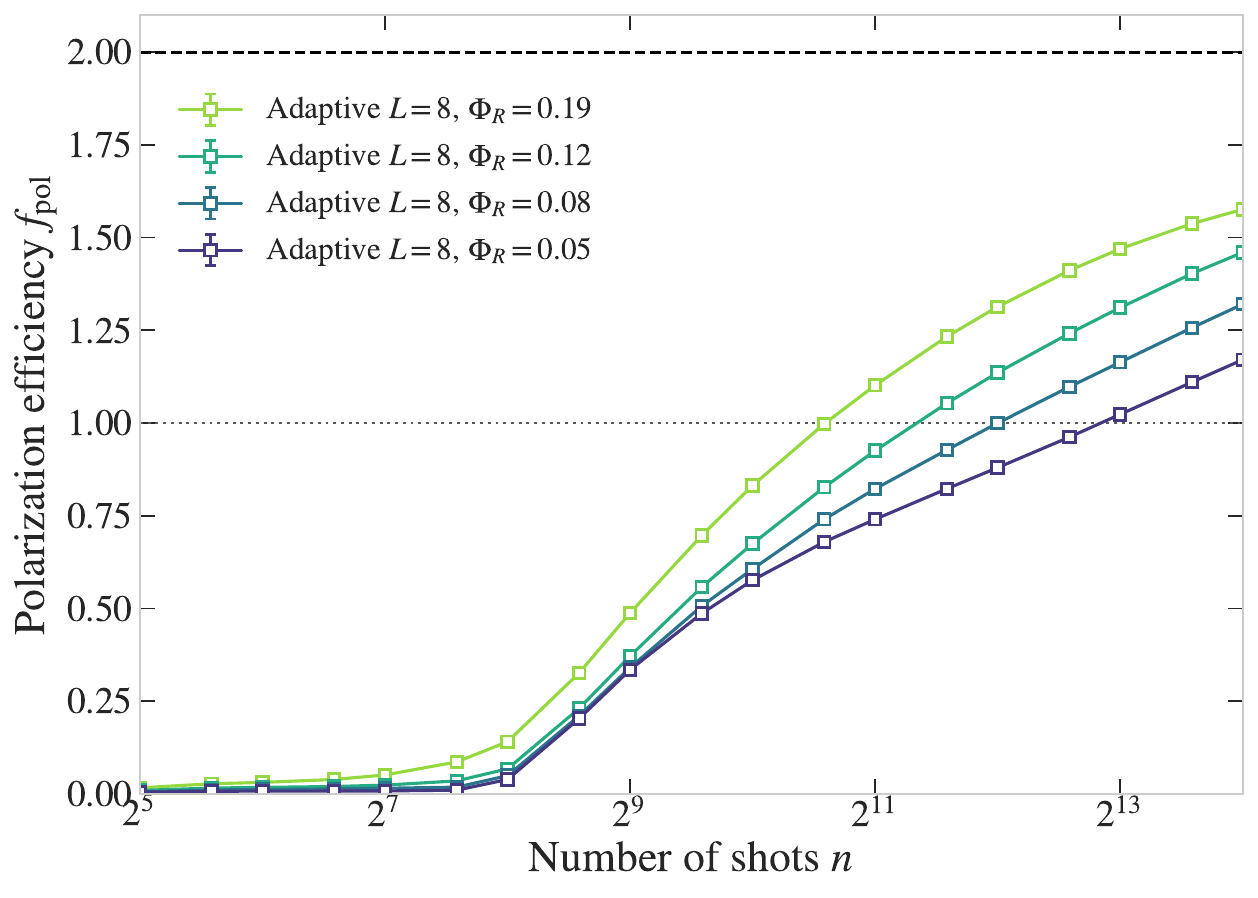}
\caption{Evolution of the geometric alignment diagnostic $f_{\rm pol}$ for higher fidelity $L=8$ signal records at four fixed Rabi amplitudes.
Each curve is averaged over $10^3$ pseudoexperiments with uniformly sampled injected phase.
The dotted and dashed horizontal lines mark the fixed-$x/y$ value $f_{\rm pol}=1$ and the known direction transverse value $f_{\rm pol}=2$, respectively.}
\label{fig:rabi_polarization_efficiency_evolution}
\end{figure}

To summarize transverse alignment over signal ensembles, we define the geometric diagnostic for these resonant records as
\begin{align}
    f_{\rm pol}
    &=
    \frac{2}{n}\sum_{k=0}^{n-1}
    \sin^2\theta_k
    \sin^2(\tilde{\varphi}-\vaphi_k),
    \label{eq:fpol_definition}
\end{align}
where $\tilde{\varphi}$ is the injected phase and $(\theta_k,\vaphi_k)$ specify the selected readout axis.
This quantity lies between $0$ and $2$: fixed-$z$ gives $0$, alternating fixed-$x/y$ gives $1$ for even $n$, and a perfectly aligned transverse oracle gives $2$.
As shown in \cref{fig:rabi_polarization_efficiency_evolution}, stronger signals acquire transverse alignment earlier, and all four curves eventually exceed the fixed-$x/y$ value.
The gap from $2$ reflects the cumulative contribution of initial exploration, periodic information gain shots, and residual departures from transverse oracle alignment.

\begin{figure}[t]
\centering
\includegraphics[width=\columnwidth]{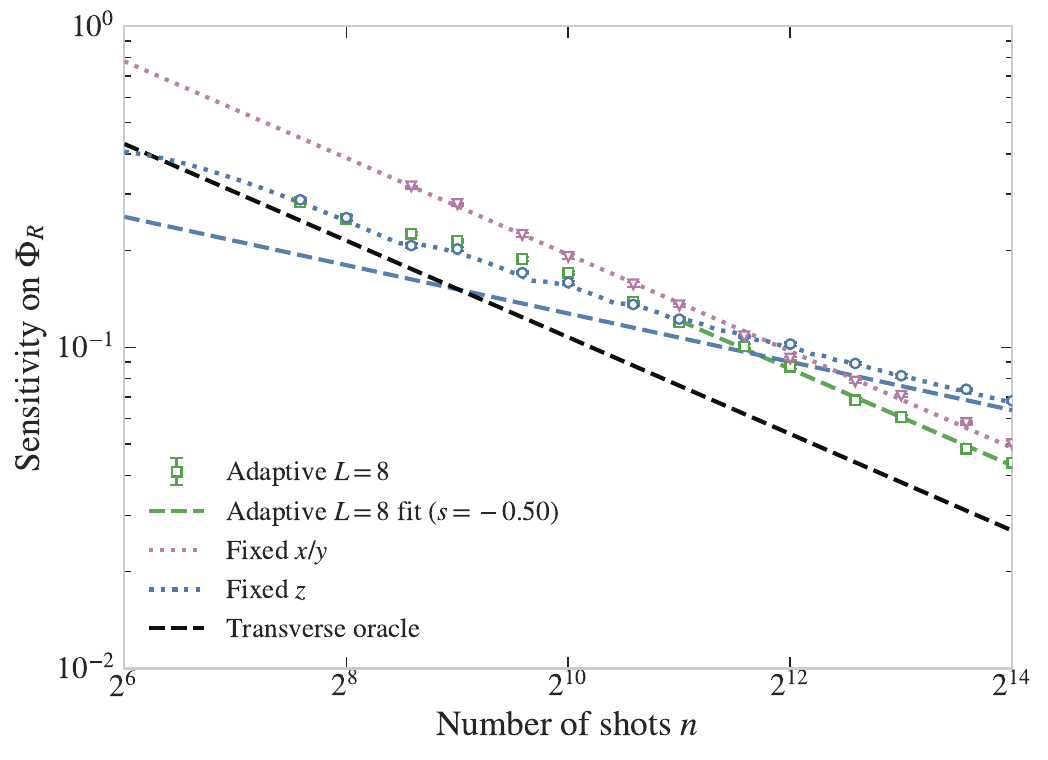}
\caption{Sensitivity to the Rabi amplitude at phase-averaged power $\overline{P}=0.7$ and $\alpha\simeq0.05$ for the higher fidelity detector profile in \cref{eq:higher_fidelity_detector_profile}.
Markers show Monte Carlo sensitivities for the $L=8$ hybrid, fixed-$z$, and fixed-$x/y$ policies, and the green dashed line is the adaptive tail fit.
The blue dashed and dotted curves are the Gaussian and finite background Poisson fixed-$z$ predictions; the purple dotted and black dashed curves are the fixed-$x/y$ and transverse oracle benchmarks.
Error bars show central $68\%$ intervals from the finite signal ensembles at fixed null thresholds.}
\label{fig:rabi_high_fidelity_sensitivity_p07}
\end{figure}

\cref{fig:rabi_high_fidelity_sensitivity_p07} shows the corresponding sensitivity at phase-averaged power $\overline{P}=0.7$ and $\alpha\simeq0.05$.
Compared with the baseline result in \cref{fig:rabi_discovery_sensitivity_p07}, the lower readout error keeps fixed-$z$ competitive up to shot counts of order $10^3$.
At larger $n$, the hybrid becomes more sensitive than both fixed-$z$ and fixed-$x/y$, with a fitted tail slope of $-0.50$ consistent with the $n^{-1/2}$ scaling.

\subsection{Detuned trajectory}
\label[appendix]{appsubsec:detuned_trajectory}

For this diagnostic at fixed $n$, we deliberately use a narrow detuning window to limit the posterior grid cost while retaining several phase wraps across the record; this is a computational choice rather than a physical prior for a broadband search.
For $n=4096$, define the half width of the detuning prior by
\begin{align}
    \Delta_{\max}
    &=
    \frac{50}{nT}
    \simeq
    0.0122\,T^{-1}.
    \label{eq:detuning_prior_half_width}
\end{align}
The three parameter prior used in this run is
\begin{align}
    \pi_0^{(3)}(\Phi_R,\Delta,\tilde{\varphi})
    &=
    \frac{\pi_0(\Phi_R,\tilde{\varphi})}{2\Delta_{\max}},
    \qquad
    |\Delta|\le\Delta_{\max},
    \label{eq:detuned_numerical_prior}
\end{align}
with zero density outside the detuning interval and with the two parameter prior $\pi_0(\Phi_R,\tilde{\varphi})$ given in \cref{eq:numerical_signal_prior}.
The full width $W_\Delta\equiv2\Delta_{\max}$ is approximately $0.39\%$ of the full Nyquist interval $2\pi T^{-1}$, but spans $100$ inverse record duration units $(nT)^{-1}$, and we discretize it using $480$ uniformly spaced detuning points.
The injected detuning is the representative interior value
\begin{align}
    \Delta_{\rm true}
    &=
    \frac{\Delta_{\max}}{2}.
    \label{eq:detuning_injected_value}
\end{align}
This benchmark produces an accumulated phase drift $\Delta_{\rm true}(nT)=25$, corresponding to approximately four full phase wraps over the record.
The complete injected point is $\btheta_{\rm true}=(\Phi_R,\Delta,\tilde{\varphi})=(0.19,\Delta_{\rm true},0.7)$, and the posterior is three dimensional.

\begin{figure}[t]
\centering
\includegraphics[width=\columnwidth]{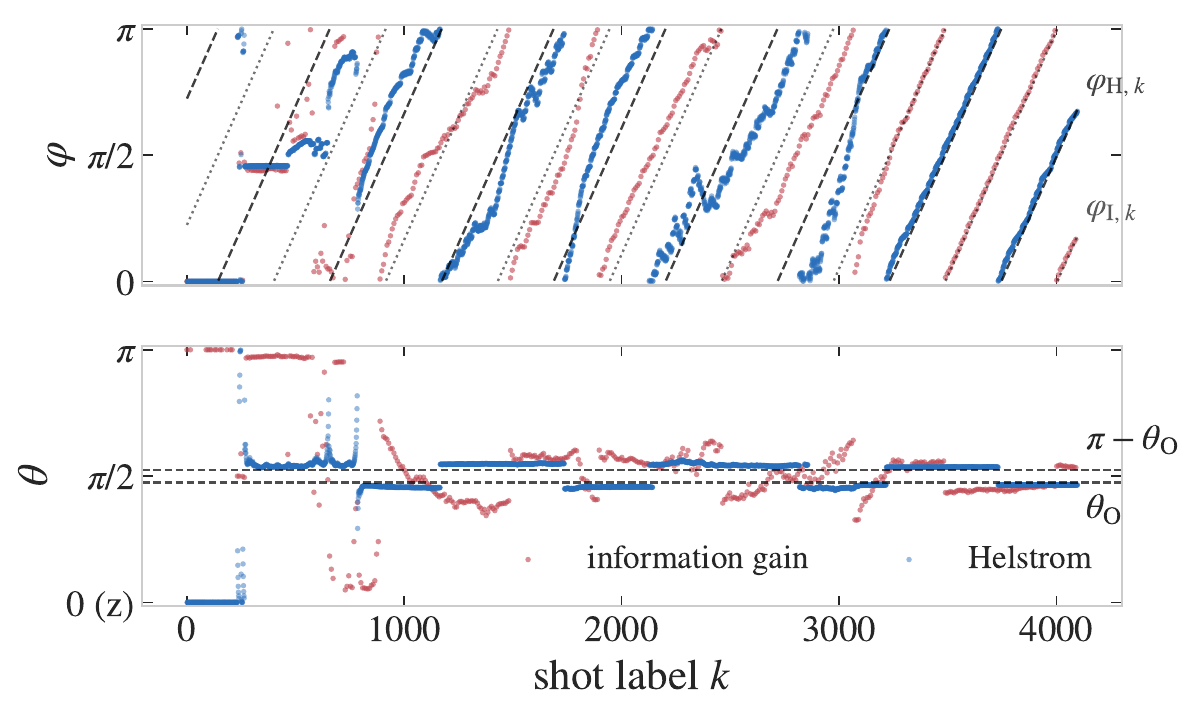}
\caption{Readout directions as in \cref{fig:rabi_posterior_learning_angles}, for a higher fidelity $L=8$ trajectory with $(\Phi_R,\Delta,\tilde{\varphi})=(0.19,25/(nT),0.7)$ and $n=4096$.
The dotted and dashed curves in the upper panel show the drifting target azimuths $\vaphi_{{\rm I},k}$ and $\vaphi_{{\rm H},k}$, respectively; the dashed horizontal lines in the lower panel mark the equivalent polar branches $\theta_{\rm O}$ and $\pi-\theta_{\rm O}$.}
\label{fig:rabi_posterior_learning_detuned_angles}
\end{figure}

At shot $k$, the effective phase is $\tilde{\varphi}_{{\rm eff},k}=\tilde{\varphi}_{\rm eff}(t_k,T;\btheta)$ as defined in \cref{eq:effective_rabi_phase_angle}.
The two localized target azimuths generalizing \cref{eq:localized_helstrom_azimuth,eq:localized_information_azimuth} are
\begin{align}
    \vaphi_{{\rm H},k}
    &=
    \tilde{\varphi}_{{\rm eff},k}+\frac{\pi}{2}
    \pmod{\pi},
    \\
    \vaphi_{{\rm I},k}
    &=
    \tilde{\varphi}_{{\rm eff},k}
    \pmod{\pi}.
    \label{eq:detuned_target_azimuths}
\end{align}
The late adaptive readouts in \cref{fig:rabi_posterior_learning_detuned_angles} track these drifting directions rather than approaching constant resonant azimuths.
A projective readout axis is unoriented: $\bm{n}$ and $-\bm{n}$ define the same PVM up to interchange of the two outcome labels, so its spherical coordinates obey $(\theta,\vaphi)\sim(\pi-\theta,\vaphi+\pi)$.
Consequently, when the azimuth is represented modulo $\pi$, the known signal Helstrom reference has the two equivalent polar branches $\theta_{\rm O}$ and $\pi-\theta_{\rm O}$.
For the injected detuning, replacing $\Phi_R$ in \cref{eq:finite_signal_oracle_polar_angle} by $\Phi_{\rm eff}=\Phi_R\operatorname{sinc}(\Delta_{\rm true}T/2)$ gives $\theta_{\rm O}\simeq1.49$.

\bibliography{bib}
\bibliographystyle{apsrev4-2}

\end{document}